\documentclass[letterpaper,11pt]{article}

\usepackage{jheppub} 
\usepackage[T1]{fontenc}
\usepackage{lmodern}
\usepackage{amsmath}
\usepackage{amssymb}
\usepackage{graphicx}
\usepackage{xspace}
\usepackage[small]{subfigure}
\usepackage{color}
\usepackage{multirow}
\usepackage{empheq}
\usepackage[absolute]{textpos}

\newcommand{\nn}{\nonumber} 
\newcommand{\bn}{{\bar n}}
\newcommand{\nz}{n_z}
\newcommand{\bnz}{\bar{n}_z}
\newcommand{\mcdot}{\!\cdot\!}

\newcommand{\be}{\begin{equation}}
\newcommand{\ee}{\end{equation}}

\newcommand{\sing}{\text{sing} }
\newcommand{\ns}{\text{ns}}

\newcommand{\bra}[1]{\left\langle #1\right\rvert}
\newcommand{\ket}[1]{\left\lvert #1\right\rangle}

\newcommand{\minus}{\!-\!}
\newcommand{\plus}{\!+\!}

\newcommand{\df}{\mathrm{d}}
\newcommand{\Lqcd}{\Lambda_{\text{QCD}}}

\newcommand{\GeV}{\text{ GeV}}

\newcommand{\as}{\alpha_s}
\newcommand{\MSbar}{\overline{\text{MS}}}

\newcommand{\cO}{\mathcal{O}}
\newcommand{\cM}{\mathcal{M}}

\newcommand{\cH}{\mathcal{H}}
\newcommand{\cF}{\mathcal{F}}
\newcommand{\cL}{\mathcal{L}}
\newcommand{\cA}{\mathcal{A}}

\newcommand{\cB}{\mathcal{B}}

\newcommand{\cK}{\mathcal{K}}

\newcommand{\e}{\epsilon}
\newcommand{\ve}{\varepsilon}

\newcommand{\eq}[1]{Eq.~\eqref{#1}}
\newcommand{\eqs}[2]{Eqs.~\eqref{#1} and \eqref{#2}}
\newcommand{\eqss}[3]{Eqs.~\eqref{#1}, \eqref{#2}, and \eqref{#3}}
\newcommand{\eqsss}[4]{Eqs.~\eqref{#1}, \eqref{#2}, \eqref{#3}, and \eqref{#4}}
\renewcommand{\sec}[1]{Sec.~\ref{sec:#1}}
\newcommand{\ssec}[1]{Sec.~\ref{ssec:#1}}

\newcommand{\appx}[1]{App.~\ref{app:#1}}
\newcommand{\fig}[1]{Fig.~\ref{fig:#1}}
\newcommand{\figs}[2]{Figs.~\ref{fig:#1} and \ref{fig:#2}}

\newcommand{\m}{{a}}
\newcommand{\B}{{b}}
\newcommand{\CM}{{c}}
\newcommand{\taun}{\tau_1}
\newcommand{\taum}{\tau_{1}^\m}
\newcommand{\tauB}{\tau_{1}^\B}
\newcommand{\tauCM}{\tau_{1}^\CM}

\newcommand{\hattau}{\hat\tau}
\newcommand{\taumax}{\tau^\text{max}}

\newcommand{\qB}{q_B}
\newcommand{\qJ}{q_J}

\allowdisplaybreaks[1]

\DeclareMathOperator{\Li}{Li}

\begin{document}

\begin{textblock}{2}(12.7,3.1)%
  MIT-CTP 4567
\end{textblock}%
\begin{textblock}{2}(12.7,3.4)%
  LA-UR-14-25039
\end{textblock}%
\begin{textblock}{3}(12.7,3.7)%
  NSF-KITP-14-079
\end{textblock}%

\title{ \Large Analytic Calculation of 1-Jettiness in DIS at $\cO(\as)$}

\author[a]{Daekyoung Kang,}
\author[b]{Christopher Lee,}
\author[a]{and Iain W. Stewart}

\affiliation[a]{Center for Theoretical Physics,  Massachusetts Institute of
Technology, Cambridge, MA 02139, USA}
\affiliation[b]{Theoretical Division, MS B283, Los Alamos National Laboratory, 
Los Alamos, NM 87545, USA }

\emailAdd{kang1@mit.edu}
\emailAdd{clee@lanl.gov}
\emailAdd{iains@mit.edu}

\abstract{ 
We present an analytic $\cO(\as)$ calculation of cross sections in
deep inelastic scattering (DIS) dependent on an event shape,
1-jettiness, that probes final states with one jet plus initial state
radiation. This is the first entirely analytic calculation for a DIS
event shape cross section at this order.  We present results for the
differential and cumulative 1-jettiness cross sections, and express
both in terms of structure functions dependent not only on the usual
DIS variables $x,Q^2$ but also on the 1-jettiness $\tau$.  Combined
with previous results for log resummation, predictions are obtained
over the entire range of the 1-jettiness distribution.  }

\maketitle

\section{Introduction}
\label{sec:Intro}

In high energy colliders, jet production plays an important role in
probing the strong interaction, hadron structure, dense media, and new
particles beyond the Standard Model.  Thus predicting jet production
cross sections and jet structure is one of the important tasks of
Quantum Chromodynamics (QCD).  Jet algorithms
\cite{Catani:1991hj,Catani:1993hr,Ellis:1993tq,Dokshitzer:1997in,Salam:2007xv,Cacciari:2008gp}
allow exclusive study of jets and definitions of cross sections with a
definite number of jets. However, they also introduce various
parameters like jet radii or sizes and jet vetoes, which require more
effort to predict accurately in analytic calculations in QCD. Event
shapes \cite{Dasgupta:2003iq} provide a simple, inclusive way to
identify final states that are jet-like, and can often be predicted to
very high accuracy in QCD.  Thrust in $e^+e^-$ collisions
\cite{Farhi:1977sg} is a classic example of a two-jet event shape that
has been extensively studied in both theory and experiment.  Thrust
cross sections in $e^+e^-$ have been predicted to very high accuracy,
N$^3$LL$+\cO(\as^3)$ in resummed and fixed-order perturbation theory
\cite{GehrmannDeRidder:2007bj,GehrmannDeRidder:2007hr,Weinzierl:2008iv,Weinzierl:2009ms,Becher:2008cf,Abbate:2010xh},
along with rigorous treatments of nonperturbative power corrections
\cite{Lee:2006nr,Abbate:2010xh,Mateu:2012nk}, that have led to
unprecedented 1\%-level precision in determinations of the strong
coupling constant $\as$ from $e^+e^-$ event shape data
\cite{Becher:2008cf,Chien:2010kc,Abbate:2010xh}.

Event shapes in DIS have also been studied but not as extensively as
in $e^+e^-$, and the theoretical accuracy has yet to catch up to the
same level. Two versions of DIS thrust have been defined and measured
in H1 and ZEUS experiments at HERA
\cite{Adloff:1997gq,Adloff:1999gn,Aktas:2005tz,Breitweg:1997ug,Chekanov:2002xk,Chekanov:2006hv}
and they have been calculated up to next-leading-logarithmic accuracy
(NLL) at resummed order and numerically to $\cO(\as^2)$ at fixed order
\cite{Antonelli:1999kx,Dasgupta:2002dc}.  The measured DIS thrusts
involve non-global logarithms (NGLs), which present a theoretical
obstacle to higher order accuracy
\cite{Dasgupta:2001sh,Dasgupta:2002dc}.

Versions of thrust such as $e^+e^-$ thrust and the DIS thrust $\tau_Q$
defined in \cite{Antonelli:1999kx} do not suffer from NGLs.  A class
of event shapes called $N$-jettiness $\tau_N$ \cite{Stewart:2010tn} is
a generalization of these versions of thrust and are applicable in
different collider environments, including $e^+e^-$, lepton-hadron,
and hadron-hadron collisions. $\tau_N$ measures the degree of
collimation of final-state hadrons along $N$ light-like directions in
addition to any initial-state radiation (ISR) along the incoming beam
directions.  In a number of recent papers
\cite{Kang:2013nha,Kang:2012zr,Kang:2013wca}, factorization theorems
for various versions of 1-jettiness $\tau$ in DIS have been derived by
using soft collinear effective theory (SCET)
\cite{Bauer:2000ew,Bauer:2000yr,Bauer:2001ct,Bauer:2001yt,Bauer:2002nz}.
To date, this has enabled log resummation up to NNLL accuracy
\cite{Kang:2013nha,Kang:2012zr,Kang:2013wca}, which is one order
higher in resummed accuracy than earlier results
\cite{Antonelli:1999kx,Dasgupta:2002dc}.

The SCET results \cite{Kang:2013nha,Kang:2012zr,Kang:2013wca}
correctly capture and resum all logarithmic terms (singular), while
non-logarithmic terms (nonsingular) can be obtained from fixed-order
computations in full QCD. The full cross section is the sum of
singular and nonsingular parts and can be written as
\be\label{xsection}
\sigma^\text{full}(\tau) 
=\sigma^\sing (\tau) +\sigma^\ns(\tau)\,.
\ee
The singular part is factorized in terms of hard, jet, beam, and soft
functions each of which depends on the relevant energy scale for each
mode \cite{Kang:2013nha,Kang:2012zr,Kang:2013wca}. This separation of
scales and renormalization group (RG) evolution between them allows
for resummation of the large logarithms in the fixed-order expansion
of the cross section. When the RG evolution is turned off in the
singular part, the full cross section reduces to the ordinary
fixed-order result. The nonsingular part is obtained by subtracting
the fixed-order singular part from the fixed-order cross section.
 
For an accurate prediction over the entire range of an event shape
distribution, both fixed-order and resummed calculations should be
consistently improved.  While NNLL resummation of the singular part in
\eq{xsection} has been performed for three different versions of DIS
1-jettiness in \cite{Kang:2013nha} and another version in
\cite{Kang:2012zr,Kang:2013wca}, no analytic computations of the
non-singular part at $\cO(\as)$ or above have yet been performed.  In
\cite{Kang:2013lga} an $\cO(\as)$ result has been numerically obtained
for a version of 1-jettiness that requires a jet algorithm to
determine the jet momentum. Such a numerical approach is appropriate
for such cases and allows for the flexibility of using different jet
algorithms.

In this paper, we carry out the first analytic $\cO(\as)$ calculation
for a DIS event shape.  We choose the version of 1-jettiness called
$\tau_1^b$ in~\cite{Kang:2013nha}, which groups final-state particles
into back-to-back hemispheres in the Breit frame and is the same as
the DIS thrust called $\tau_Q$ in Ref.~\cite{Antonelli:1999kx}.  It
can be written as
\be \label{tau1b}
\tau = \frac{2}{Q^2}\sum_{i\in X}\min\{ \qB^b\mcdot p_i,\qJ^b\mcdot p_i\}
        \stackrel{\text{Breit}}{=}  1-\frac2Q \sum_{i\in \cH_J} p_{i\,z}
 \,,
\ee
where $p_i$ is the momentum of the $i$th particle in the final state,
and $Q^2\equiv -q^2$ is determined by the momentum transfer $q$ in the
event. The reference vectors are defined by $q_B^b = xP$ and $q_J^b =
q+xP$, where $P$ is the proton momentum. In the Breit frame these
vectors point exactly back-to-back. The second definition in
\eq{tau1b} is valid in the Breit frame, and requires measuring the $z$
components $p_{i \,z}$ of momenta of particles only in the jet hemisphere
(current hemisphere) $\cH_J$. The definition in \eq{tau1b} differs
from the measured version $\tau^{\text{H1}}=1-T_\gamma^{\text{ZEUS}}$
\cite{Aktas:2005tz,Chekanov:2006hv} in normalization (replacing $2/Q$
by $1/E_{\rm hemi}$ where $E_{\rm hemi}$ is the hemisphere energy).

We present our results in terms of fixed-order singular and
nonsingular parts of the cross section as in \eq{xsection}. They can
be put in a simple form which can easily be implemented in other
analyses.  The main new results of this paper are the nonsingular
1-jettiness structure functions given by \eq{AiBi}.

We also show numerical results with perturbative uncertainties by
varying scales at the HERA energy.  Our results could be compared to
existing HERA data
\cite{Adloff:1997gq,Adloff:1999gn,Aktas:2005tz,Breitweg:1997ug,Chekanov:2002xk,Chekanov:2006hv}
or to future EIC data \cite{Accardi:2012hwp}. In \cite{Kang:2013nha},
by comparing our resummed singular cross section to the known
fixed-order total cross section, we estimated that the nonsingular
corrections would amount to several percent of the total cross
section, and this expectation is borne out by our computations here.

The paper is organized as follows: In \sec{jettiness} we briefly
review the relevant kinematic variables in DIS and our definition of
1-jettiness, and express the cross section in terms of structure
functions.  In \sec{setup}, we outline the basic steps of the
$\cO(\as)$ computation including the phase space for 1-jettiness and
perturbative matching of the hadronic tensor onto parton distribution
functions (PDFs).  \sec{Fi} contains our main results, analytic
$\cO(\as)$ expressions for 1-jettiness structure functions.  Details
of the fixed-order calculation are given in \appx{plus},
\appx{Wparton} and \appx{Wns}.  In \sec{results} numerical results are
given for structure functions at $\cO(\as)$ fixed-order accuracy and
cross sections at NLL$'+\cO(\as)$ resummed accuracy. Basic details
entering the resummation of the singular terms are reviewed in
\appx{profile} and \appx{resum} for convenience. Finally, we will
conclude in \sec{conclusions}.

\section{1-Jettiness in DIS} \label{sec:jettiness}

In this section we review DIS kinematic variables that will be used
throughout the paper and the definition of the 1-jettiness $\tau$
cross section in DIS, whose computation will be the main prediction of
our paper.

\subsection{Kinematic Variables}
\label{ssec:kinematics}

In DIS, an incoming electron with 4-momentum $k$ scatters off a proton
with momentum $P$ by exchanging a virtual photon\footnote{For
  simplicity we do not include $Z$ boson exchange in this paper. See
  \cite{Kang:2013nha} for appropriate modifications.}  with a large
momentum transfer $q=k-k'$, where $k'$ is the momentum of the outgoing
electron.  Because the photon has spacelike momentum it has a negative
virtuality, and one can define the positive definite quantity
\be
Q^2 \equiv -q^2\,.
\ee
$Q$ sets the momentum scale of the scattering.  We will be interested
in hard scattering, where $Q\gg \Lqcd$.  A dimensionless quantity $x$
called the Bj\"{o}rken scaling variable is defined by
\be
x \equiv -\frac{q^2}{2P\mcdot q}
  = \frac{Q^2}{2P\mcdot q},
\label{x}
\ee
which ranges between $0\le x \le 1$.  Another dimensionless quantity
$y$ is defined by $y \equiv \frac{2P\cdot q}{2P\cdot k}$, which ranges
between $0\le y\le 1$.  This variable $y$ represents the energy loss
of the electron in the proton rest frame.  The three variables $x$,
$y$, and $Q^2$ are related to one another via $Q^2=xys$, where
$s=(P+k)^2$ is the total invariant mass of the incoming particles.
The total momentum of the final state $X$ is $p_X=q+P$ and the
invariant mass is given by $p_X^2 = \frac{1-x}{x}Q^2$. For large $x$
very near 1, the final state consists of a single tightly collimated
jet of hadrons. This region has been analyzed in SCET in, \emph{e.g.},
\cite{Manohar:2003vb,Chay:2005rz,Becher:2006mr,Chen:2006vd,Fleming:2012kb}. We
will instead be interested in different region where two or more
energetic jets can occur. This occurs in the ``classic'' region where
$x$ has a generic size $x\sim1-x\sim 1$ such that $p_X^2\sim Q^2$.

Although the cross section we compute is frame independent, there is a
convenient frame in which to perform the intermediate steps of the
calculation.  This is the Breit frame, where the virtual photon with
momentum $q^\mu$ is purely spacelike, and collides with the proton
with momentum $P^\mu$ along the $z$ direction.  In this frame the
virtual photon and the proton have momenta
\be\label{qP}
q^\mu = Q\frac{\nz^\mu-\bnz^\mu}{2}\,,\quad
P^\mu = \frac{Q}{x}\frac{\bnz^\mu}{2}\,,
\ee
where $\nz=(1,0,0,1)$ and $\bnz=(1,0,0,-1)$.

\subsection{$1$-Jettiness}
\label{ssec:1jettiness}

To probe the number of jets in the final state produced at a given
value of $x$ and $Q$, an additional measurement needs to be made. A
simple event shape that accomplishes this is the $N$-jettiness
\cite{Stewart:2010tn}, a generalization of the thrust
\cite{Farhi:1977sg}.  It is defined by the sum of projections of
final-state particle momenta onto whichever axis is closest among $N$
jet and $N_b$ beam axes, where $N_b=0$ for $e^+e^-$ collisions, $1$
for $ep$ DIS, and 2 for $pp$ collisions. The $N$-jettiness $\tau_N$ is
designed so that it becomes close to zero for an event with $N$
well-collimated jets in the final state away from any hadronic beam
axes.  For example, 1-jettiness in DIS is defined by one jet and one
beam axis:
\be \label{tau1def}
\taun \equiv \frac{2}{Q^2}\sum_{i\in X}\min\{ \qB\mcdot p_i,\qJ\mcdot p_i\}\,,
\ee
where $\qB,\qJ$ are lightlike four-vectors along the beam and jet
directions.  It is natural to choose $q_B$ along the proton direction.
One can consider several options for choosing $\qJ$.  In
\cite{Kang:2013nha}, we defined three versions of 1-jettiness $\taum$,
$\tauB$, and $\tauCM$ distinguished by different choices for $\qJ$:
(a) $\qJ^a$ \emph{a}ligned along the jet axis determined by a jet
algorithm, (b) $\qJ^b$ along the $z$ axis in the \emph{B}reit frame,
and (3) $\qJ^c$ along the $z$ axis in the \emph{c}enter-of-momentum
(CM) frame.

In this paper we consider $\tauB$ for which $\qB^b$ and $\qJ^b$ are
given by
\be
\label{Breitvectors}
 {\qB^b}^\mu = xP^\mu \,, \quad {\qJ^b}^\mu = q^\mu +xP^\mu \,.
\ee
As shorthand, we drop both superscript and subscript in $\tauB$
 throughout the remainder of the paper.
\be
\label{tau}
\tau\equiv\tauB.
\ee
In the Breit frame, the vectors $q_{B,J}^b$ point exactly back-to-back
with equal magnitude: \be q_B^b \overset{\text{Breit}}{=} Q
\frac{\bn_z}{2} \,,\qquad q_J^b \overset{\text{Breit}}{=} Q
\frac{n_z}{2}\,, \ee and divide particles in the final state into two
equal hemispheres. One is the ``beam'' or ``remnant'' hemisphere
$\cH_B$ in the $-z$ direction and the other is the ``jet'' or
``current'' hemisphere $\cH_J$ in the $+z$ direction.

The 1-jettiness $\tau$ in \eq{tau} has an experimental advantage in
that it can be determined by measuring only one of the hemispheres,
namely $\cH_J$. This avoids having to measure the whole final state
including the beam remnants, a technical difficulty in experiments
such as H1 and ZEUS at HERA. By using $q$ and $P$ in the Breit frame
in \eq{qP}, the 1-jettiness can be written in the form
\be
\label{tauQ}
\tau \stackrel{\text{Breit}}{=} 
 \frac{1}{Q}\sum_{i\in X}\min\{ \bnz\mcdot p_i, \nz\mcdot p_i\}
=1-\frac{2}{Q}\sum_{i\in \cH_J} p_{i\,z}
\,.
\ee
 We used momentum conservation $p_B=p_X-p_J$, where $p_B=\sum_{i\in
   \cH_B} p_{i}$ and $p_J=\sum_{i\in \cH_J} p_{i}$.  The definition
 \eq{tauQ} directly corresponds to the thrust $\tau_Q$ in DIS defined
 in \cite{Antonelli:1999kx}.  We can obtain the physical upper limit
 on $\tau$ using the kinematic constraints that the jet momentum
 $p_{J\,z}\ge 0$ has to be positive, and that the beam momentum's $z$
 component is negative, so that $p_{J\,z} = p_{X\,z} - p_{B\,z} \ge
 p_{X\,z}=Q\,(2x-1)/(2x)$. These conditions imply the upper limits on
 $\tau$:
\begin{align}
\label{tmax}
& \taumax=
\begin{cases}
1 & x \le 1/2
\,,\\
\frac{1-x}{x} & x \ge 1/2
\,.\end{cases}
\end{align}

\subsection{1-Jettiness Cross Section}
\label{ssec:cs}

The 1-jettiness cross section can be expressed in terms of leptonic
and hadronic tensors:
\be
\label{tauCS}
\frac{d\sigma}{dx \, dQ^2 \, d\tau} = L_{\mu\nu}(x,Q^2)W^{\mu\nu}(x,Q^2,\tau)\,,
\ee
where the lepton tensor for a photon exchange is given by
\be
\label{Ltensor}
L_{\mu\nu}(x,Q^2)= - \frac{\alpha^2 }{2x^2 s^2} 
\bigg[g_{\mu\nu}-2\frac{k^\mu k'^{\nu} +k'^{\mu} k^{\nu}}{Q^2}
\bigg]
\,,
\ee
where $k$ and $k'$ are incoming and outgoing electron momenta and
$\alpha\equiv \alpha_{\text{em}}$.  The hadronic tensor is the
current-current correlator in the proton state,

\be \label{WxQ2tau}
 W^{\mu\nu}(x,Q^2,\tau) 
 = \int \, d^4 x\,  e^{iq\cdot x} \!\bra{P} \! J^{\mu\dag} (x) 
 \delta(\tau - \hattau) J^\nu(0)\! \ket{P},
\ee
where $\hattau$ is a 1-jettiness operator that measures 1-jettiness
when it acts on the final states, which we defined in
\cite{Kang:2013nha}, based on the construction of event shape
measurement operators from the energy-momentum tensor in
\cite{Sveshnikov:1995vi,Cherzor:1997ak,Belitsky:2001ij,Bauer:2008dt}. In
this paper, we consider only the vector current $J^\mu= \sum_f Q_f
\bar q_f\gamma^\mu q_f$. Previously we worked with both vector and
axial-vector currents, see \cite{Kang:2013nha} for the appropriate
generalizations.\footnote{Here we will include the quark charges $Q_f$
  in the hadronic current, whereas in \cite{Kang:2013nha} they were in
  $L_{\mu\nu}$.}  Because the hadronic tensor depends only on the two
momenta $P$ and $q$, it can be decomposed into products of tensors
constructed with $g_{\mu\nu}$, $P^\mu$, $q^\mu$ and structure
functions depending on $x$, $Q$, and $\tau$.  In our conventions,
\be
\label{Wdecomposition}
W^{\mu\nu}(x,Q^2,\tau) = 4\pi \biggl[ T_1^{\mu\nu} \cF_1(x,Q^2,\tau) + T_2^{\mu\nu} \frac{\cF_2(x,Q^2,\tau)}{P\cdot q}\biggr]\,,
\ee
where the two tensor structures that appear are:
\be
T_1^{\mu\nu} = -g^{\mu\nu} + \frac{q^\mu q^\nu}{q^2} \,,\quad T_2^{\mu\nu} = \Bigl( P^\mu - q^\mu \frac{P\cdot q}{q^2}\Bigr) \Bigl( P^\nu - q^\nu \frac{P\cdot q}{q^2}\Bigr) \,,
\ee
which arise from parity conservation and the Ward identity $q_\mu
W^{\mu\nu} = W^{\mu\nu} q_\nu = 0$.  If we considered parity-violating
scattering, \emph{e.g.} with neutrinos, a third tensor $T_3^{\mu\nu} =
-i\epsilon^{\mu\nu\alpha\beta}q_\alpha P_\beta$ would also appear.

In terms of the structure functions appearing in \eq{Wdecomposition},
the cross section \eq{tauCS} can be expressed
\begin{align}
\label{sigmaF1L}
 \frac{d\sigma}{dxdQ^2d\tau}
&=
\frac{4\pi \alpha^2}{Q^4}\left[ (1+(1-y)^2)\cF_1 +\frac{1-y}{x}\cF_L\right]\,,
\end{align}
where $\cF_L \equiv \cF_2 - 2x\cF_1$. We use calligraphic font for the
structure functions in the differential $\tau$ cross section. We will
use Roman font $F_{1,L}$ for the structure functions in the integrated
cross section, see \eqs{cumulant}{Fc}.

The structure functions $\cF_i$ can be obtained by contracting the
hadronic tensor with the metric tensor or the proton momentum $P^\mu$:
\begin{align}
\label{Fi-def}
\cF_1(x,Q^2,\tau)&=\frac{1}{8\pi (1-\epsilon)} \bigg(-g_{\mu\nu} W^{\mu\nu}+\frac{4x^2}{Q^2} P_\mu P_\nu W^{\mu\nu}\bigg)
\,,\nn\\
\cF_L(x,Q^2,\tau)&=\frac{2x^3}{\pi Q^2} P_\mu P_\nu W^{\mu\nu}
 \,.
\end{align}
We choose to always work with expressions in $D=4-2\epsilon$
dimensions for the vector indices $\mu$ and $\nu$, so the factor of
$1/(1-\epsilon)$ in $\cF_1$ comes from taking the contraction
$g_{\mu\nu}T_1^{\mu\nu}$. The contraction $P_\mu P_\nu W^{\mu\nu}$
turns out to be finite as $\epsilon\to 0$.  The standard structure
functions depend just on $x$ and $Q^2$, while those in \eq{Fi-def} are
additionally differential in $\tau$.  The structure functions can be
written in terms of singular and nonsingular parts as we did for the
cross section in \eq{xsection}.  We will present singular and
nonsingular parts of the structure functions in \sec{Fi}, from which
one easily obtains the corresponding parts of the cross section via
\eq{sigmaF1L}.

\section{Setup of the Computation}
\label{sec:setup}
In this section we outline the basic steps in the $\cO(\as)$
computation of the 1-jettiness cross section in DIS. First, we
describe the standard perturbative matching procedure for the hadronic
tensor onto PDFs, which allows us to compute the matching coefficients
using partonic external states. Then we set up the phase space
integrals in the Breit frame in which the intermediate steps of the
computation are simpler.  The final results are frame independent. The
reader who wishes to skip these details may turn directly to the final
results in \sec{Fi} and \sec{results}.

\subsection{Perturbative Matching}
\label{ssec:matching}

Here, we describe the matching procedure to determine the
short-distance coefficients that match the hadronic tensor
$W_{\mu\nu}(x,Q^2,\tau)$ onto PDFs. By using the operator product
expansion (OPE) the hadronic tensor can be written in the factorized
form
\label{Wtensor}
\begin{align}
W^h_{\mu\nu}(x,Q^2,\tau)
=\sum_{i\in \{q,\bar q,g\}} \int_x^1 \frac{d\xi}{\xi} f_{i/h}(\xi,\mu)\, w^{i}_{\mu \nu}\Bigl ( \frac{x}{\xi},Q^2,\tau,\mu\Bigr)
\,\,\biggl[1+\cO \Bigl(\frac{\Lqcd}{Q\tau}\Bigr)\biggr]
\label{W-factor}
\,,\end{align}
where $f_{i/h}$ is the PDF for a parton $i\in \{q,\bar q, g\}$ in a
hadron $h$, and $w_{\mu \nu}^i$ is the short-distance coefficient that
we will determine by perturbative matching.\footnote{The first power
  correction $\sim \Lambda_{\rm QCD}/(Q\tau)$ in Eq.~(\ref{Wtensor}),
  as well as higher-order terms $\sim [\Lambda_{\rm QCD}/(Q\tau)]^k$,
  are all described by the leading-order soft function in the
  small-$\tau$ factorization theorem~\cite{Kang:2013nha}.  For $\tau
  \sim \Lambda_{\rm QCD}/Q$ the leading power corrections not
  contained in the factorization theorem are ${\cal O}(\Lambda_{\rm
    QCD}/Q)$, while for large $\tau \sim 1$ the leading power
  corrections are ${\cal O}(\Lambda_{\rm QCD}^2/Q^2)$. }
The sum over $q,\bar q$ goes over all light flavors
$f\in\{u,d,s,c,b\}$ at the collision energies we consider. On the
left-hand side of \eq{W-factor}, the superscript $h$ specifies the
hadron in the initial state. The coefficients $w_{\mu\nu}^i$ however
can be computed in perturbation theory using any appropriate initial
state including partonic ones. This is what we shall describe in this
subsection.

The factorization theorem for an initial parton $j$ is given by
\begin{align}\label{Wi-factor}
W_{\mu\nu}^j
&=\sum_{i\in \{q,\bar q,g\}}  f_{i/j}\otimes w_{\mu\nu}^{i}\,,
\end{align}
where the arguments are implicit for simplicity and the convolution
integral over $\xi$ in \eq{Wi-factor} is replaced by the symbol
$\otimes$.  By comparing \eq{Wi-factor} to \eq{W-factor}, all implicit
notations can easily be recovered.  We determine the coefficients
$w^i$ by computing the $W_{\mu\nu}^j$, which are defined by
\eq{WxQ2tau} but with a quark or antiquark $j=q,\bar q$ or gluon $j=g$
in the initial state, and subtracting out the partonic PDFs, which are
IR divergent and require a regulator. We perform this computation
using dimensional regularization and defining PDFs in the $\MSbar$
scheme.

Working to $\cO(\as)$, and denoting the order $\as^n$ piece of each
function by a superscript $^{(n)}$, \eq{Wi-factor} becomes
\be
\label{Wi-matching}
\begin{split}
W_{\mu\nu}^{j\,(0)} &= \sum_{i\in \{q,\bar q,g\}}f_{i/j}^{(0)} \otimes w_{\mu\nu}^{i\,(0)} \,,\\
W_{\mu\nu}^{j\,(1)} &= \sum_{i\in \{q,\bar q,g\}}\Bigl[f_{i/j}^{(1)} \otimes w_{\mu\nu}^{i\,(0)} 
    + f_{i/j}^{(0)} \otimes w_{\mu\nu}^{i\,(1)}\Bigr]\,.
\end{split}
\ee
Using $\MSbar$ and using $\e$ to regulate IR divergences, the partonic
PDFs to $\cO(\as)$ are given by (see, \emph{e.g.},
\cite{Stewart:2010qs} for related discussion)
\be
\label{PDFparton}
\begin{split}
f_{i/j}^{(0)} = \delta_{ij} \delta(1-z) \,, \qquad f_{i/j}^{(1)} = -\frac{1}{\epsilon}\frac{\as}{2\pi}C_{ij}P_{ij}(z)\,,
\end{split}
\ee
where the color factors and splitting functions are given by
\begin{align}\label{PDFq}
&C_{qq'}=C_{\bar q \bar q'}=C_F\,,\quad
P_{qq'}(z)=P_{\bar q \bar q'}(z)=\delta_{q q'}P_{qq}(z)
\,,\\
&C_{q g}=C_{\bar q g}=T_F\,,\quad\ \
P_{\bar q g}(z)=P_{q g}(z)\,,\quad
\label{PDFg}
\end{align}
with $P_{qq}(z)$ and $P_{qg}(z)$ given in \eq{Pij} below. There are no
contributions containing the splitting functions $P_{gq},P_{g\bar
  q},P_{gg}$ since the tensor $W_{\mu\nu}^j$ in \eq{WxQ2tau} we
compute contains only the quark current.

For the 1-jettiness structure functions \eq{Fi-def}, we only need the
projections $-g_{\mu\nu}W^{\mu\nu}$ and $P_\mu P_\nu W^{\mu\nu}$ of
the hadronic tensor in \eq{Wtensor}; hence, we obtain the projected
coefficients $-g^{\mu\nu}w^i_{\mu\nu}$ and $P^\mu P^\nu
w^i_{\mu\nu}$. We compute the contracted tensors $-g^{\mu\nu}
W_{\mu\nu}^i$ explicitly in \appx{Wparton}.  Including the factor of
$(1-\e)$ coming from the tensor contractions in $D$ dimensions, we
obtain:
\begin{align}
\label{gWform}
\frac{-g^{\mu\nu}W_{\mu\nu}^q(x,Q^2,\tau)}{1-\e} 
&= 4\pi Q_f^2 \delta(1-x)\delta(\tau) 
+ 2\as C_F Q_f^2 \Bigl[ -\frac{1}{\epsilon} P_{qq}(x) \delta(\tau) + w_G^q(x,Q^2,\tau)\Bigr] \,, \\
\frac{-g^{\mu\nu}W_{\mu\nu}^g(x,Q^2,\tau)}{1-\e} 
&= 4\as T_F\sum_f Q_f^2 \Bigl[ - \frac{1}{\epsilon}P_{qg}(x) \delta(\tau) + w_G^g(x,Q^2,\tau)\Bigr]\,,
\end{align}
where $w_G^{q,g}$ are finite as $\epsilon\to 0$. For the quark tensor
we consider one flavor $f$ at a time, since it will get convolved with
a different PDF for each quark flavor, while for the gluon tensor we
include the sum over all flavors.  The contractions $P^\mu P^\nu
W_{\mu\nu}^i$ begin at $\cO(\as)$, and take the form
\be
P^\mu P^\nu W_{\mu\nu}^i(x,Q^2,\tau) =
\begin{cases}
\as C_F Q_f^2 \,w_{P}^q(x,Q^2,\tau) \hspace{2cm} &i=q \,,
\\
\as T_F \sum_f Q_f^2 \,w_{P}^g(x,Q^2,\tau) \hspace{2cm} &i=g \,,
\end{cases}
\ee
and are finite as $\epsilon\to 0$. Plugging these forms into the
matching conditions \eq{Wi-matching}, we find the $1/\epsilon$ IR
divergences cancel between the PDFs in \eq{PDFparton} and the computed
tensors in \eq{gWform}, leaving the finite matching coefficients
\begin{align}
& -g^{\mu\nu}w_{\mu\nu}^{q\,(0)} = 4\pi Q_f^2 \delta(1-x)\delta(\tau) \,,\qquad \quad \ \
-g^{\mu\nu}w_{\mu\nu}^{g\,(0)}=P^\mu P^\nu w_{\mu\nu}^{q,g\,(0)} = 0 
\,,\\[0.5cm]
\label{matchingcoeffs}
&
- g^{\mu\nu} w_{\mu\nu}^{i\,(1)} =
\begin{cases}
2\as C_F Q_f^2\, w_G^q 
\\
4\as T_F \sum_f Q_f^2 \, w_G^g
\end{cases}
, \qquad
P^\mu P^\nu w_{\mu\nu}^{i\,(1)} =
\begin{cases}
\as C_F Q_f^2 \,w_{P}^q \hspace{1.5cm} &i=q \,,
\\
\as T_F \sum_f Q_f^2 \,w_{P}^g \hspace{1.5cm} &i=g \,,
\end{cases}
\,.
\end{align}
We compute the finite coefficients $w_{G,P}^{q,g}$ explicitly in
\appx{Wparton}, and they are given in \eqsss{wPq}{wGq}{wPg}{wGg}.

\subsection{Phase Space}
\label{ssec:phase}
In this section, we evaluate some of the phase-space integrals for 1-
and 2-body final states. In the partonic computation of the tensor
$W^{\mu\nu}$ given in \eq{WxQ2tau} or \eq{Wtensor}, we sum over all
the possible $n$-body final partonic states,
\be
\label{WdPhin}
\begin{split}
W^j_{\mu\nu}(x,Q^2,\tau) &=\frac{1}{s_j} \sum_n \int d\Phi_n \cM_\mu^*(j(P)\to p_1\dots p_n) \cM_\nu(j(P)\to p_1\dots p_n) \\
&\qquad\times (2\pi)^D \delta^D( P + q - \sum_i p_i) \delta\Big(\tau - \tau(\{p_1\dots p_n\})\Big) \equiv \sum_n W_{\mu\nu}^{j[n]}\,,
\end{split}
\ee
where the $1/s_i$ factor is from averaging over the spins or
polarizations of the initial parton $j$: $s_q=s_{\bar q}=2$ and
$s_g=2(1-\e)$, and in the last equality we define the $n$-body
contribution $W_{\mu\nu}^{j[n]}$ to $W$. We sum over the spins or
polarizations of all the final-state partons. In \eq{WdPhin},
\be
\cM_\mu \equiv \bra{p_1\dots p_n} J_\mu \ket{j(P)}
\ee
is the amplitude for the initial parton $j$ with momentum $P$ to
scatter off the current $J_\mu$ and produce the final-state partons
with momenta $p_1\dots p_n$, and the $n$-body phase space integration
measure is given by \be
\label{dPhin}
\int d\Phi_n \equiv \prod_{i=1}^n \biggl( \int \frac{d^D p_i}{(2\pi)^D} 2\pi \delta(p_i^2)\biggr)\,.
\ee

The $n=1$ term in the sum in \eq{WdPhin} is given by
\begin{align}
\label{dPhi1}
W_{\mu\nu}^{j[1]} = \frac{1}{s_j}\int d\Phi_1 \cM_\mu^* \cM_\nu (2\pi)^D\delta^D(P+q-p_1) \delta(\tau- \tau(p_1)) &=
\frac{1}{s_j}\frac{2\pi}{Q^2}\delta(1-x)\delta(\tau) \cM_\mu^*\cM_\nu\,,
\end{align}
where the arguments of $\cM_\mu$ are implicit.
The $n=2$ term is given by
\be
\label{dPhi2}
\begin{split}
W_{\mu\nu}^{j[2]}&= \frac{1}{s_j}\int d\Phi_2 \cM_\mu^* \cM_\nu (2\pi)^D \delta^D(P+q-p_1-p_2) \delta\Bigl(\tau-\tau(p_1,p_2)\Bigr) \\
&= \frac{1}{s_j}\frac{1}{8\pi Q} \frac{(4\pi)^\epsilon}{ \Gamma(1-\epsilon)}
\,\int_0^Q \frac{dp_2^- }{(p_2^+ p_2^-)^\epsilon} \delta\Bigl(\tau-\tau(p_1,p_2)\Bigr)\cM_\mu^* \cM_\nu
\,, 
\end{split}
\ee
where the lightcone components are $(p_2^+,p_2^-)=(\nz\cdot p_2, \bnz
\cdot p_2)$. We have chosen to do the integrals using the
momentum-conserving delta function in \eq{dPhi2} and the mass-shell
delta functions in \eq{dPhin} in such an order that the $p_2^-$
integral is left over in \eq{dPhi2} to be done last. Where $p_2$ and
$p_1$ appear in \eq{dPhi2}, they take values given by the formulas:
\be
\label{p1p2}
\begin{split}
p_1^\mu &= (Q-p_2^-) \frac{\nz^\mu}{2} + \frac{1-x}{x} p_2^- \frac{\bnz^\mu}{2} - p_\perp^\mu\,, \\
p_2^\mu &= p_2^- \frac{\nz^\mu}{2} + \frac{1-x}{x}(Q-p_2^-) \frac{\bnz^\mu}{2} + p_\perp^\mu \,,
\end{split}
\ee
where $p_\perp^2 = - (Q-p_2^-)p_2^- (1-x)/x$. The integrand in \eq{dPhi2} is independent of the azimuthal angle $\phi$ of $p_\perp$. For example, the $p_2^+$ in the denominator of \eq{dPhi2} is $p_2^+=(1-x)(Q-p_2^-)/x$.

\begin{figure}[t]{
\begin{center}
\vspace{-1em}
     \includegraphics[width=.8\columnwidth]{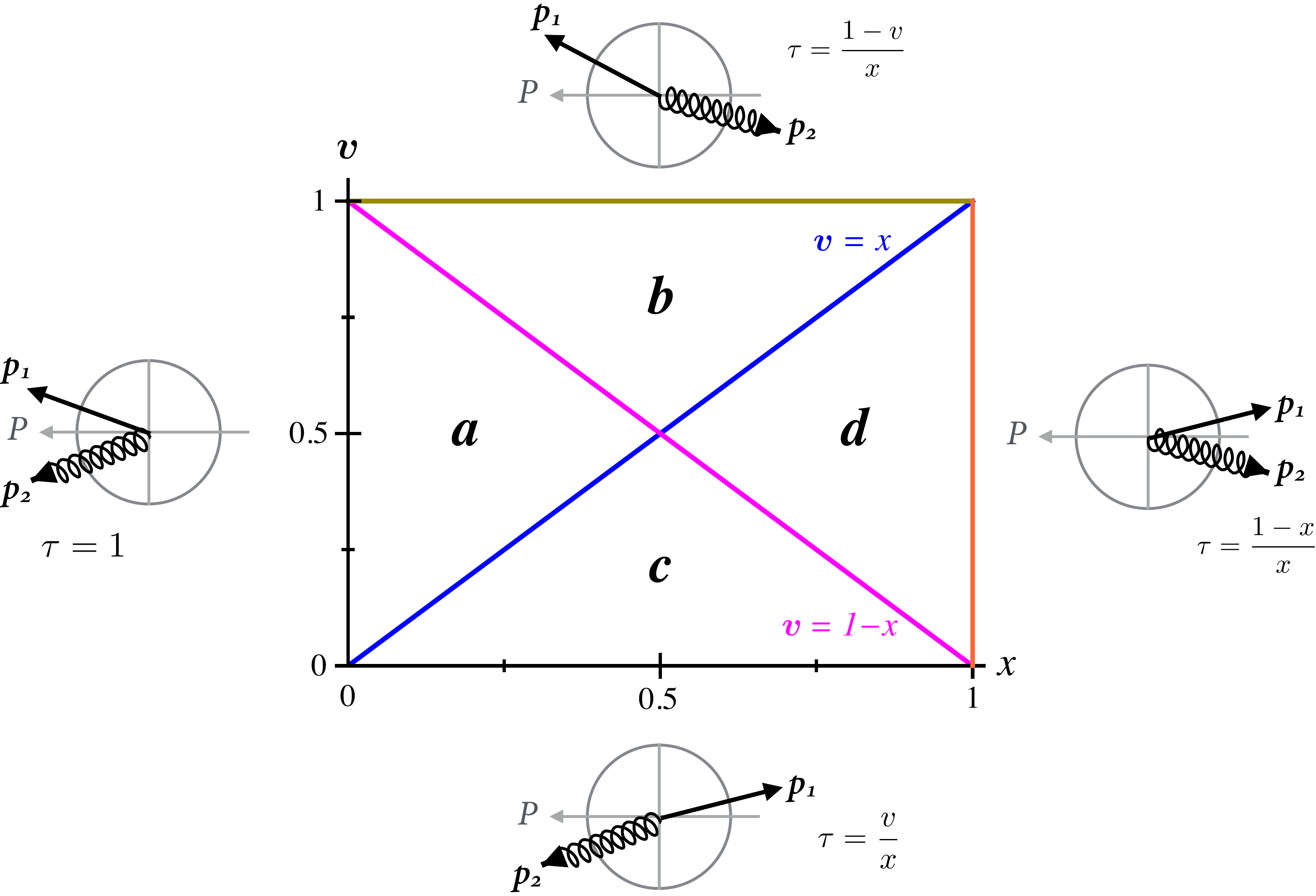}   
     \vspace{-2em} 
\end{center}
    { \caption[1]{ 
Regions of two-body phase space in the Breit frame. In this frame the
incoming proton has momentum along the $-z$ direction given by $P =
Q\bnz/(2x)$. The figure shows a quark and a gluon in the final state,
corresponding to the Feynman diagrams in \fig{qgam}. For the diagrams
in \fig{ggam} there would be a quark and an antiquark in the final
state. The 1-jettiness $\tau$ groups particles into back-to-back
hemispheres in this frame, in the $\pm z$ directions. There are four
distinct regions in $x$ and $v\equiv p_2^-/Q$ space in which the
particles are grouped differently, making $\tau$ a function
$\tau(x,v)$. In regions (a) and (d) both particles end up in the same
region, giving a constant value of $\tau$. In regions (b) and (c) the
two particles are in opposite regions, and $\tau$ varies according to
the projection of the two particles' momenta onto the $\pm z$
axes. The values of $\tau$ in these four regions are given in
\eq{tauiThetai}, and enter the phase space integral in \eq{dPhi2-v}.
      }
  \label{fig:regions-v}} }
  \vspace{-1em}
\end{figure}

We find it convenient to rewrite the phase space in \eq{dPhi2} in
terms of a dimensionless variable $v\equiv p_2^-/Q$,
\begin{align}
\label{dPhi2-v}
W_{\mu\nu}^{j[2]} = \frac{1}{s_j}\frac{1}{8\pi} \Bigl(\frac{4\pi}{Q^2}\Bigr)^\epsilon\frac{1}{\Gamma(1-\epsilon)}
      \bigg(\frac{x}{1-x}\bigg)^\epsilon
\,\int_0^1 \frac{dv }{v^\epsilon (1-v)^\epsilon} \cM_\mu^* \cM_\nu
 \delta\Bigl(\tau-\tau(x,v)\Bigr)
\,.
\end{align}
The 1-jettiness $\tau$ is now expressed as a function of $x$ and $v$.
Two particles in the final state can be assigned in four different
ways to the two hemispheres, and the formula for $\tau(x,v)$ differs
in each of these regions. These four regions $(a)$ to $(d)$ are
illustrated in \fig{regions-v}. The function $\tau(x,v)$ can be broken
down into four pieces,
\be
\tau(x,v) = \sum_{i\in \{a,b,c,d\}} \Theta^{(i)}(x,v) \tau^{(i)}(x,v)\,,
\ee
where the two-dimensional step function $\Theta^{(i)}$ covers each
region $(i)$ and $\tau^{(i)}$ is the value of 1-jettiness in the
corresponding region:
\begin{align}
&\tau^{(i)}(x,v)=
\begin{cases}
1 \hspace{2cm} 
\\
\frac{1-v}{x}
\\
\frac{v}{x}  
\\
\frac{1-x}{x} 
\end{cases}
\hspace{-2cm}\,,\hspace{1cm}
\Theta^{(i)}(x,v)=
\begin{cases}
\theta(-v+(1-x))\,\theta(v-x) \hspace{2cm} & i=a \,,
\\
\theta(v-(1-x))\, \theta(v-x)\hspace{2cm} & i=b \,,
\\
\theta(-v+(1-x))\, \theta(-v+x)\hspace{2cm} & i=c \,,
\\
\theta(v-(1-x))\, \theta(-v+x)\hspace{2cm} & i=d \,.
\end{cases}
\label{tauiThetai}
\end{align}
Note that in regions $i=a,d$ the value of $\tau$ is constant in $v$
and thus the delta function comes outside the integral in
\eq{dPhi2-v}. As illustrated in \fig{regions-v}, in these two regions
both final state particles are in the same hemisphere, and $\tau$
takes the $v$ independent value shown in \eq{tauiThetai} over the
entire region, which corresponds to the maximum values of $\tau$ given
in \eq{tmax}. In regions $i=b,c$, the value of $\tau$ varies with $v$
and thus the delta function remaining in \eq{dPhi2-v} can be used to
evaluate the $v$ integral.

We evaluate the integral \eq{dPhi2-v} over the four regions in
\fig{regions-v} using the expressions \eq{tauiThetai} in
\appx{Wprojected}.

\section{Analytic Results for DIS 1-Jettiness Structure Functions}
\label{sec:Fi}

In this section we present our analytic results for the structure
functions in \eq{Fi-def} that determine the 1-jettiness cross section
in \eq{sigmaF1L}. We will present our results in terms of the
structure functions appearing in the cumulative (integrated) cross
section,
\be
\label{cumulant}
\sigma^c(x,Q^2,\tau) \equiv \int_0^\tau d\tau' \frac{d\sigma}{dx\,dQ^2\,d\tau'}\,,
\ee
which decomposes into structure functions $F_i$ exactly like \eq{sigmaF1L}, where
\be
\label{Fc}
F_i(x,Q^2,\tau) = \int^{\tau}_0  d\tau'\,  \cF_i (x,Q^2,\tau')\,,
\ee
where the $\cF_i$ are given by \eq{Fi-def}. The results for the
integrated structure functions $F_i$ are more compact to write down
than for $\cF_i$. We give the results for the differential structure
functions $\cF_i$ in \appx{Wns}.

As the cross section in \eq{xsection} is written in terms of singular
and nonsingular parts, we express the structure functions as:
\begin{align}
\label{Fi}
F_i&=F_i^\sing+F_i^\ns
 \,.\end{align}
The fixed-order structure functions are obtained from the calculation
of projected hadronic tensors in \eq{Fi-def} that are calculated in
\appx{Wparton} and \appx{Wns}.  The singular part of the cross section
was calculated in \cite{Kang:2013nha}. Our main new results here are
the nonsingular parts of the structure functions that are obtained by
subtracting off the known singular parts from the full expressions.

We will present our final expressions for the singular and nonsingular
parts of $F_1$ and $F_L$ in \eq{Fi} in the following form:
\begin{subequations}
\label{FAB}
\begin{empheq}[box=\fbox]{align}
\label{F1AB}
F_1 (x,Q^2,\tau)
&=\sum_{i\in\{q,\bar q,g\}} (A_i+B_i)
\,,\\
\label{FLAB}
F_L (x,Q^2,\tau)
&= \sum_{i\in\{q,\bar q,g\}} 4x \,A_i
 \,.
\end{empheq}
\end{subequations}
The singular parts of these can be extracted from the singular cross
section in \cite{Kang:2013nha}, and are given in \eq{AiBi-sing}. Our
main new results here are for the nonsingular parts.  The functions
$A^\ns_i$ and $B^\ns_i$ are given by the nonsingular parts of $P^\mu
P^\nu W^i_{\mu\nu}$ and $-g^{\mu\nu} W^i_{\mu\nu}$, respectively. They
are obtained by integrating the differential structure functions in
\eq{cAicBi}. We find
\begin{subequations}\label{AiBi}
\begin{empheq}[box=\fbox]{align}
\label{Aq}
A^\ns_q &= \sum_f  Q_f^2 \frac{\as C_F}{4\pi}
\biggl\{ \Theta_0 \!\! \int_x^{\frac{1}{1+\tau}} dz \, f_{q} (\tfrac{x}{z}) (2z \tau -1)  + \int_x^1 dz\, f_{q}(\tfrac{x}{z})\biggr\}
\,,\\
\label{Ag}
A^\ns_g &= \sum_f Q_f^2 \frac{\as T_F}{\pi} 
\biggl\{ \Theta_0\!\! \int_x^{\frac{1}{1+\tau}} dz f_g(\tfrac{x}{z}) (2z \tau - 1) (1-z)  + \int_x^1 dz\, f_g(\tfrac{x}{z}) (1-z)\biggr\}
\,,\\\label{Bq}
B^\ns_q &=\sum_f Q_f^2 \frac{\as C_F}{4\pi} 
 \biggl\{
\Theta_0\!\! \int_{x}^{\frac{1}{1+\tau}}\!\! \frac{dz}{z} f_q(\tfrac{x}{z})
\Bigl[ \frac{1-4z}{2(1-z)} (2z\tau-1) +P_{qq}(z)\ln\frac{z\tau}{1-z\tau}\Bigr] 
\nn\\
&\qquad
 +  f_q(x)  \bigl(  3 \ln\tau +2\ln^2\tau\bigr) 
+ \int _x^1 \frac{dz}{z} f_q(\tfrac{x}{z}) \Bigl[\cL_0(1-z) \frac{1-4z}{2} - P_{qq}(z)\ln z\tau\Bigr]  
\biggr\}
\,,\\ \label{Bg}
B^\ns_g &= \sum_f Q_f^2 \frac{\as T_F}{2\pi} 
\biggl\{ \Theta_0\!\! \int_x^{\frac{1}{1+\tau}}\frac{dz}{z} f_g(\tfrac{x}{z})
\biggl[-(2z\tau-1) + P_{qg}(z)\ln\frac{z\tau}{1-z\tau}\biggr]
 \nn \\&\hspace{7cm}
 - \int_x^1 \frac{dz}{z}f_g(\tfrac{x}{z}) \bigl[ 1 + P_{qg}(z)\ln z\tau\bigr] \biggr\}
\,,
\end{empheq}
\end{subequations}
which is one of our main results. Here, we have defined the theta
function
\be
\label{Theta0}
\Theta_0\equiv \Theta_0(\tau,x)\equiv \theta(\tau)\theta(1-\tau)\theta\biggl(\frac{1-x}{x}-\tau\biggr)
\,,\ee
which turns on inside the physically-allowed region $0<\tau<\taumax$
given by \eq{tmax} and turns off outside.  The plus distribution
$\cL_n(z)$ is defined in \appx{plus}.  The standard splitting
functions $P_{qq}$ and $P_{qg}$ are given by
\begin{subequations}
\label{Pij}
\begin{align}
P_{qq}(z) &\equiv\left[\theta(1-z)\frac{1+z^2}{1-z}\right]_+  = (1+z^2) \cL_0(1-z) + \frac{3}{2}\delta(1-x) \\
P_{qg}(z) &\equiv \theta(1-z)[(1-z)^2+z^2].
\end{align}
\end{subequations}
The formulas for $B_q$ and $B_g$ in \eqs{Bq}{Bg} appear to contain
terms which are still divergent as $\tau\to 0$, but these divergences
cancel in the sum of all terms. Formulas for $B_{q,g}$ given as sums
of explicitly nonsingular terms can be found in \eq{BNS}.

One may recognize that the $\Theta_0$ terms in \eq{AiBi} introduce a
discontinuity in the cumulative cross section at $\tau=1$.  This
feature is associated with asymmetric initial momentum in the $z$
direction, which can give rise to an event with one of the hemispheres
containing all final-state particles and the other being empty.  As
illustrated in \fig{regions-v}, this occurs in regions $(a)$ and
$(d)$, where $\tau$ takes on its maximum allowed values in \eq{tmax},
$1$ for $x<1/2$ and $(1-x)/x$ for $x>1/2$.  For $x<1/2$, this appears
at $\tau=1$ as a delta function in the differential structure
functions \eq{cAicBi} and a discontinuity in the integrated structure
functions \eq{AiBi}.  However, this feature is not seen for $x>1/2$ at
$\tau=(1-x)/x$, because we see that the integrals proportional to
$\Theta_0$ in \eq{AiBi} go to zero for $\tau=(1-x)/x$, the range of
integration shrinking to zero.

The singular part of the cross section has been computed in
\cite{Kang:2013nha}, from which the singular part of the structure
functions can be extracted. $F_1^\sing$ is simply half of the cumulant
cross section given in Eq.~(174) in \cite{Kang:2013nha}, and
$F_L^\sing=0$.  The singular parts $A^\sing_i$ and $B^\sing_i$ of the
functions in \eq{FAB} are given by
\begin{subequations}\label{AiBi-sing}
\begin{align}
\label{Asqg}
A^\sing_{q,g} &= 0
\,,\\\label{Bsq}
B^\sing_q &=\sum_f Q_f^2 \biggl\{  
f_q(x) \bigg[\frac12
 -   \frac{\as C_F}{4\pi}  \biggl(\frac92+\frac{\pi^2}{3} + 3 \ln\tau +2\ln^2\tau\biggr) \bigg]
\nn\\&\qquad\quad
+ \frac{\as C_F}{4\pi} \int _x^1 \frac{dz}{z} f_q(x/z) 
\biggl[\cL_1(1-z) \,(1+z^2)+(1-z) + P_{qq}(z)\ln \frac{Q^2 \tau}{\mu^2}\biggr]  
\biggr\}
\,,\\ \label{Bsg}
B^\sing_g &= \sum_f Q_f^2 \frac{\as T_F}{2\pi} 
\int_x^1 \frac{dz}{z}f_g(x/z) \biggl[ 1-P_{qg}(z) + P_{qg}(z)\ln \frac{Q^2\tau (1-z)}{\mu^2}\biggr] 
\,.\end{align}
\end{subequations}

The sum of \eqs{AiBi}{AiBi-sing} gives the complete fixed-order
$\cO(\as)$ result for the DIS 1-jettiness structure functions.  When
we take values of $\tau$ beyond the physical maximum, where $\Theta_0$
terms are turned off, the result reproduces the standard inclusive
structure functions in $x$ and $Q^2$, which are given by
(e.g. \cite{Ellis:1991qj})
\begin{subequations}
\label{Ftot}
\begin{align}
\label{F1tot}
F_1(x,Q^2)&=\sum_f  Q_f^2 \int_x^1 \frac{dz}{z}\, \biggl\{
\bigg[\frac{\delta(1-z)}{2}+\frac{\as C_F}{4\pi}\,C_q(z)\bigg]\,\bigl[f_{q} \bigl(\tfrac{x}{z}\bigr)+f_{\bar q}\bigl(\tfrac{x}{z}\bigr)\bigr]  +\frac{\as T_F}{2\pi}\,C_g(z)f_{g}\bigl(\tfrac{x}{z}\bigr)
\biggr\}
\,, \\
\label{FLtot}
F_L(x,Q^2)&=4x\sum_f  Q_f^2 \int_x^1 dz\, \bigg\{
\frac{\as C_F}{4\pi}\,\bigl[f_{q} \bigl(\tfrac{x}{z}\bigr)+f_{\bar q} \bigl(\tfrac{x}{z}\bigr)\bigr]
+\frac{\as T_F}{\pi}(1-z)\,f_{g} \bigl(\tfrac{x}{z}\bigr)
\bigg\}
\,,
\end{align}
\end{subequations}
where we have defined the two functions
\begin{subequations}
\begin{align}
C_q(z)&\equiv -\bigg(\frac92+\frac{\pi^2}{3}\bigg)\delta(1-z)
-\frac32 \cL_0(1-z)+2\cL_1(1-z)\\
&\hspace{4cm}
+3-(1+z)\ln(1-z) +P_{qq}(z)\,\ln\frac{Q^2}{\mu^2\,z}
\,,\nn\\
C_g(z)&\equiv 1+ P_{qg}(z)\bigg[-2+\ln\bigg(\frac{Q^2}{\mu^2}\frac{1-z}{z}\bigg)\bigg]
\,.\end{align}
\end{subequations}

In this section we have presented the complete $\cO(\as)$ results for
the fixed-order structure functions in the DIS 1-jettiness cross
section. The expressions \eq{AiBi} for the nonsingular contributions
to the structure functions in \eq{FAB} are the primary new results of
this paper.

\section{Numerical Results}
\label{sec:results}

In this section, we present numerical results for the structure
functions $\cF_{1,L}$ that appear in the differential 1-jettiness
cross section in \eq{sigmaF1L} and the corresponding $F_{1,L}$ in
\eq{Fc} that appear in the integrated cross section \eq{cumulant}. We
computed these structure functions to $\cO(\as)$ in \sec{Fi} and
\appx{Wns}. We also present predictions for the $\tau$ cross sections
themselves.  For structure functions, we show the fixed-order
$\cO(\as)$ results for the singular part (in $\tau$), the nonsingular
part and their sum.  For the cross section, we show resummed results
at NLL$'+\cO(\as)$ accuracy as well as the pure fixed-order
results. At this order of accuracy we have the fixed-order parts of
the hard, jet, beam, and soft functions in the singular part
\eq{resummedtaumB} at the same order in $\cO(\as)$ as in the
nonsingular part.\footnote{Resummation of the singular terms in the
  $\tau$ cross section is in fact available up to NNLL accuracy
  \cite{Kang:2013nha}. For simplicity, we choose to illustrate results
  only at NLL$'$ resummed accuracy in this paper (see
  \cite{Abbate:2010xh,Almeida:2014uva} for definition of primed
  accuracy). As described in Ref.~\cite{Almeida:2014uva}, formulae for
  resummed differential and integrated cross sections at unprimed
  orders of accuracy may suffer from a mismatch in the actual
  logarithmic accuracy achieved, depending on how the formulae are
  written. One can ensure that the differential distribution at
  N$^k$LL matches the accuracy of the corresponding integrated cross
  section by differentiating the integrated cross section including
  the $\tau$ dependence in the scales $\mu_i(\tau)$.  However, in the
  large $\tau$ (``far tail'') region, Ref.~\cite{Abbate:2010xh}
  observed that this procedure leads to unrealistically large
  uncertainties, and recommends that the $\tau$ dependence in
  $\mu_i(\tau)$ \emph{not} be differentiated in going from the
  integrated to the differential cross section. It is possible to
  write the differential cross section in a way that interpolates
  between the two approaches for small and large $\tau$, but this task
  does not lie within the scope of this paper. As observed in
  \cite{Almeida:2014uva}, equivalent accuracy between differential and
  integrated cross sections is in fact maintained if one works at
  primed orders, whether one differentiates $\mu_i(\tau)$ or not. Thus
  we will work here at NLL$'$ accuracy and evaluate the differential
  cross section by \emph{not} differentiating $\mu_i(\tau)$ in the
  integrated cross section, see \eq{cumulantderivative}. This avoids
  the potential negative issues pointed out in both
  \cite{Abbate:2010xh} and \cite{Almeida:2014uva}. Some recent
  progress (\emph{e.g.} \cite{Gaunt:2014xga}) has been made in
  obtaining ingredients needed for NNLL$'$ or N$^3$LL accuracy
  \cite{N3LL} for the related version of 1-jettiness $\tau_1^a$
  defined in \cite{Kang:2013nha,Kang:2013wca}.\label{footnote}}

For our numerical results plotted here, we set the collision energy to
be $\sqrt{s}=300\GeV$, which corresponds to the H1 and ZEUS
experiments, and choose $Q=80\GeV$. We adopt MSTW2008 PDF sets at NLO
\cite{Martin:2009iq} with five light quark and antiquark flavors and
run $\as(\mu)$ with the 2-loop beta function in \eq{alphas} starting
at the values $\as(m_Z)=0.1202$ used in NLO PDFs.

\begin{figure}[t!]{
\vspace{-1ex}
 \includegraphics[width=.47\columnwidth]{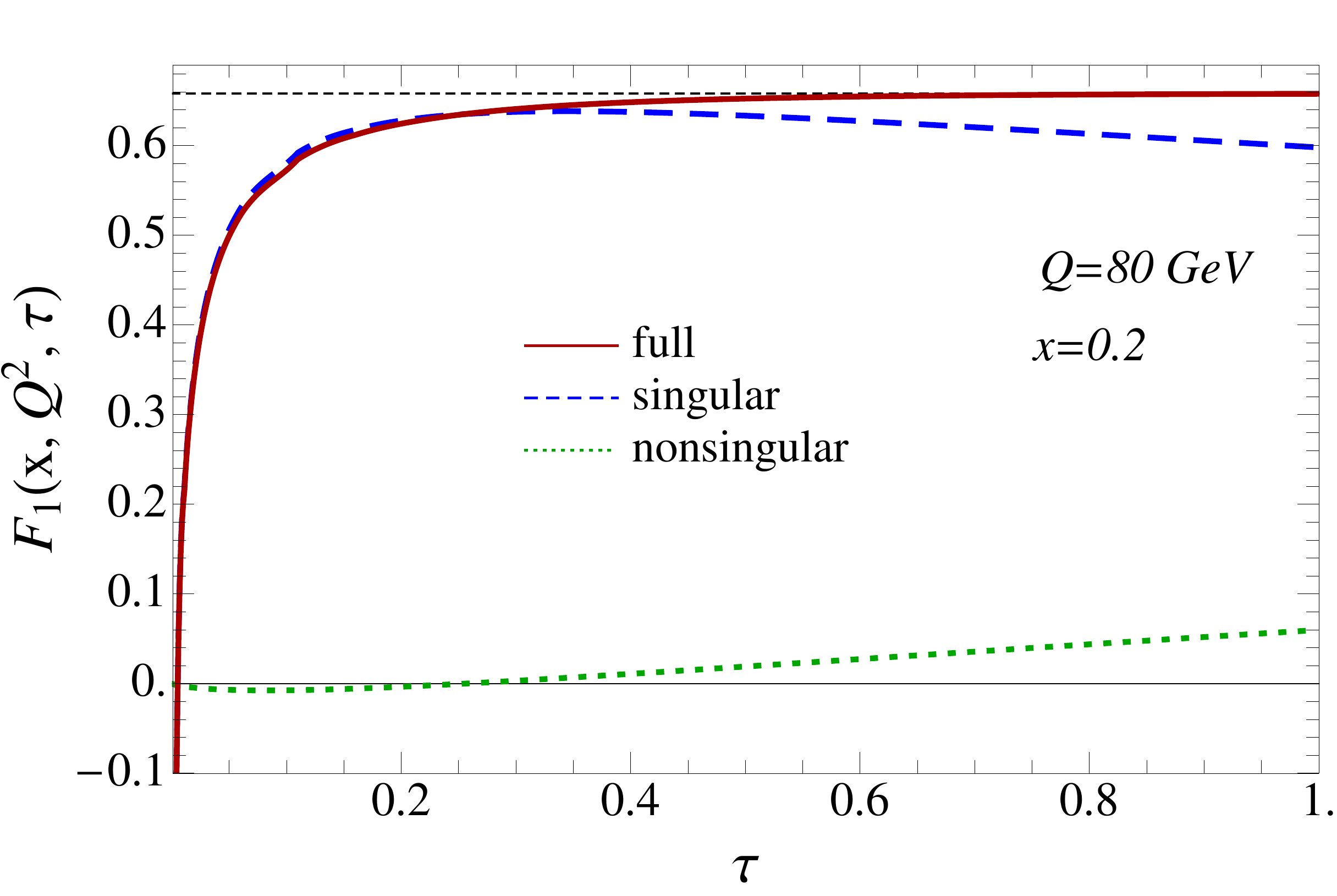}\hspace{4ex}
 \includegraphics[width=.47\columnwidth]{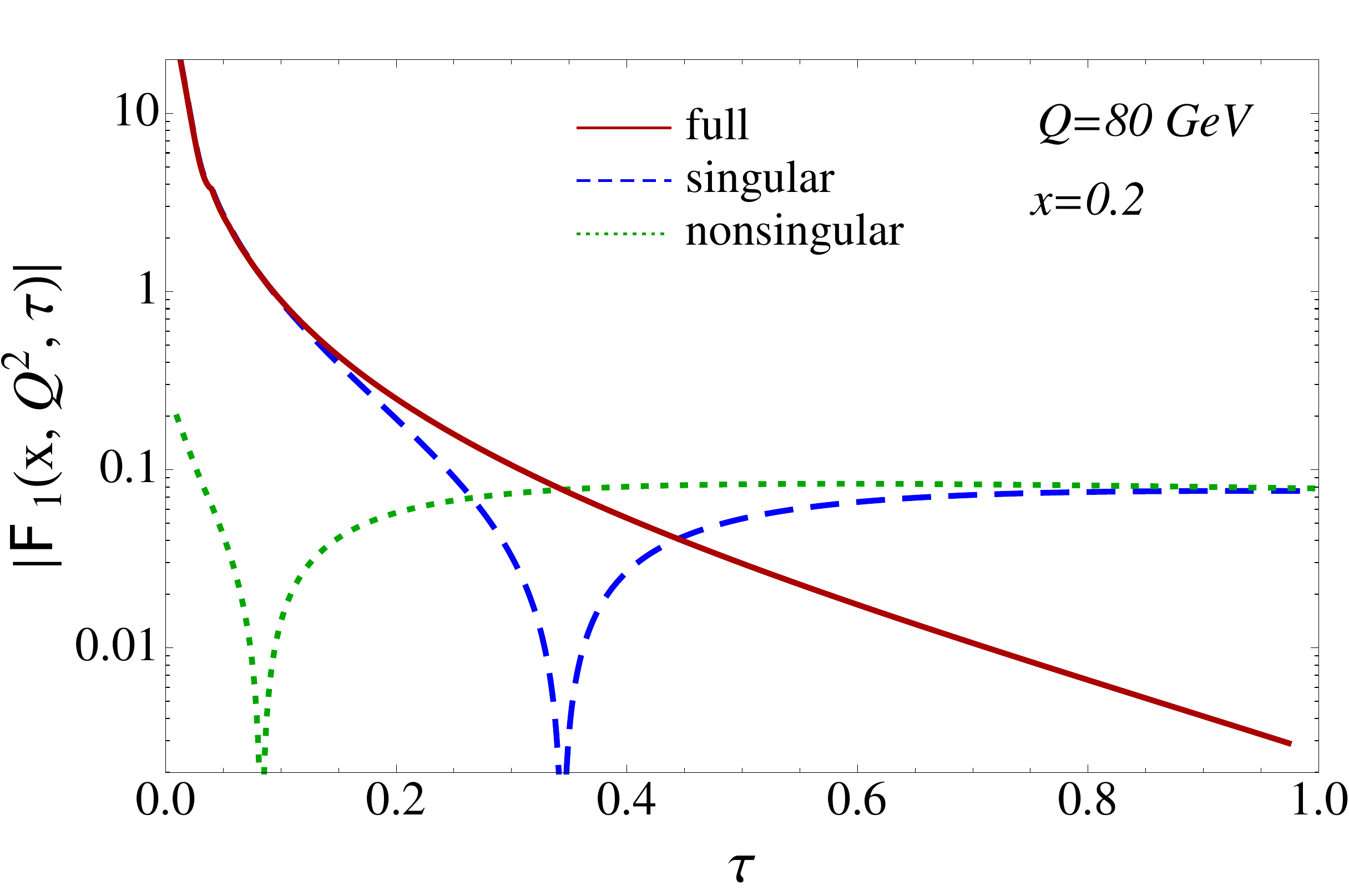} 
 \includegraphics[width=.47\columnwidth]{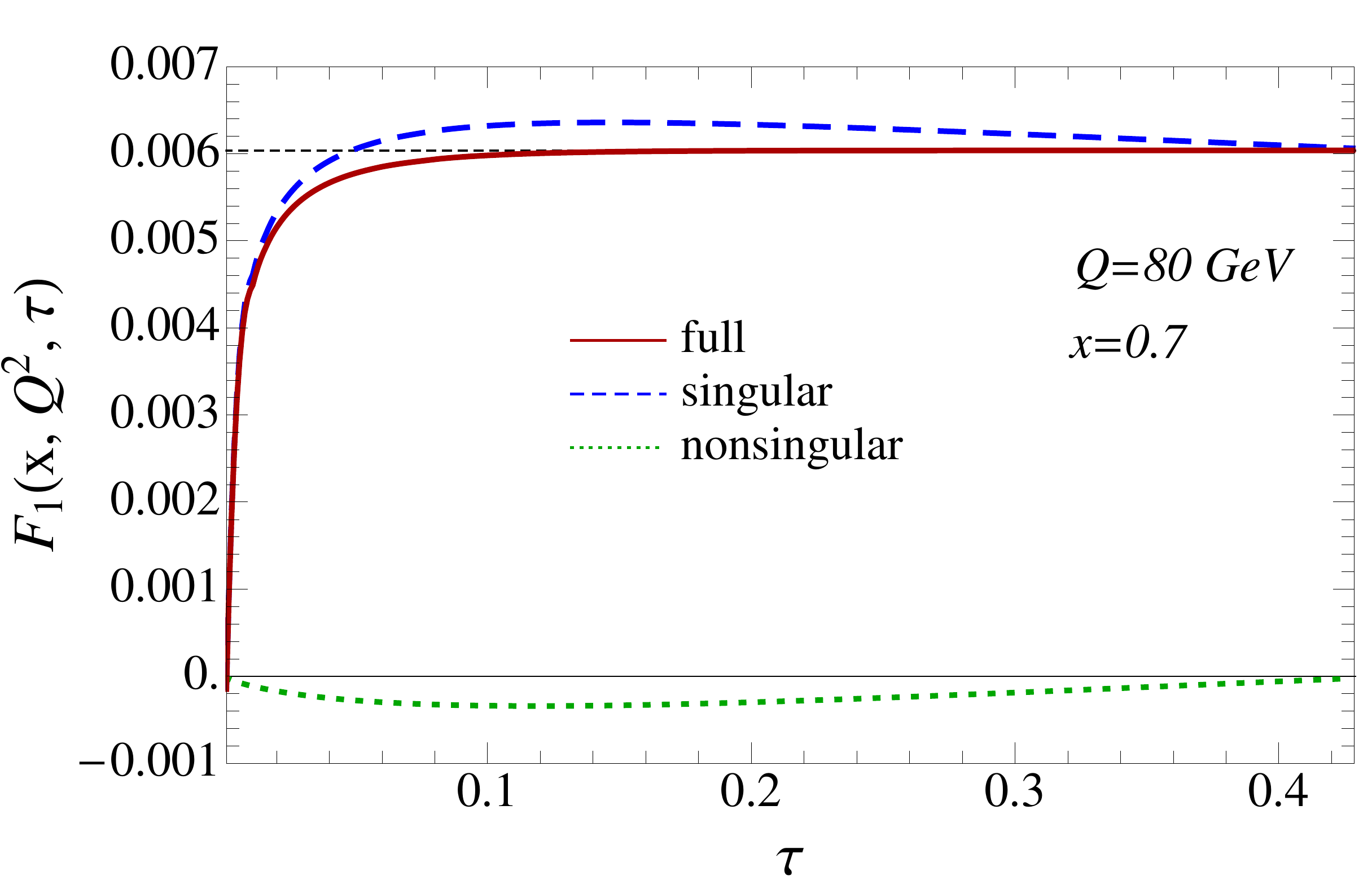}\hspace{4ex}
 \includegraphics[width=.47\columnwidth]{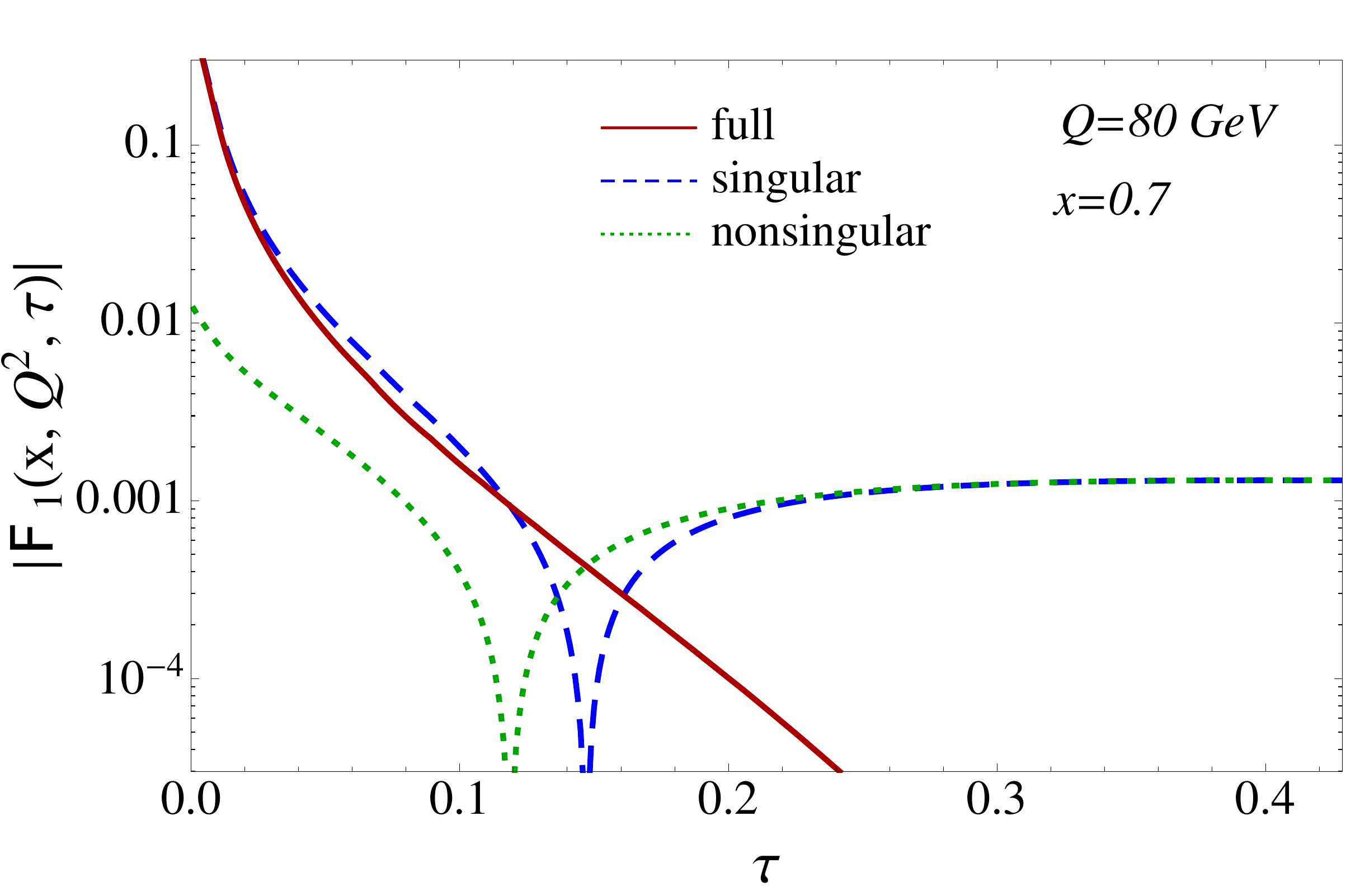} 
 \vspace{-1em}
{ \caption{
Fixed order components of structure function $F_1( x, Q^2, \tau)$
(left) in the integrated $\tau$ cross section and $\cF_1(x, Q^2,
\tau)$ (right) in the differential $\tau$ cross section at $Q=80\GeV$
and $x=0.2$ and 0.7.  Full (red solid), singular (blue dashed), and
nonsingular (green dotted) contributions. Horizontal dashed line in
left plots indicates total $F_1(x, Q^2)$ in \eq{F1tot}.}
  \label{fig:F1}} }
\end{figure}

\begin{figure}[t!]{
\vspace{-1ex}
    \includegraphics[width=.47\columnwidth]{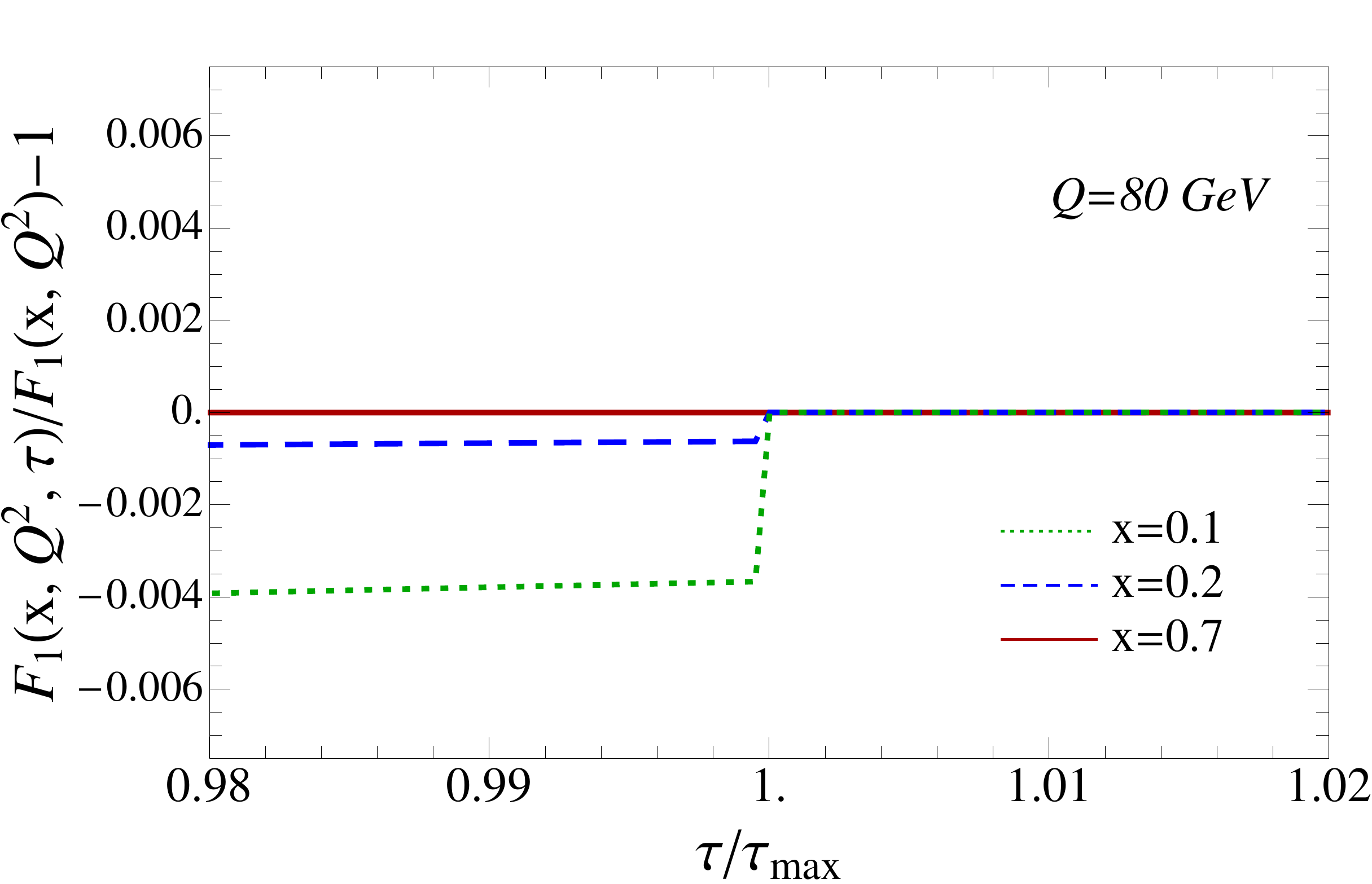}\hspace{4ex}
    \includegraphics[width=.47\columnwidth]{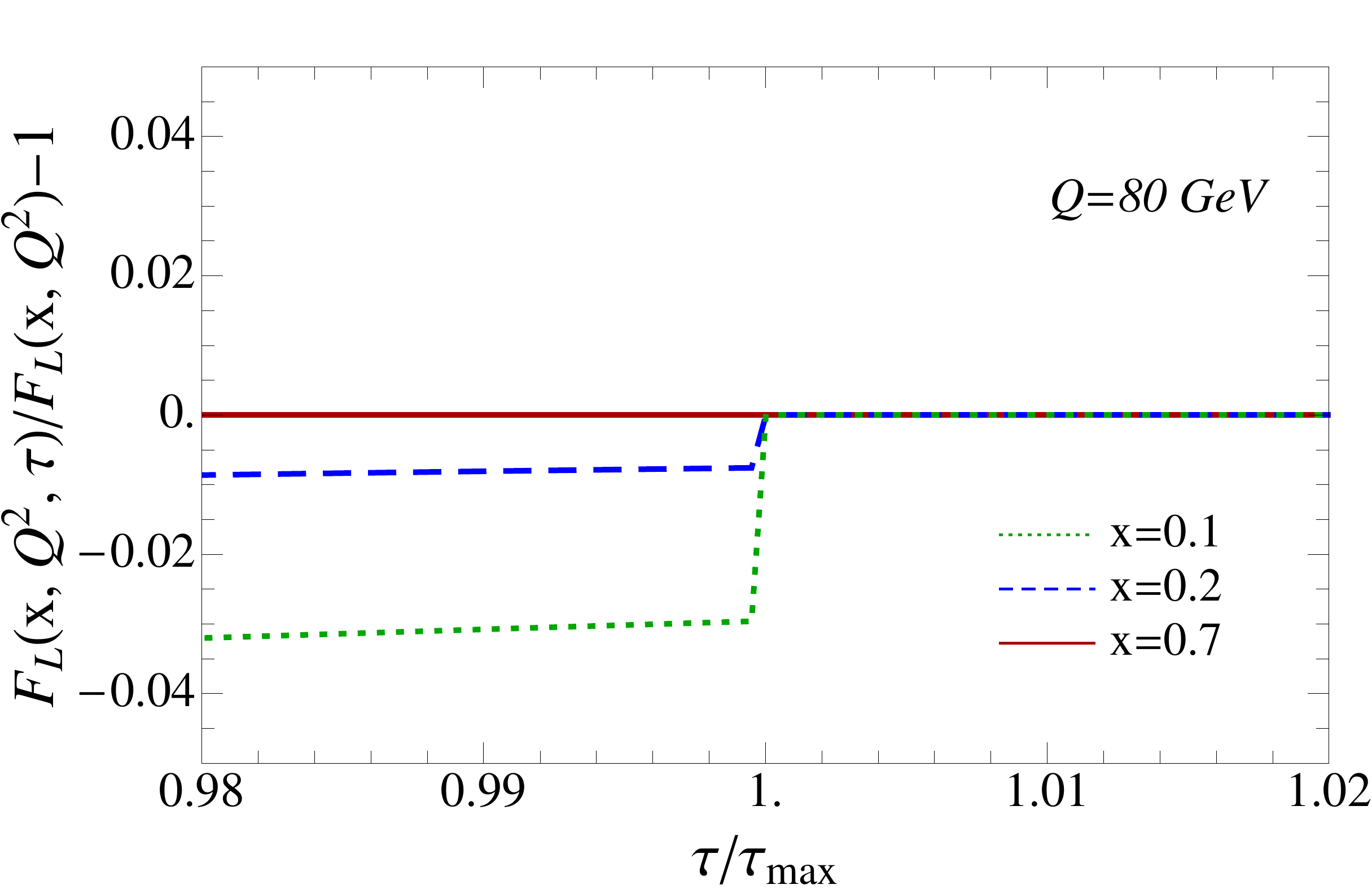}    
      \vspace{-1em}
    { \caption[1]{Discontinuities of normalized cumulants $F_1$ and $F_L$ 
       at $Q=80\GeV$ and $x=0.1,\, 0.2,\, 0.7$.}
  \label{fig:disc}} }
\end{figure}

\fig{F1} shows the components of the fixed-order results for the
structure function $F_1( x, Q^2, \tau)$ in the integrated cross
section, given by \eqss{F1AB}{AiBi}{AiBi-sing}, and of the $\cF_1(x,
Q^2, \tau)$ in the differential distribution, given by
\eqss{cF1cFL}{cAicBi-sing}{cAicBi}, at two values $x=0.2$ and 0.7. We
set all scales to be $\mu=Q=80\GeV$. In the integrated structure
function $F_1$, the sum of singular (dashed line) and nonsingular
(dotted line) contributions give the full result (solid line). The
full result approaches the total result $F_1(x,Q^2)$ (horizontal
dashed line) in \eq{F1tot} as $\tau$ approaches 1. For $x=0.2$, the
singular part alone undershoots the total, and the nonsingular part
makes up the difference. For $x=0.7$, the singular part overshoots the
total, and the corresponding nonsingular part is mostly
negative. Although it is imperceptible in \fig{F1}, there is actually
a small discontinuity in the $x=0.2$ plot at $\tau=1$, and the total
(solid red) $F_1$ does not reach the full result (dashed black) until
above $\tau=1$.  We will zoom in on this feature in \fig{disc}.

For the differential $\cF_1$ in \fig{F1}, we plot the absolute value
on a log scale. The results illustrate that there is a large
cancellation between the singular and nonsingular pieces in the large
$\tau$ region so that the total goes to zero in this tail.  This same
cancellation was discussed for $e^+e^-$ thrust in
Ref.~\cite{Abbate:2010xh}, and appears in various other cross sections
that have singular and nonsingular components.  The tail falls faster
for larger $x$ because $\tau$ dependence enters into PDFs in a form
like $f_q(x(1+\tau))$, as seen in \eq{cAicBi}, which falls faster as
$x$ increases.  The overall normalization also becomes smaller for
larger $x$ due to the PDFs falling off.

\begin{figure}[t!]{
     \includegraphics[width=.47\columnwidth]{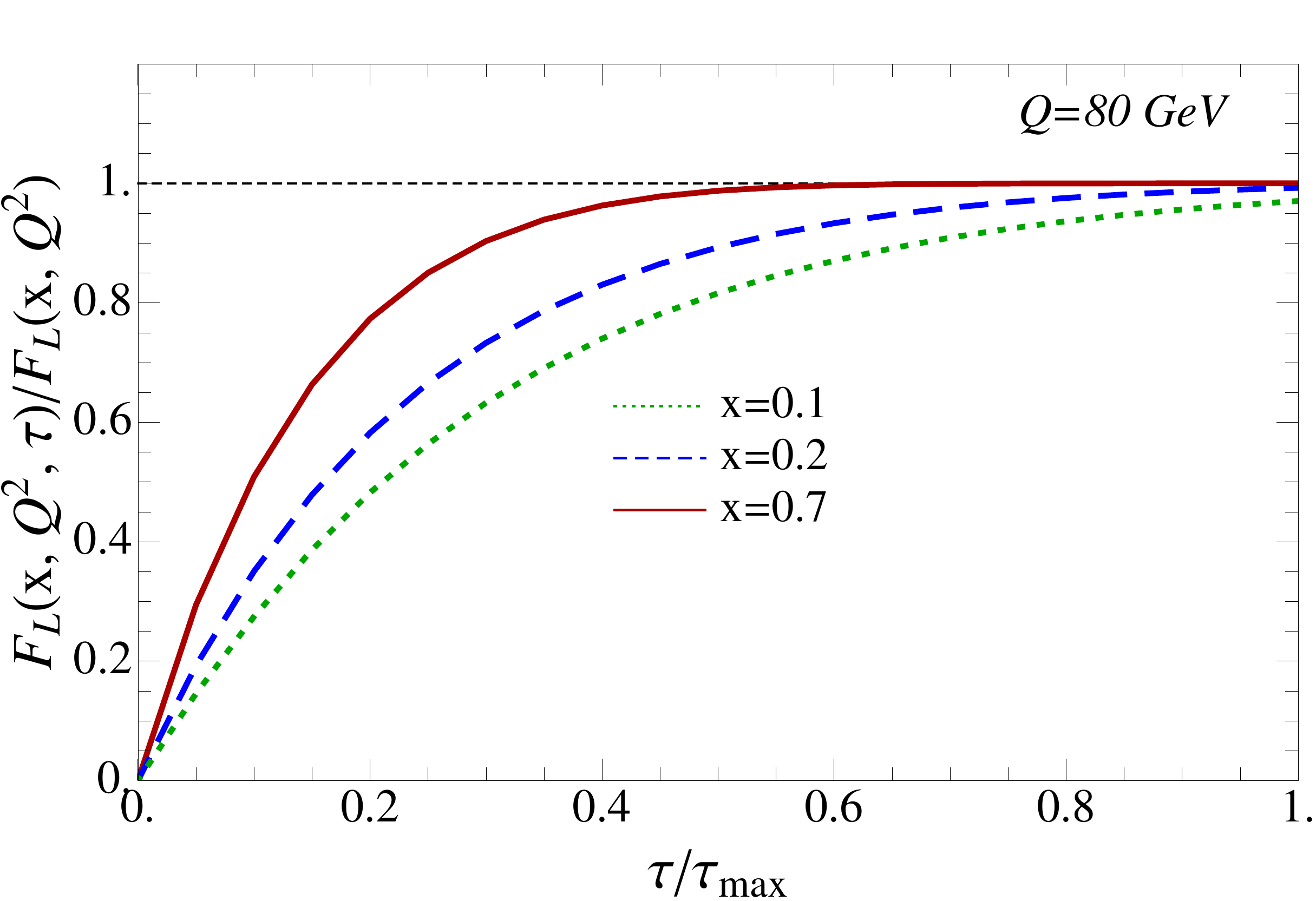}\hspace{4ex}
      \includegraphics[width=.47\columnwidth]{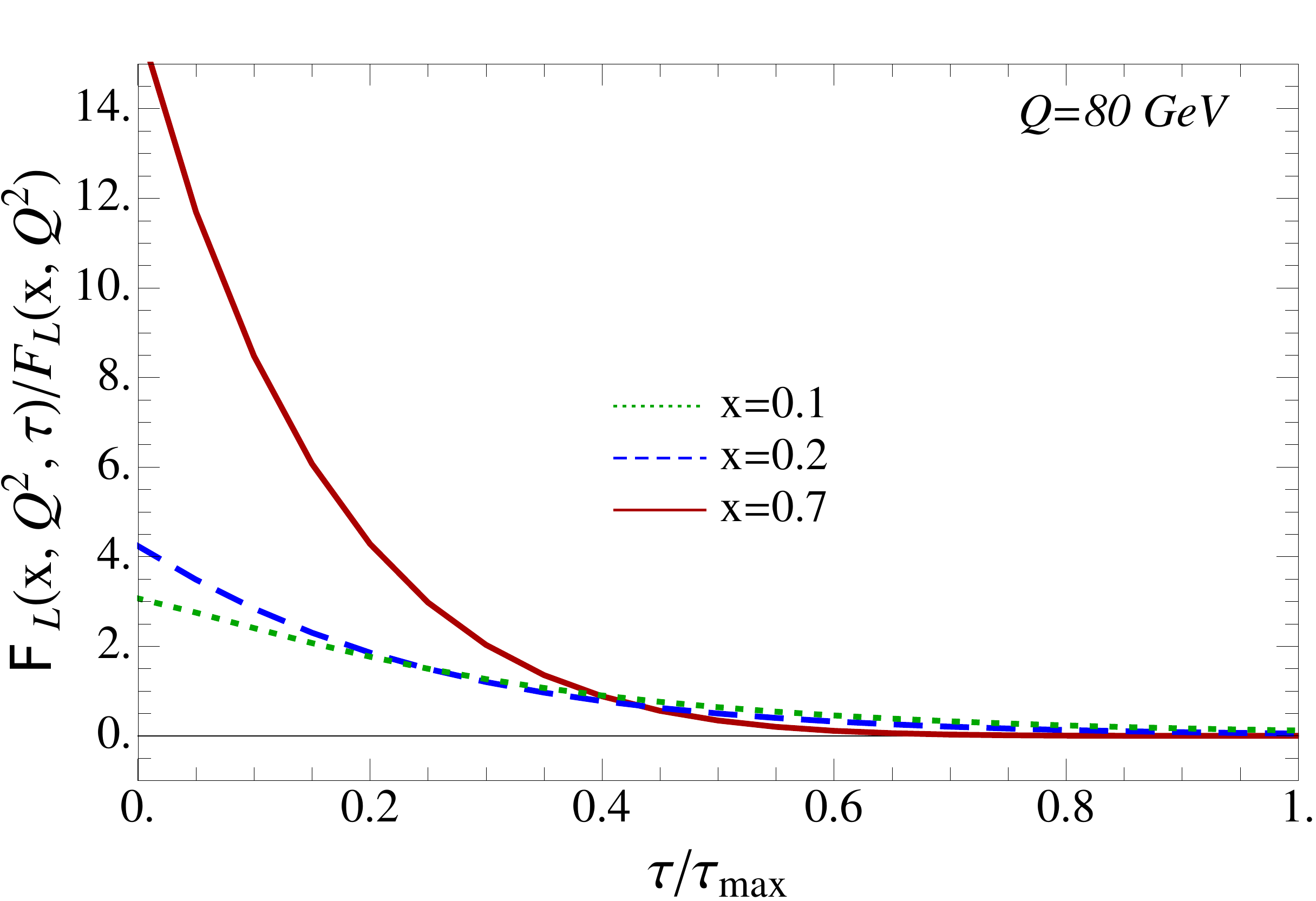} 
      \vspace{-1em}
{ \caption[1]{Longitudinal structure function $F_L(x,Q^2,\tau)$ (left)
    for the integrated $\tau$ cross section and $\cF_L(x,Q^2,\tau)$
    (right) for the differential $\tau$ cross section, divided by the
    total $F_L(x,Q^2)$ in \eq{FLtot} at $Q=80~\GeV$ and
    $x=0.1,~0.2~\text{and},~0.7$. }
  \label{fig:FL}} }
\end{figure}

\fig{FL} shows the fixed-order results for the longitudinal structure
function $F_L( x, Q^2, \tau)$ for the integrated cross section, given
by \eqs{FLAB}{AiBi}, and $\cF_L(x, Q^2, \tau)$ for the differential
distribution, given by \eqs{cF1cFL}{cAicBi}, at $x=0.1\,,0.2,$ and
0.7. These are purely nonsingular in $\tau$. The plots are normalized
to the total $F_L( x, Q^2)$ in \eq{FLtot}.  Note that $\cF_L$ is
finite at $\tau=0$ at $\cO(\as)$. The distribution monotonically
decreases with $\tau$. For the left plot at $x=0.1,0.2$, there is a
perceptible gap from the total (straight dashed line) at $\tau=1$
before the curves reach the value 1. This jump is explored in
\fig{disc}.

\fig{disc} illustrates the discontinuities in the cumulative $F_1$ and
$F_L$ near $\tau^\text{max}$. The jump is smaller than 1\% in $F_1$
and is about a few percent in $F_L$. These discontinuities are reduced
for increasing $x$ and disappear at $x=1/2$ and beyond. As described
in \sec{Fi}, these discontinuities are associated with events where
the jet hemisphere is empty and the beam hemisphere contains all
final-state particles as seen in the Breit frame, so whole regions of
phase space end up contributing to the same fixed value of $\tau$ (see
\fig{regions-v}).  Such events do not occur in the observables defined
in the partonic CM frame such as $e^+ e^-$ thrust.  This discontinuity
is infrared safe, and though its magnitude is very small, it is in
principle measurable.

\begin{figure}[t]{
\vspace{-1em}
\begin{center}
 \includegraphics[width=.6\columnwidth]{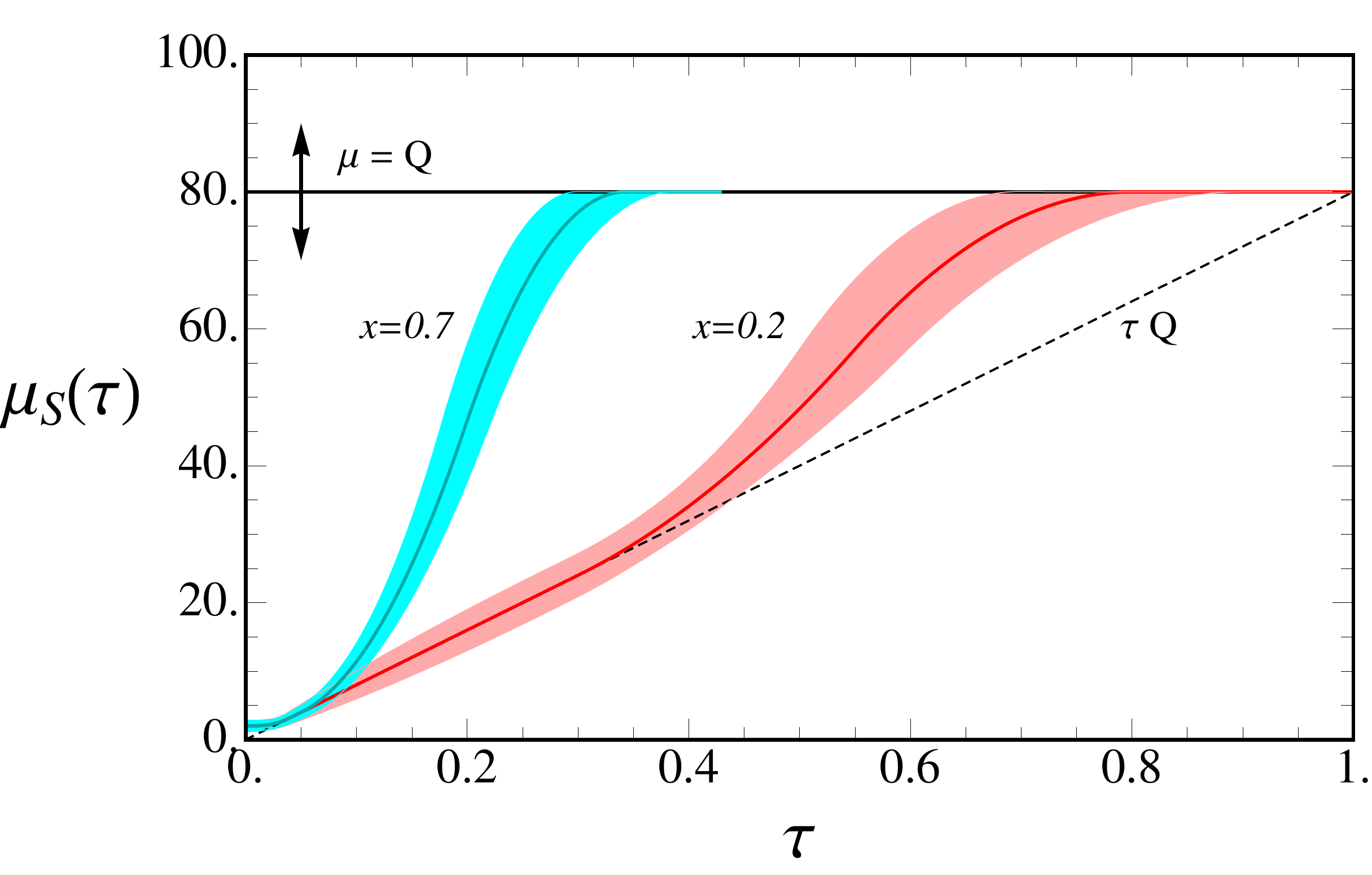}
\end{center}
\vspace{-2em} { \caption{Profile $\mu_S(\tau)$ for $x=0.2$ and $0.7$
    for $Q=80\GeV$.  Uncertainty bands are sum of all variations in
    \eq{scalevariations} in quadrature.  The dashed line indicates the
    canonical choice $\mu_s(\tau)=\tau Q$ and the vertical arrow
    implies that the scale $\mu$ is varied up and down by a factor of
    2.  }
  \label{fig:profile}} }
\end{figure}

The cross section in \eq{xsection} with all the scale dependencies
made explicit in singular and nonsingular parts can be written
\be\label{xsection2}
\sigma^\text{full}(\tau;\mu_H,\mu_J,\mu_B,\mu_S,\mu_\ns) 
=\sigma^\sing (\tau;\mu_H,\mu_J,\mu_B,\mu_S) +\sigma^\ns(\tau;\mu_\ns)\,,
\ee
and is given in \eq{resummedtaumB}.  The singular part depends on the
scales $\mu_H$, $\mu_J$, $\mu_B$, $\mu_S$ associated with hard, jet,
beam, and soft radiation, respectively, and the nonsingular part
depends on $\mu_\ns$ as in conventional fixed-order results.  For the
full calculation, all scales should be specified.  In the region $\tau
\ll 1$, there are large logarithms in the singular part and the
logarithms can be resummed by RG evolution of the functions between
$\mu$ and their individual canonical scales: $\mu_H\sim Q$,
$\mu_{B,J}\sim \sqrt{\tau}Q$, $\mu_S\sim\tau Q$.  (For more details on
resummation of the singular part, see \cite{Kang:2013nha}. Basic
results are reviewed in \appx{resum}.)  However, $\mu_S$ cannot be
arbitrary small and it should freeze above the nonperturbative regime
that lives below $1$~GeV.  On the opposite end, where $\tau\sim
\cO(1)$ and logs of $\tau$ are not large, the resummation should be
turned off by setting all $\mu_i\approx Q$.  In \cite{Kang:2013nha} we
used profile functions $\mu_i(\tau)$ satisfying above constraints and
estimated perturbative uncertainties by varying parameters in the
profile functions \cite{Ligeti:2008ac,Abbate:2010xh,Berger:2010xi}.
However, the profile defined in \cite{Kang:2013nha} has scales away
from the canonical scales when $x$ increases.  Here we use improved
profiles given in \appx{profile}, which set canonical scales in the
resummation region that are independent of $x$.  \fig{profile} shows
the soft scale $\mu_S(\tau)$ as a function of $\tau$ at $x=0.2$ and
$0.7$ as well as the canonical choice $\tau Q$ (dashed line).

\begin{figure}[t!]{
    \includegraphics[width=.47\columnwidth]{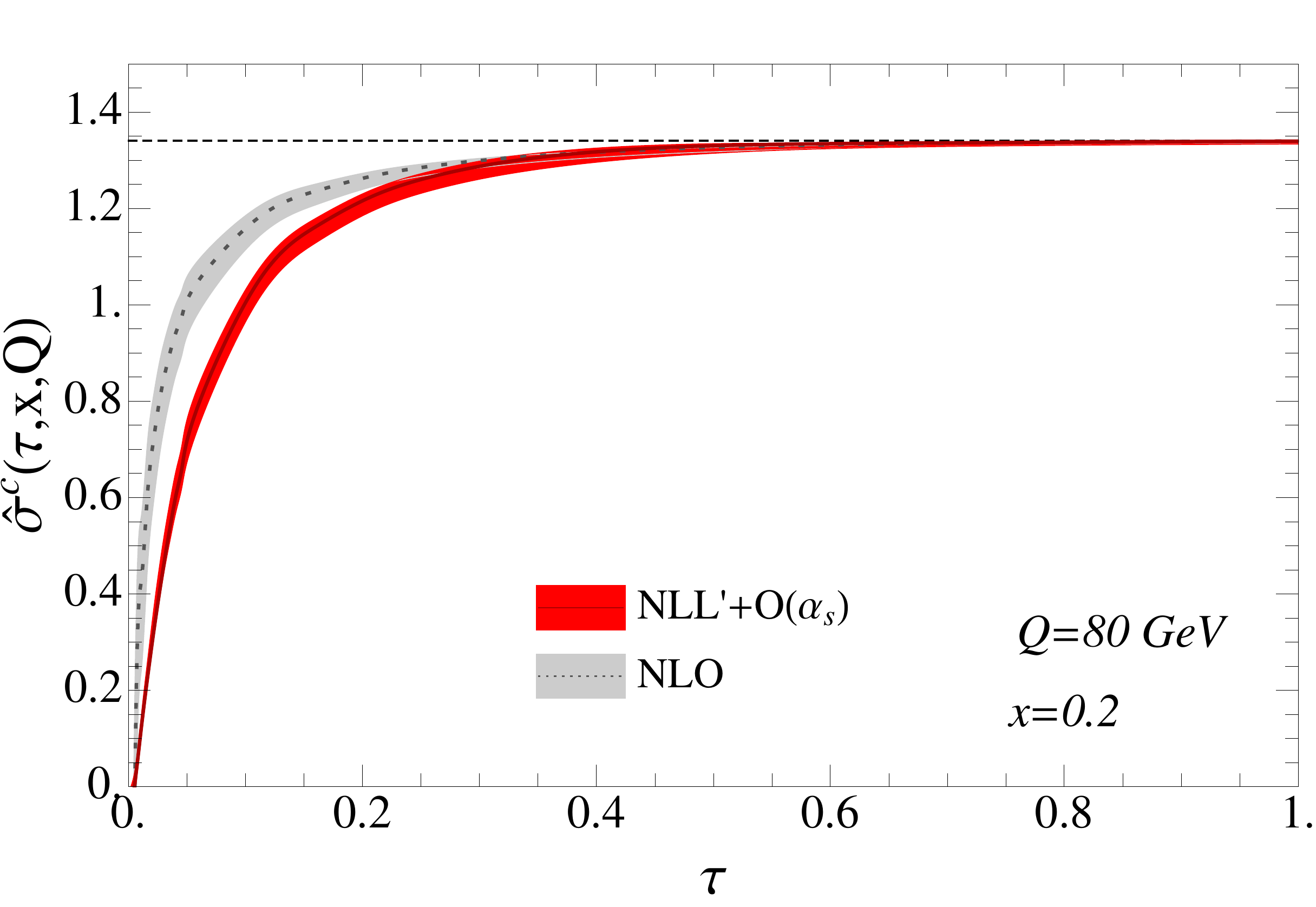}\hspace{4ex}
    \includegraphics[width=.47\columnwidth]{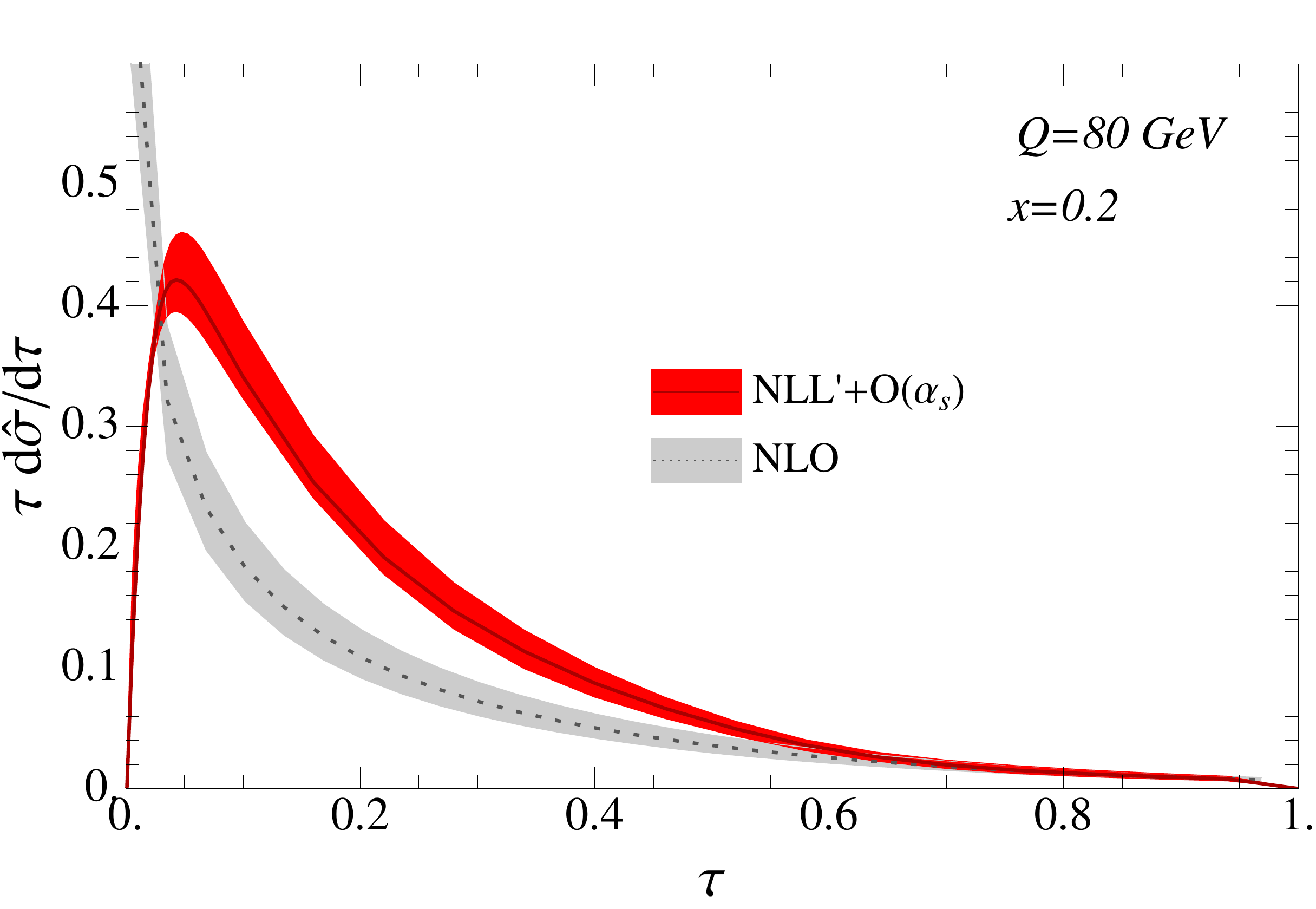}
    \includegraphics[width=.47\columnwidth]{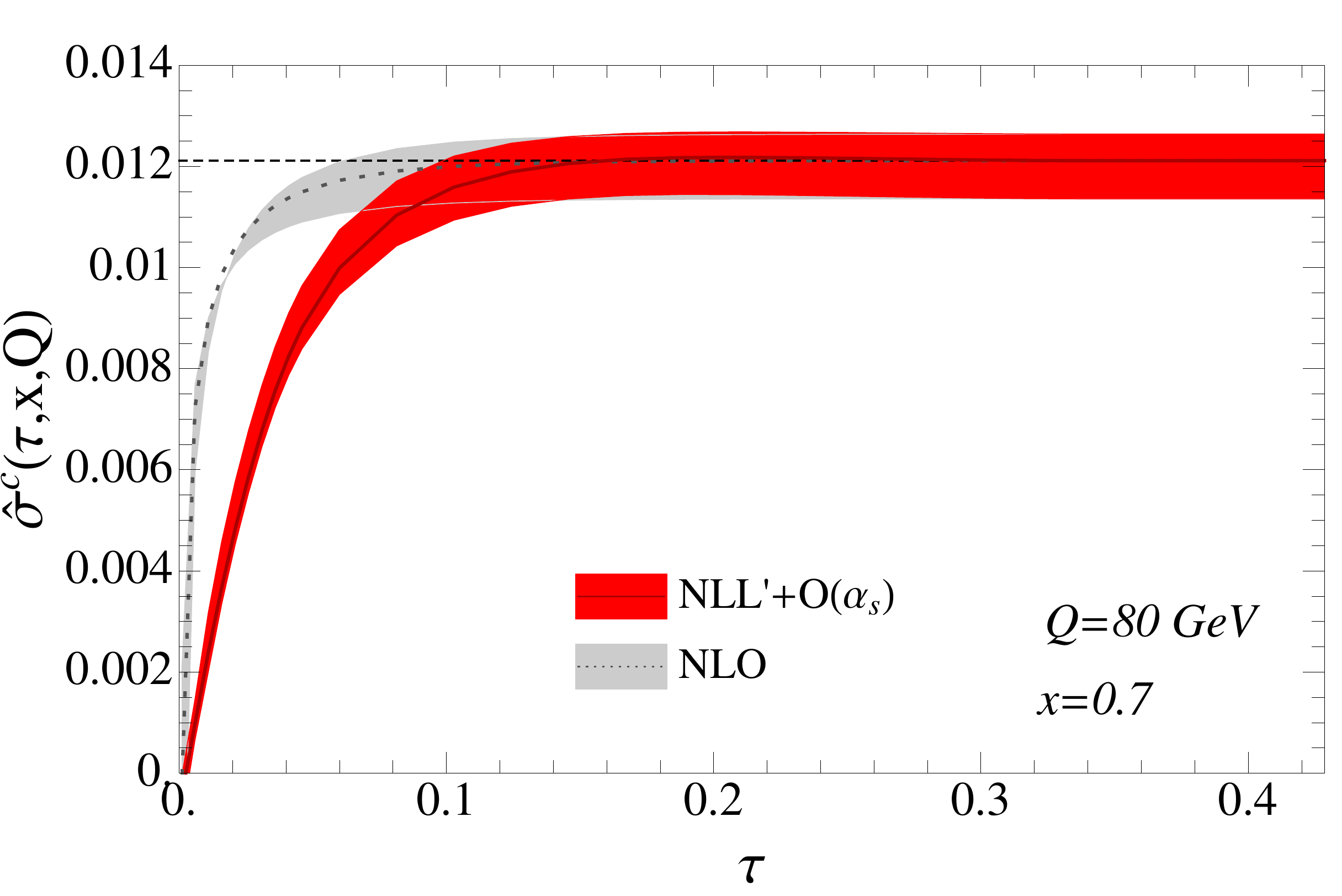}\hspace{4ex}
    \includegraphics[width=.47\columnwidth]{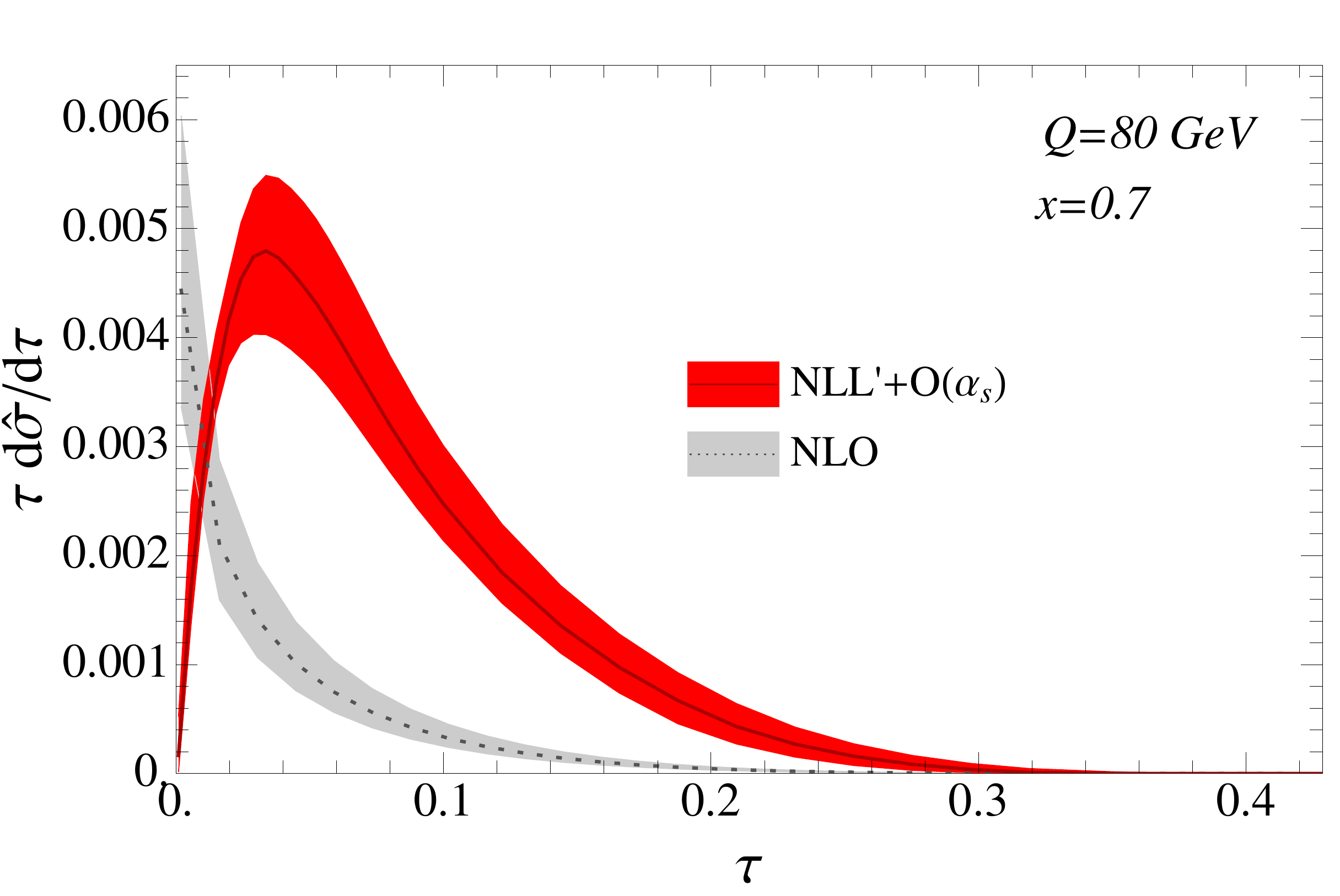}
      \vspace{-1em}
{ \caption[1]{Cumulant and differential cross section at
    NLL$'+\cO(\as)$ for $Q=80\GeV$ and $x=0.2$ and 0.7. The
    uncertainty bands for the resummed results are obtained by summing
    all scale variations described in \eqs{muNS}{scalevariations} in
    quadrature.  }
  \label{fig:sig}} }
\end{figure}

For the central values of $\mu_\ns$ and its variations, we make the
same choice as \cite{Abbate:2010xh},
\begin{align}\label{muNS}
\mu_\ns &= \mu_J\,,
& \mu_\ns &= \Big[ \frac{\mu_J+\mu_s}{2}  ,\mu_H \Big]
\,.\end{align}
The scales are chosen to estimate theory uncertainties from
un-resummed subleading logarithms in the nonsingular part.

\fig{sig} shows resummed integrated and differential cross sections at
NLL$'+\cO(\as)$ as well as the purely fixed-order result at
$\cO(\as)$.  We plot normalized cross sections defined as
\be\label{rescale}
\hat \sigma^c(x,Q^2,\tau) =\frac{\sigma^c(x,Q^2,\tau)}{\sigma_0} \,,\qquad 
\tau \frac{d \hat \sigma}{d\tau}=\frac{\tau }{\sigma_0}\,\frac{d\sigma}{dx\, dQ^2\, d\tau}
\,.\ee
where $\sigma_0=2\pi\alpha^2[1+(1-y)^2]/Q^4$. We multiply the
differential distribution by $\tau$ for ease of displaying the whole
$\tau$ region.

The uncertainty bands for the resummed results are obtained by summing
all scale variations described in \eqs{muNS}{scalevariations} in
quadrature.  The uncertainty bands for the fixed-order results are
obtained from varying the central scale $\mu=Q$ up and down by factors
of 2.  As seen in \fig{F1} the tail of the distribution becomes
shorter with increasing $x$.  Relative uncertainties about the central
value are larger for larger $x$ because of slower convergence of the
perturbative corrections associated with the PDF for increasing $x$
(as can be seen from the fact that the residual scale dependence of
the PDF increases with $x$).

\fig{sig} only includes purely perturbative results.  Nonperturbative
effects in 1-jettiness are power suppressed by $\Lqcd/(\tau Q)$ for
$\tau\gg \Lqcd/Q$, and the leading power correction can be expressed
in terms of a single nonperturbative parameter $\Omega_1$. The
parameter is universal for different versions of 1-jettiness in DIS
defined in \cite{Kang:2013nha}, and even appears in the power
corrections for certain jet observables in $pp\to H/Z+\text{jet}$ with
a small jet radius \cite{Stewart:2014nna}.  Alternatively, a shape
function that takes nonperturbative behavior into account in the
nonperturbative region as well as the power correction region
\cite{Hoang:2007vb}, can be used as in \cite{Kang:2013nha}. In this
paper, we omit implementing these nonperturbative effects.

\section{Conclusions}
\label{sec:conclusions}

Events with one or more jets plus initial state radiation dominate the
population of final states in DIS for typical values of $x$. These
events can be further probed by the inclusive event shape 1-jettiness
$\tau$. Events with small values of $\tau$ contain only one non-ISR
jet, while multiple jets populate the large $\tau$ region. In this
paper, we obtained analytically the $\cO(\as)$ cross section for all
values of $\tau$, and combined it with NLL$'$ resummation of the
singular terms at small $\tau$ to obtain results accurate over the
entire range of $\tau$.  This is the first analytic calculation of a
DIS event shape at this order.

We wrote the results in terms of structure functions $F_1(x,Q^2,\tau)$
and $F_L(x,Q^2,\tau)$ which generalize the usual DIS structure
functions $F_{1,L}(x,Q^2)$. We gave structure functions for both the
cumulative or integrated $\tau$ distribution as well as the
differential $\tau$ distribution. Our predictions for the cumulative
distribution agree with the total $F_{1,L}(x,Q^2)$ for $\tau>\taumax$.

The cumulative cross section displays an interesting feature, a small
discontinuity at $\tau=1$, which is a consequence of asymmetric
initial momentum that can lead to one of hemispheres (in the Breit
frame) being empty in the final state. This does not happen in
$e^+e^-$ thrust defined in the partonic CM frame.

We presented numerical results with perturbative uncertainties by
varying scales at the HERA energy.  In general the uncertainties grow
with $x$ due to the convergence of perturbative corrections in the
cross section that are connected with the PDFs through their scale
dependence. The tail of the $\tau$ distribution falls off faster as
$x$ grows.  The size of the nonsingular terms is consistent with our
expectations from \cite{Kang:2013nha} where we compared the resummed
singular cross section with the total QCD cross section at $x,Q^2$.

Our results represent a significant improvement in precision in the
prediction of DIS event shape cross sections. The groundwork is in
place to go to higher resummed \cite{N3LL} and fixed-order accuracy
which we will pursue in the near future, and bring the science of
event shapes in DIS to the same level of precision as has been
achieved in $e^+e^-$. These predictions can be tested with existing
HERA data and future EIC data, which should yield determinations of
the strong coupling and hadron structure to unprecedented accuracy.

\begin{acknowledgments}
The work of DK and IS is supported by the Office of Nuclear Physics of
the U.S. Department of Energy under Contract DE-SC0011090, and the
work of CL by DOE Contract DE-AC52-06NA25396 and by the LDRD office at
Los Alamos.  We thank the organizers of the 2013 ESI Program on ``Jets
and Quantum Fields for LHC and Future Colliders'' in Vienna, Austria,
where part of this work was performed. DK would like to thank the
Nuclear Theory group at LANL and CL would like to thank the MIT Center
for Theoretical Physics and the KITP at UCSB for hospitality during
portions of this work. This research was supported in part by the
National Science Foundation under Grant No.~PHY11-25915.
\end{acknowledgments}

\appendix
\section{Plus Distributions}
\label{app:plus}
In this section, we define plus distributions that we use and collect
some useful identities involving them.  The standard set of plus
distributions $\cL_n(z)$ are defined by (see, \emph{e.g.},
\cite{Ligeti:2008ac})
\begin{align}\label{plus-def}
\cL_n(z)\equiv \lim_{\ve\to0}
\frac{d}{dz}\bigg[\frac{\theta(z-\ve)\ln^{n+1}z}{n+1}\bigg]
=\bigg[\frac{\theta(z)\ln^n(z)}{z}\bigg]_+
\,.\end{align}
Integrating against a well-behaved test function $g(z)$ gives the
familiar rule,
\begin{align}\label{int-tst}
&\int^{z}_0 dz'\cL_n(z') g(z')
=\int^{z}_0 dz'\frac{\ln^n z'}{z'} [g(z')-g(0)]
   +g(0) \frac{\ln^{n+1} z}{n+1}.
\end{align}
We also define a distribution function with two arguments, which can
be used when the presence of the divergence in a variable $z$ is
controlled by the value of a second variable $z_0$,
\begin{align}\label{plus2}
\cL_n(z,z_0)&\equiv \lim_{\ve\to 0} \frac{d}{dz}
                     \bigg[ \frac{\theta(z-z_0-\ve) \ln^{n+1}z}{n+1} \bigg]
\hspace{2cm}z_0\ge 0
\,,\\
\int_0^{z} dz' \cL_n(z',z_0) g(z')&=\int^z_{z_0} dz'\,\frac{\ln^n z'}{z'}[g(z')-g(z_0)]
+g(z_0)\frac{\ln^{n+1}z}{n+1}
\,,\end{align}
where $g(z)$ is a test function. In the standard distribution
$\cL_n(z)$ the subtraction of the singularity occurs at the singular
point $z=0$, while in $\cL(z,z_0)$ the subtraction occurs at $z=z_0$
even if there is no singularity when $z_0\neq 0$. $\cL(z,z_0)$ reduces
to $\cL(z)$ at $z_0=0$.

Evaluating phase space or loop integrals at $\cO(\as)$ or higher in
dimensional regularization, we encounter singular terms like
\begin{align}\label{plusID1}
\frac{\theta(z)}{z^{1+\e}}&=
-\frac{\delta(z)}{\e}+\cL_0(z)-\epsilon \cL_1(z)+\cO(\e^2)
\,,\end{align}
which have been expanded in powers of $\epsilon$ making use of the
plus distributions in \eq{plus-def}.  We also encounter more
complicated doubly-singular expressions, \emph{e.g.} in
\eqs{gW-c}{Ifin-ns}, which can be expanded in $\epsilon$ using both
\eqs{plus-def}{plus2}, such as:
\begin{align}\label{plusID2}
&\frac{\theta(\tau)}{\tau^{1+\e}} 
\frac{\theta\big(z-\tfrac{\tau}{1+\tau}\big)}{z^{1+\e}}
= \frac{\delta(z)\delta(\tau)}{2\e^2}-\frac{\delta(\tau)\cL_0(z)}{\e}
 + I^s(\tau,z)+ I^{\ns}(\tau,z)+\cO(\e) \,,
\end{align}
where the ${\cal O}(\epsilon^0)$ terms are
\begin{align}
&I^s(\tau,z)
=\delta(\tau)\cL_1(z)
+\cL_0(z)\cL_0(\tau)-\cL_1(\tau)\delta(z)
\,,\label{Is}\\
&I^{\ns}(\tau,z)
=-\big[\delta\big(z-\tfrac{\tau}{1+\tau}\big)-\delta(z)\big]
    \frac{\theta(\tau)\ln\tau}{\tau}
+\big[\cL_0\big(z,\tfrac{\tau}{1+\tau}\big)-\cL_0(z)\big]\frac{\theta(\tau)}{\tau}
+\delta\big(z-\tfrac{\tau}{1+\tau}\big) \frac{\ln(1+\tau)}{\tau}
\nn\\&\hspace{2cm}
-\frac{\delta'\big(-\tau+\tfrac{z}{1-z}\big)}{(1-z)^2}
 \theta(\tau)\bigg[-\Li(z)+\Li\Big(\frac{\tau}{1+\tau}\Big)+\frac{\ln^2(1+\tau)}{2} 
 -\frac{ \ln^2(z/\tau)}{2}\bigg]
\label{Ins}
\,,\end{align}
where $\Li(z)$ is the dilogarithm, defined by 
\be
\Li(z)= -\int_0^z dz' \frac{\ln(1-z')}{z'}\,.
\ee
The function $I^s(\tau,z)$ is singular both in $\tau$ and $z$,
depending on both ${\cal L}(\tau)$ and ${\cal L}(x)$, while
$I^{\ns}(\tau,z)$ is not singular in $\tau$ (though still singular in
$z$).  Note that the term on the last line of \eq{Ins}, which has a
$\delta'\big(-\tau+\tfrac{z}{1-z}\big)$, will not contribute to any of
our perturbative structure functions because the expression in
brackets that it multiplies and its derivative respect to $\tau$ are
both zero at $\tau=\tfrac{z}{1-z}$.

\section{Hadronic Tensor at Parton Level}
\label{app:Wparton}
\begin{figure}[t]{
\begin{center}
 \includegraphics[height=.30\columnwidth]{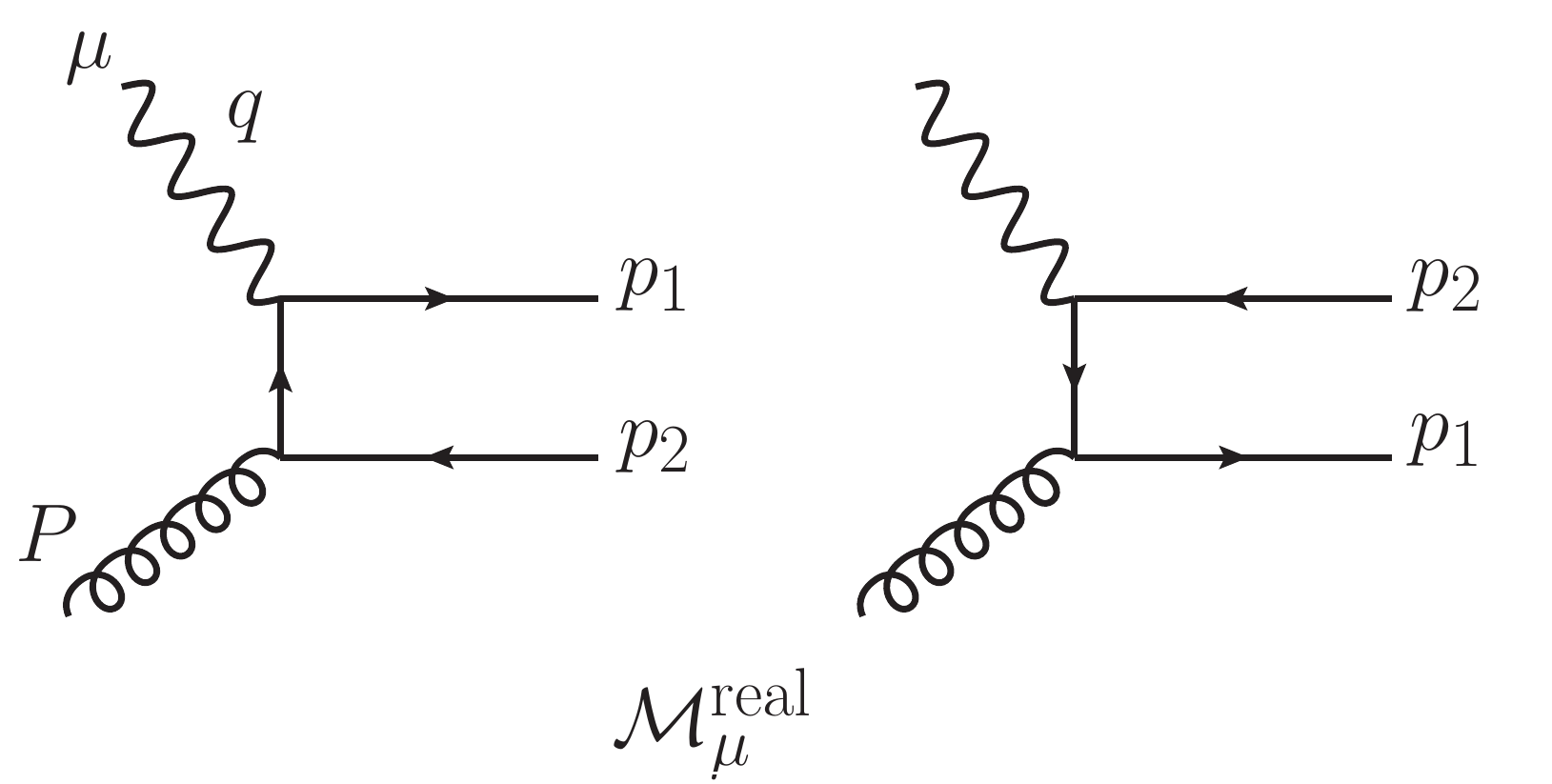}
\end{center}
\vspace{-2em}
{ \caption{Diagrams for $\gamma^* + \text{gluon}$ processes at
    $\cO(\as)$. They contribute to the real amplitudes in
    \eqs{gMM-g}{ppMM-g}. We sum the two diagrams that are related by
    interchanging the quark and antiquark to obtain
    \eqs{gMM-g}{ppMM-g}, thus it is necessary only to sum over the
    five light flavors $u,d,s,c,b$ once, not twice.}
  \label{fig:ggam}} }
\end{figure}

In this section we calculate the hadronic tensor $W_{\mu\nu}$ defined
in \eq{WxQ2tau} where the proton initial state is replaced with a
partonic (quark or gluon) state.  Such a computation allows us to
extract the short-distance matching coefficients $w_{\mu\nu}^{q,g}$ in
\eq{W-factor} onto PDFs, as described in \ssec{matching}.  We denote
the tensor for a quark initial state as $W^q_{\mu\nu}$ and for a gluon
initial state as $W^g_{\mu\nu}$.  Up to $\cO(\as)$, $W^q_{\mu\nu}$
involves a tree-level contribution and the one-gluon diagrams in
\fig{qgam}, and can be decomposed into
\begin{align}
\label{Wq-decompose}
 W^{q}_{\mu\nu}=W_{\mu\nu}^{(0)}+W_{\mu\nu}^\text{vir}+W_{\mu\nu}^\text{real}
\,.\end{align}
Meanwhile, $W^g_{\mu\nu}$ is given just by tree-level real diagrams
at $\cO(\as)$, shown in \fig{ggam}.

The partonic tensor $W_{\mu\nu}^i$ can be computed from \eq{WdPhin},
which is a phase space integral over the squared amplitude. In this
section we compute the squared amplitudes; in the next section we will
evaluate the complete phase space integrals.  \figs{qgam}{ggam}
represent the $\cO(\as)$ amplitudes for initial quark and initial
gluon states. In \appx{MMq} and \appx{MMg} we evaluate the squared
amplitudes built from these diagrams.  The real diagrams have two-body
final states with momenta $p_1$ and $p_2$ and as described in
\fig{regions-v}, they can enter two back-to-back hemispheres in four
different ways, and the formula for the 1-jettiness $\tau$ in terms of
$p_{1,2}$ in each configuration differs.

\subsection{Squared Amplitudes for $\gamma^* +q$}
\label{app:MMq}

\begin{figure}[t]{
\begin{center}
 \includegraphics[height=.25\columnwidth]{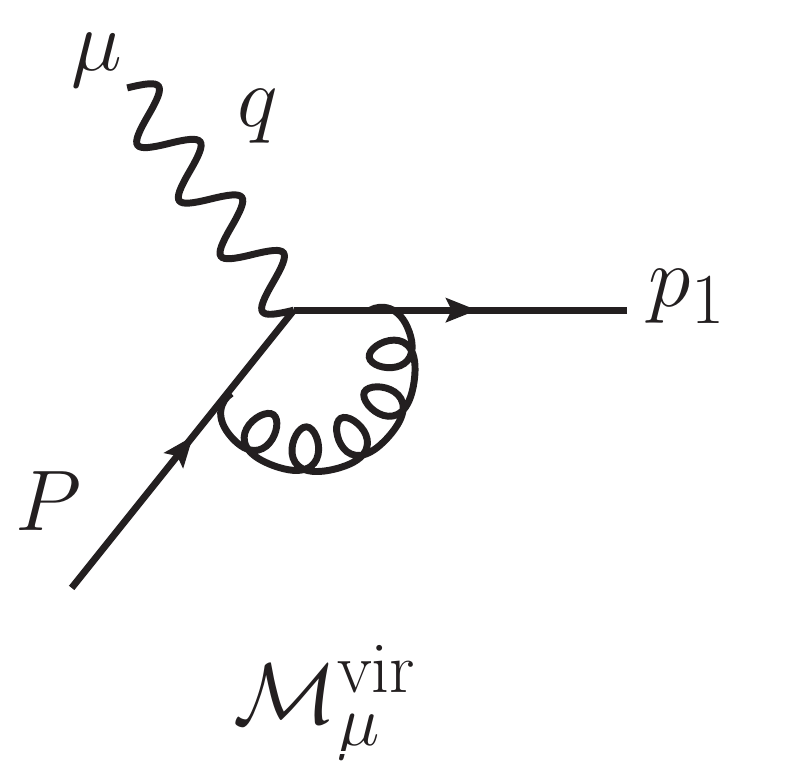}\hspace{1cm}
 \includegraphics[height=.25\columnwidth]{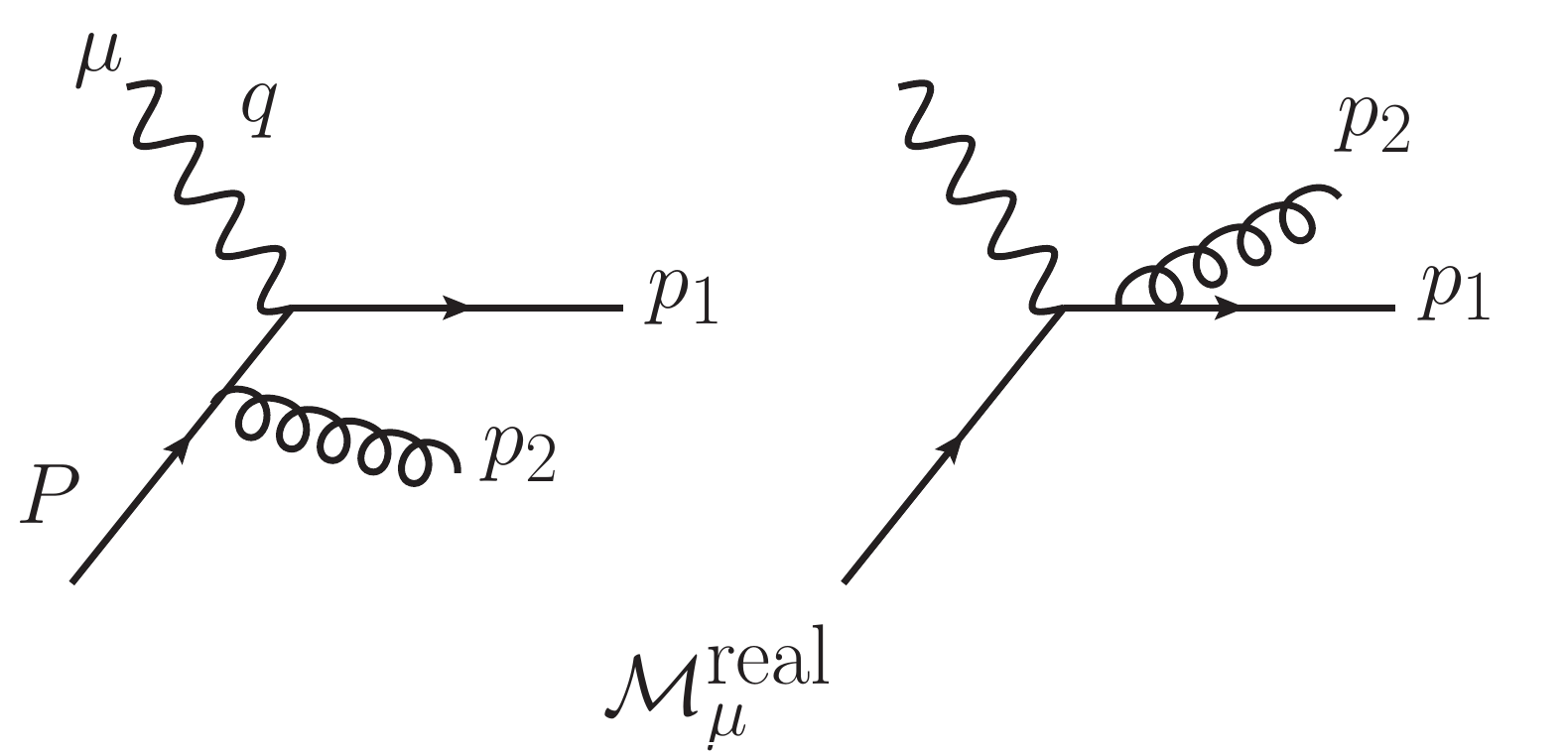} 
\end{center}
\vspace{-2em}
{ \caption{Virtual and real diagrams for $\gamma^* + \text{quark}$
    processes at $\cO(\as)$. They contribute to the virtual amplitudes
    in \eqs{gMM-vir}{ppMM-vir} and the real amplitudes in
    \eqs{gMM-real}{ppMM-real}. There are also corresponding diagrams
    for incoming antiquarks.}  \label{fig:qgam}} }
\end{figure}

For the process $\gamma^*(q^\mu)+q(P^\mu) \to q(p_1^\mu)$ at tree
level, the amplitude is $\bar{u}(p_1)\gamma_\mu u(P)$. To obtain the
structure functions \eq{Fi-def} one needs projected squared amplitudes
as
\begin{align}
\label{gMM-tree}
-g^{\mu\nu}\, \cM^{(0)}_\mu \cM^{(0)\,*}_\nu
&= 4Q_f^2\,Q^2
\,,\\
\label{ppMM-tree}
P^\mu P^\nu\,\cM^{(0)}_\mu \cM^{(0)\,*}_\nu &= 0
\,,\end{align}
where we have also summed over all quark spins. The projection in
\eq{ppMM-tree} is zero because of the Dirac equation
$P\!\!\!\!\slash\, u(P)=0$.  Here, $Q_f$ is the electromagnetic charge
of quark with flavor $f$. We do not sum over flavors for the quark
tensors until we convolve with PDFs.

The virtual contribution can be extracted from the literature, see,
\emph{e.g.}, Eq.~(14.19) in \cite{Sterman:1994ce}. At $\cO( \as)$ we
obtain the cross terms between the tree-level and the virtual diagram
shown in \fig{qgam}:
\begin{align}
\label{gMM-vir}
&-g^{\mu\nu}\,\big[ \cM^{\text{vir}}_\mu \cM^{(0)\,*}_\nu + \cM^{(0)}_\mu \cM^{\text{vir}\,*}_\nu\big]
 \nn \\&\hspace{1cm}
= -8Q^2\frac{\as C_F Q_f^2}{2\pi} (1-\epsilon) \bigg(\frac{4\pi\mu^2}{Q^2}\bigg)^\e
\frac{\Gamma(1+\epsilon) \Gamma(1-\epsilon)^2}{\Gamma(1-2\epsilon)}
\bigg(\frac{1}{\epsilon^2} +\frac{3}{2\epsilon}+4\bigg) 
\,,\nn\\& \hspace{1cm}=
4Q^2 \frac{\as C_F Q_f^2}{2\pi} (1-\epsilon)
\bigg[ -\frac{2}{\e^2}-\frac{1}{\e}\bigg(2\ln \frac{\mu^2}{Q^2}+3 \bigg) 
-\ln^2 \frac{\mu^2}{Q^2} - 3\ln\frac{\mu^2}{Q^2}+\frac{\pi^2}{6}-8
\bigg]
\,,  \\
\label{ppMM-vir}
&P^\mu P^\nu\,\big[ \cM^{\text{vir}}_\mu \cM^{(0)\,*}_\nu + \cM^{(0)}_\mu \cM^{\text{vir}\,*}_\nu\big] = 0
\,,\end{align}
again summed over all quark spins. We have kept the factor
$(1-\epsilon)$ out front because it is to be cancelled by the same
factor in \eq{Fi-def}.  In the second step of \eq{gMM-vir}, we
converted to the $\MSbar$ scheme, making the replacement:
\be
\label{MSbar}
\mu^2 \to \frac{\mu^2 e^{\gamma_E}}{4\pi}\,.
\ee
Note that the finite part in \eq{gMM-vir} is the $\as$ term of the
hard function, which already appeared in our discussion in
Ref.~\cite{Kang:2013nha}.  For discussion of the hard function in SCET
see \cite{Bauer:2003di,Manohar:2003vb}.  \eq{ppMM-vir} is zero again
by the Dirac equation $P\!\!\!\!\slash\, u(P)=0$.

The real contribution to $W_{\mu\nu}^q$ in \eq{Wq-decompose} at $\cO
(\alpha_s)$ comes from two diagrams for $\gamma^*(q^\mu)+q(P^\mu)\to
q(p_1^\mu)+g(p_2^\mu)$ shown in \fig{qgam}.  The projected amplitudes
for the diagrams, summed over quark spins and gluon polarizations, are
given by
\begin{align}
\label{gMM-real}
-g^{\mu\nu}\cM_\mu^{\text{real}} \cM_\nu^{\text{real}\,*}
&= 32\pi\as C_FQ_f^2  (1-\epsilon) \bigg(\frac{\mu^2 e^{\gamma_E}}{4\pi}\bigg)^\e
\nn\\&\hspace{2cm}
\times\bigg[
 (1-\epsilon) \bigg( \frac{1-x}{v}
+ \frac{v}{1-x} \bigg)
+2 \frac{x}{1-x}\frac{1-v}{v}
+2\epsilon
\bigg]
\,,\\
\label{ppMM-real}
P^\mu P^\nu\,\cM_\mu^{\text{real}} \cM_\nu^{\text{real}\,*} &= 
16\pi\as  C_F Q_f^2 Q^2(1-\epsilon) \bigg(\frac{\mu^2 e^{\gamma_E}}{4\pi}\bigg)^\e\frac{1-v}{x}
\,,\end{align}
where $v=p_2^-/Q$ as in \eq{dPhi2-v} or \fig{regions-v}.
\eq{gMM-real} can be found from Eq.~(14.23) in \cite{Sterman:1994ce}.

\subsection{Squared Amplitudes for $\gamma^* +g$}
\label{app:MMg}
The tree-level process with an initial gluon
$\gamma^*(q^\mu)+g(P^\mu)\to q (p_1^\mu)\,\bar{q}(p_2^\mu)$ starts at
$\cO(\as)$, illustrated in \fig{ggam}.  The projected amplitudes for
the process, summed over gluon polarizations and quark spins, are
given by
\begin{align}
\label{gMM-g}
-g^{\mu\nu}\cM_\mu^{\text{real}} \cM_\nu^{\text{real}\,*}
&= 32\pi\as T_F \sum_f Q_f^2 (1-\e) \bigg(\frac{\mu^2 e^{\gamma_E}}{4\pi}\bigg)^\e
\nn\\&\hspace{2cm}\times
\bigg[
 (1-\epsilon) \bigg( \frac{1-v}{v}
+ \frac{v}{1-v} \bigg)
-2 \frac{x(1-x)}{v(1-v)}
-2\epsilon
\bigg]
\,,\\
\label{ppMM-g}
P^\mu P^\nu\,\cM_\mu^{\text{real}} \cM_\nu^{\text{real}\,*} &= 
32\pi\as  T_F \sum_f Q_f^2Q^2(1-\epsilon)  \bigg(\frac{\mu^2 e^{\gamma_E}}{4\pi}\bigg)^\e
\frac{1-x}{x}
\,.\end{align}
Note that these are symmetric under $v\to 1-v$, which just switches
the final-state quark and antiquark in \fig{ggam}. Note also that the
projection in \eq{ppMM-g} is independent of $v$, making the
phase-space integral in \eq{dPhi2-v} particularly simple. Here we go
ahead and include the sum over quark flavors $f\in \{u,d,s,c,b\}$
since the gluon PDF with which we will convolve these results is
independent of quark flavors produced in the final states in
\fig{ggam}. Since both possibilities of the photon interacting with the quark or the antiquark are already included in the sum of the two diagrams in \fig{ggam}, we need sum over the five light flavors only once, and not repeat the sum for antiquark flavors $\bar f$.

\subsection{Projected Hadronic Tensor}
\label{app:Wprojected}
In this subsection we obtain hadronic tensors by integrating the
squared amplitudes obtained in \appx{MMq} and \appx{MMg} using the
phase space integrations in \eqs{dPhi1}{dPhi2-v}. The latter goes over
the four regions in \fig{regions-v} with a different formula for
$\tau(x,v)$ depending on which hemispheres the two final-state
particles enter.

\subsubsection{Quark tensor}

For $P^\mu P^\nu\, W^q_{\mu\nu}$, only the real emission contribution
\eq{ppMM-real} is nonzero, and it contains no IR divergence, so we can
safely set $\epsilon=0$.  Using \eq{dPhi2-v} to integrate
\eq{ppMM-real} over the four regions in \fig{regions-v}, we obtain the
contributions
\begin{subequations}
\label{PPWq}
\begin{align}
\label{PPWq-a}
P^\mu P^\nu W_{\mu\nu}^{(a)} &= \as C_F Q_f^2 Q^2 \frac{1-2x}{2x}\theta\Bigl(\frac{1}{2}-x\Bigr) \delta(\tau-1) \,,\\
\label{PPWq-b}
P^\mu P^\nu W_{\mu\nu}^{(b)} &= \as C_F Q_f^2 Q^2 x\tau \,\Theta_0(\tau,x) \,,\\
\label{PPWq-c}
P^\mu P^\nu W_{\mu\nu}^{(c)} &= \as C_F Q_f^2 Q^2 (1-x\tau) \,\Theta_0(\tau,x) \,,\\
\label{PPWq-d}
P^\mu P^\nu W_{\mu\nu}^{(d)} &= \as C_F Q_f^2 Q^2 \frac{2x-1}{2x}\theta\Bigl(x-\frac{1}{2}\Bigr) \delta\Bigl(\tau-\frac{1-x}{x}\Bigr)\,,
\end{align}
\end{subequations}
where the generalized theta function $\Theta_0$ is defined in
\eq{Theta0}. The sum of the four contributions in \eq{PPWq} gives the
result:
\begin{subequations}
\begin{align}
& P^\mu P^\nu\, W^q_{\mu\nu}=\as C_F Q_f^2\,w_P^q\,,
\label{ppW-q}\\
& w_P^q =   Q^2\,\Theta_0(\tau,x)
\left[1 +\delta\Bigl(\frac{1}{1+\tau}-x\Bigr)\frac{1-\tau}{2(1+\tau)^2}  
        +\delta(\tau-1)  \frac{1-2x}{2x}
\right]
\label{wPq}
\,,\end{align}
\end{subequations}
where in the middle term of \eq{wPq} we rescaled variables in the
delta function in \eq{PPWq-d}. This result gives the matching
coefficient $w_P^q$ in \eq{matchingcoeffs}.

The tree-level and virtual contributions to $-g^{\mu\nu}\,
W_{\mu\nu}^{q}$ are given by inserting \eqs{gMM-tree}{gMM-vir} into
the formula for a one-body final-state phase space in \eq{dPhi1}:
\be
\label{gW-0v}
\begin{split}
-g^{\mu\nu} (W_{\mu\nu}^{(0)} + W_{\mu\nu}^{\text{vir}}) &= 
\biggl\{ 4\pi Q_f^2 
+ 2\as C_F Q_f^2 (1-\epsilon) \biggl[ -\frac{2}{\e^2}-\frac{1}{\e}\bigg(2\ln \frac{\mu^2}{Q^2}+3 \bigg)  \\
 &\hspace{3cm} 
-\ln^2 \frac{\mu^2}{Q^2} - 3\ln\frac{\mu^2}{Q^2}  +\frac{\pi^2}{6}-8\biggr]\biggr\}
\delta(1-x)\delta(\tau)\,.
\end{split}
\ee 
The contribution from the real diagrams in \eq{gMM-real} is more
involved. We must integrate \eq{gMM-real} over the two-body phase
space using \eq{dPhi2-v}. We consider in turn the four contributions
$-g^{\mu\nu}\, W_{\mu\nu}^{\text{real}\,(a,b,c,d)}$ corresponding to
the four regions in \fig{regions-v}.

In region $(a)$, where $x<v<1-x$ and $x<1/2$, the integrand in
\eq{gMM-real} is finite and we can set $\epsilon=0$ in
\eqs{dPhi2-v}{gMM-real}, giving
\begin{align}
-g^{\mu\nu}\, W_{\mu\nu}^{\text{real}\,(a)}
&=2\as C_F Q_f^2\,\delta(\tau-1)\,\theta\Bigl(\frac{1}{2}-x\Bigr)
\bigg[ \frac{(1-2x)(1-4x)}{2(1-x)} +\frac{1+x^2}{1-x}\ln\frac{1-x}{x}
\bigg]
\label{gW-a}
\,.\end{align}
In region $(b)$, $v>x$ and $v>1-x$ and $v=1-x\tau$. Because the
$\Theta_0(\tau,x)$ in \eq{Theta0} sets $x <1/(1+\tau)$ the term
$1/(1-x)$ in \eq{gMM-real} is finite for the region of $x$.  So, $\e$
can be set to zero in \eqs{gMM-real}{dPhi2-v} and we have
\begin{align}
-g^{\mu\nu}\, W_{\mu\nu}^{\text{real}\,(b)}
&=
2\as C_FQ_f^2
\Theta_0(\tau,x)
\,x\bigg(
\frac{1-x}{1-x\tau}
+\frac{1-x\tau}{1-x}
+2 \frac{x^2}{1-x} \frac{\tau}{1-x\tau}
\bigg)
\label{gW-b}
\,.\end{align}
In region $(c)$, where $v<x$ and $v<1-x$ and $v=x\tau$, there are two
IR divergent terms in \eq{gMM-real} that go like $1/\tau$ and $1/[\tau
  (1-x)]$, which can be expanded by using the identities in
\eqs{plusID1}{plusID2}.  Then, we have
\begin{align}
-g^{\mu\nu}\, W_{\mu\nu}^{\text{real}\, (c)}
&=
2\as C_F Q_f^2\bigg( \frac{\mu^2}{Q^2} \bigg)^\e
 (1-\e) 
\Theta_0(\tau,x)
\nn\\&\quad\times
\bigg[
\bigg(\frac{1}{\e^2} +\frac{3}{2\e} \bigg) \delta(\tau)\delta(1-x) -\frac{\delta(\tau) P_{qq}(x)}{\e}
+E^s(\tau,x)+E^{\ns}(\tau,x)
\bigg]
\label{gW-c}
\,,\end{align}
where we converted to the $\MSbar$ scheme using \eq{MSbar}, and the
singular and nonsingular parts of the finite terms are given by
\begin{align}\label{Ifin-s}
E^s(\tau,x) 
&= -2\cL_1(\tau)\delta(1-x) 
+\cL_0(\tau) [P_{qq}(x)-\frac32\delta(1-x)] 
\nn\\
&\qquad 
+\delta(\tau)\bigg[ (1+x^2)\cL_1(1-x)-\frac{\pi^2}{12}\delta(1-x) +1-x \bigg] 
\,,\\\label{Ifin-ns}
E^{\ns}(\tau,x) &=-(2-\tau)\frac{x^2}{1-x} +2x\, I^{\ns}(\tau,1-x)
\,,\end{align}
where $\cL_n$ and $I^{\ns}(\tau,1-x)$ are given above in
\eqs{plus-def}{Ins}.  The splitting function $P_{qq}(x)$ is given by
\eq{Pij}.

In region $(d)$, where $1-x<v<x$ and $\tau=(1-x)/x$, the term
$1/(1-x)$ in \eq{gMM-real} is IR divergent because the condition
$\Theta_0(\tau,x)$ becomes $\theta(x-1/2)\theta(1-x)$.  Integrating
\eq{gMM-real} by using \eq{dPhi2-v} and expanding in $\e$ by using
\eq{plusID1}, we obtain in $\MSbar$,
\begin{align}\label{gW-d}
&-g^{\mu\nu}\, W_{\mu\nu}^{\text{real}\,(d)}
=2\as C_F Q_f^2 (1-\e)\bigg(\frac{\mu^2}{Q^2}\bigg)^\epsilon
 \delta\bigg(\frac{1}{1+\tau}-x\bigg)\theta(x-1/2)\,
\\
&\quad \times
\bigg[
\delta(\tau)\bigg( \frac{1}{\epsilon^2}+\frac{3}{2\epsilon} +\frac{7}{2}-\frac{5\pi^2}{12}\bigg)
-\frac{3}{2}\cL_0(\tau)-2\cL_1(\tau)
+\frac{(3\tau^2+8\tau+13)+2(2\tau^2+5\tau+4)\ln\tau}{2(1+\tau)^3}
\bigg] \nn
\,.\end{align}

Now we collect all pieces contributing to $-g_{\mu\nu}W_{\mu \nu}^{q}$
and sum them together.  The IR divergent $1/\e^2$ and $1/\e$ terms
appear with $\delta(1-x)\delta(\tau)$ which are all canceled when the
virtual part from \eq{gW-0v} and real parts in \eqs{gW-c}{gW-d} are
added together.  There is one additional IR divergence with
$P_{qq}(x)\delta(\tau)/\epsilon$ that is associated with the one-loop
quark PDF, and hence remains uncancelled when adding virtual and real
contributions. Summing all the terms in \eqsss{gW-a}{gW-b}{gW-c}{gW-d}
together with the tree-level and virtual contributions from
\eq{gW-0v}, we obtain the final result
\begin{subequations}
\begin{align}
&-g^{\mu\nu}\, W^q_{\mu\nu}
=4\pi Q_f^2 \delta(1-x) \delta(\tau)
+2\as  C_FQ_f^2 (1-\e)\bigg[ -\frac{ P_{qq}(x)}{\e}\delta(\tau)+w_G^q \bigg]\,,
\label{gW-q}\\
&w_G^q =
\Theta_0(\tau,x)
\bigg[
\delta(\tau) \,S_{-1}^q(x) 
+\cL_0(\tau) \,S_{0}^q(\tau,x)
+\cL_1(\tau)\,S_{1}^q(\tau,x)
\nn\\&\hspace{3cm}
+R^q(\tau,x)
+\delta(\tau-1) \,\Delta_1^q (x)
+\delta(\tfrac{1}{1+\tau}-x)\,\Delta_2^q(\tau)
\bigg]
\label{wGq}
\,.\end{align}
\end{subequations}
Here we separately write IR divergent and finite terms in \eq{gW-q} in
order to clearly show the structure of the result, which we
anticipated above in \eq{gWform}. From this result we extract the
matching coefficient $w_G^q$ in \eq{matchingcoeffs}.  The functions
$S_i^q$ are coefficients of singular terms in $\tau$, $R^q$ is regular
in $\tau$, and $\Delta_i$ are coefficients of delta functions.  They
are given by
\begin{align}
S_{-1}^q(x)&=
-P_{qq}(x) \ln\bigg(\frac{\mu^2}{Q^2}\bigg)
+ (1+x^2)\cL_1(1-x)-\bigg( \frac{9}{2}+\frac{\pi^2}{3}\bigg)\delta(1-x) +1-x 
\,,\nn\\
S_{0}^q(\tau,x)&=
2x\,\cL_0\big(1-x,\tfrac{\tau}{1+\tau}\big) -\frac{3}{2}\delta\big(\tfrac{1}{1+\tau}-x\big)+(1-x)
\,,\nn\\
S_{1}^q(\tau,x)&=
-2\frac{2+\tau}{1+\tau}\delta \big(\tfrac{1}{1+\tau}-x \big)
\,,\nn\\ 
R^q(\tau,x)&=
x\bigg[\frac{1-x}{1-x\tau}+\frac{1-x\tau}{1-x}
+2 \frac{x^2}{1-x} \frac{\tau}{1-x\tau}
\bigg]
-(2-\tau)\frac{x^2}{1-x} 
\,,\nn\\
\Delta_1^q (x)&=
\frac{(1-2x)(1-4x)}{2(1-x)} +\frac{1+x^2}{1-x}\ln\big(\frac{1-x}{x}\big)
\,,\nn\\
\Delta_2^q(\tau)&=
\frac{(3\tau^2+8\tau+13)+2(2\tau^2+5\tau+4)\ln\tau}{2(1+\tau)^3}+\frac{2}{\tau}\frac{\ln(1+\tau)}{1+\tau}
\label{coeffq}
\,.\end{align}

\subsubsection{Gluon tensor}

The calculation of the hadronic tensor for the gluon state follows the
same steps as for the quark state.  For the projection $P^\mu P^\nu
W_{\mu\nu}^g$, we insert \eq{ppMM-g} into the two-body phase space
integral \eq{dPhi2-v}, and obtain
\begin{subequations}
\begin{align}
& P^\mu P^\nu\, W^g_{\mu\nu} =\as T_F \sum_f Q_f^2\, w_P^g \,,
\label{ppW-g}\\
& w_P^g = 2   Q^2\Theta_0(\tau,x)
\frac{1-x}{x}
\left[ 2x +\delta(\tau-1)  (1-2x)
          +\delta(\tau-\tfrac{1-x}{x})(2x-1)
\right]
\label{wPg}
\,.\end{align}
\end{subequations}
The integration in \eq{dPhi2-v} for this projection is particularly
simple since the squared amplitude in \eq{ppMM-g} is independent of
$v$. So we do not give the individual contributions in regions
(a)--(d) in \fig{regions-v} separately. From the result \eq{wPg} we
obtain the matching coefficient $w_P^g$ in \eq{matchingcoeffs}.

For the projection $-g^{\mu\nu} W_{\mu\nu}^g$, we insert \eq{gMM-g}
into \eq{dPhi2-v}, and obtain in the four different regions in
\fig{regions-v},
\begin{subequations}
\begin{align}
\label{gWg-a}
-g^{\mu\nu} W_{\mu\nu}^{g(a)} &= 4\as  T_F \sum_f Q_f^2 \delta(\tau-1) \theta\bigg(\frac12 - x\bigg) I_{a}(x) \,,\\
\label{gWg-bc}
-g^{\mu\nu} W_{\mu\nu}^{g(b,c)} &= 2\as T_F \sum_f Q_f^2 \Bigl(\frac{\mu^2 e^{\gamma_E}}{Q^2}\Bigr)^\epsilon \frac{1-\epsilon}{\Gamma(1-\epsilon)} \Theta_0(\tau,x) I_b(\tau,x,\e) \,, \\
\label{gWg-d}
-g^{\mu\nu} W_{\mu\nu}^{g(d)} &= 4\as  T_F \sum_f Q_f^2 \delta\bigg(\tau-\frac{1-x}{x}\bigg) \theta\bigg(x-\frac12 \bigg) I_{d}(x) \,,
\end{align}
\end{subequations}
where
\begin{subequations}
\begin{align}
I_a(x) &= -I_d(x) \equiv 2(2x-1) + 2P_{qg}(x)\ln\frac{1-x}{x} \,,\\
I_b(\tau,x,\e) &= P_{qg}(x) \Bigl\{ \Bigl[ -\frac{1}{\epsilon} - 1 + \ln(1-x) \Bigr] \delta(\tau) + \cL_0(\tau) + \frac{x}{1-x\tau}\Bigr\} + \delta(\tau) - 2x\,.
\end{align}
\end{subequations}
We see that contributions $(a),(d)$ in \eqs{gWg-a}{gWg-d} are finite
while contributions $(b,c)$ in \eq{gWg-bc} contain an IR divergent
term associated with the gluon PDF. In \eq{gWg-bc} we work in the
$\MSbar$ scheme, see \eq{MSbar}. Summing contributions $(a)$-$(d)$ in
\eqss{gWg-a}{gWg-d}{gWg-bc}, we obtain the result
\begin{subequations}
\begin{align}
& -g^{\mu\nu}\, W_{\mu\nu}^{g}
= 4\as T_F\sum_f Q_f^2 (1-\e)
\bigg[
-\frac{P_{qg}(x)}{\e}\delta(\tau)+w_G^g
\bigg]\,,
\label{gW-g}\\
& w_G^g =
\Theta_0(\tau,x)\bigg\{
\bigg[1+P_{qg}(x)\bigg(-1+\ln (1-x)-\ln\frac{\mu^2}{Q^2}\bigg)\bigg]\,\delta(\tau)
+P_{qg}(x)\cL_0(\tau)
\nn\\&\hspace{3cm}
+R^g(\tau,x)
-\big[\delta(\tau-1)-\delta(\tau-\tfrac{1-x}{x})\big]  \Delta^g(x)
\bigg\}
\label{wGg}
\,,\end{align}
\end{subequations}
where we again separately write the IR divergent and finite terms to
reflect the structure anticipated in \eq{gWform}. This result gives
the matching coefficient $w_G^g$ in \eq{matchingcoeffs}.  The
functions $R^g$ and $\Delta^g$ are defined by
\begin{align}
\label{coeffg}
R^{g}(\tau,x) &= -x \bigg( 2-\frac{P_{qg}(x)}{1-x\tau}\bigg)
\,,\nn\\
\Delta^g(x) &= 1-2x-P_{qg}(x)\, \ln\frac{1-x}{x}
\,.\end{align}

\section{Separating Singular and Nonsingular Parts of Hadronic Tensor}
\label{app:Wns}
Here, we isolate the singular and nonsingular parts of the projections
of the hadronic tensor for quark and gluon initial states computed in
\appx{Wparton}.  The tensor is obtained by convolving short distance
coefficients determined by perturbative matching in \ssec{matching}
with PDFs as in \eq{W-factor}.  The nonsingular part is obtained by
subtracting singular part of the $W_{\mu\nu}$ tensor that has been
already calculated by using SCET in~\cite{Kang:2013nha}.

One can also separate singular and nonsingular parts by isolating the
structures $\delta(\tau)$ and $\cL_n(\tau)$ that encode the most
singular terms in the $\tau\to 0$ limit in \eqs{gW-q}{gW-g}. The
nonsingular part is then obtained by subtracting these terms from
\eqs{gW-q}{gW-g}. There is no singular term in \eqs{ppW-q}{ppW-g}.  We
can separately carry out perturbative matching for singular part and
nonsingular part and determine the short distance coefficients of each
part.

We write hadronic tensors in terms of three pieces associated with
PDFs for $q, \bar q, g$
\begin{align}\label{ppW}
P^\mu P^\nu W_{\mu\nu}
&= \frac{ 2\pi Q^2 }{x^2} \big(\cA_q+\cA_{\bar{q}}+\cA_g\big)
\,,\\
\label{gW}
-g^{\mu\nu} W_{\mu\nu}
&= 8\pi(1-\e)\big(\cB_q+\cB_{\bar q}+\cB_g\big)
\,.
\end{align}
In \eq{ppW} the factor $1/x^2$ is factored out to clarify that it
comes from the product of proton momenta $P^\mu P^\nu$.  The
differential structure functions $\cF_i$ in \eq{sigmaF1L} can be
expressed in terms of $\cA_i$ and $\cB_i$ by using \eq{Fi-def} in
similar pattern to \eqs{F1AB}{FLAB},
\be
\label{cF1cFL}
\cF_1 = \sum_{i\in\{q,\bar q,g\}} (\cA_i + \cB_i) \,,\qquad \cF_L = \sum_{i\in\{q,\bar q,g\}} 4x\cA_i\,.
\ee
As we promised we present the results in terms of singular and
nonsingular parts
\be
\cA_i=\cA_i^\sing +\cA_i^\ns\,, \quad\quad
\cB_i=\cB_i^\sing +\cB_i^\ns\,.
\ee
The singular parts $\cA_i^\sing$ and $\cB_i^\sing$ can be extracted
from the calculation of the singular cross section in
\cite{Kang:2013nha}, giving
\begin{subequations}\label{cAicBi-sing}
\begin{align}
\label{cAsqg}
\cA^\sing_{q,g} &= 0
\,,\\\label{cBsq}
\cB^\sing_q &=\sum_f Q_f^2\Bigg\{  
f_q(x)\frac{\delta(\tau)}{2}
 -   \frac{\as C_F}{4\pi} f_q(x) \biggl[\bigg(\frac92+\frac{\pi^2}{3}\bigg)\delta(\tau) + 3 \cL_0(\tau) +4\cL_1(\tau) \biggr] \\&
\hspace{-7mm} + \frac{\as C_F}{4\pi} \int _x^1 \frac{dz}{z} f_q\bigg(\frac{x}{z}\bigg) 
\bigg(\Bigl[\cL_1(1-z) \,(1+z^2)+(1-z) + P_{qq}(z)\ln \frac{Q^2}{\mu^2} \Bigr] \delta(\tau)
+ P_{qq}(z)\cL_0(\tau) \bigg)
\Bigg\} \nn
\,,\\ \label{cBsg}
\cB^\sing_g &= \sum_f Q_f^2 \frac{\as T_F}{2\pi} 
\int_x^1 \frac{dz}{z}f_g\bigg(\frac{x}{z}\bigg) 
\biggl[ \bigg(1-P_{qg}(z) + P_{qg}(z)\ln \frac{Q^2 (1-z)}{\mu^2}\bigg)\delta(\tau)+ P_{qg}(z)\cL_0(\tau)\biggr] 
\,,\end{align}
\end{subequations}
where $P_{qq}$ and $P_{qg}$ are given in \eq{Pij}. The antiquark
contributions $\cA_{\bar q}^\sing$ and $\cB_{\bar q}^\sing$ are
obtained by simply replacing $q\to\bar q$ in \eqs{cAsqg}{cBsq}. We now
include the sum over flavors in both the quark and gluon
contributions.

The nonsingular parts $\cA_i^\ns$ and $\cB_i^\ns$ are given by
\begin{subequations}\label{cAicBi}
\begin{align}
\label{cAq}
\cA_q^\ns &=
\sum_f Q_f^2 \frac{\as C_F}{4\pi} \,
\biggl\{
\Theta_0\,\bigg[\int_x^{\tfrac{1}{1+\tau}}dz\, 2z\,f_q\bigg(\frac{x}{z}\bigg)
+ \frac{1-\tau}{(1+\tau)^3}f_q\big(x(1+\tau)\big)
\bigg]
\\&\hspace{5.5cm} 
+\delta(\tau-1) \int_x^{1/2} dz(1-2z) f_q\bigg(\frac{x}{z}\bigg)
\biggr\} \nn
\,,\\
\label{cAg}
\cA_g^\ns &=
\sum_f Q_f^2 \frac{\as T_F}{\pi}  
\,\biggl\{
\Theta_0\,\bigg[
\int_x^{\tfrac{1}{1+\tau}}dz\, 2 z (1-z) f_g\bigg(\frac{x}{z}\bigg) 
+\frac{\tau(1-\tau)}{(1+\tau)^4} f_g\big(x(1+\tau)\big)
\bigg]
\\&\hspace{5cm}
+ \delta(\tau-1) \int_x^{1/2} dz\,(1-z)(1-2z) f_g\bigg(\frac{x}{z}\bigg)
\biggr\} \nn
\,,\\\label{cBq}
\cB_q^\ns &=
\sum_f Q_f^2 \frac{\as C_F}{4\pi}
\biggl\{
N_{1}(\tau,x)+ N_{0}(\tau,x) +\delta(\tau-1)\int_x^{1/2} \frac{dz}{z}f_q\bigg(\frac{x}{z}\bigg)\,\Delta_1^q (z)
\\&\hspace{2.5cm}
+\Theta_0\,\bigg[
\int_x^{\tfrac{1}{1+\tau}}\frac{dz}{z} f_q\bigg(\frac{x}{z}\bigg) R^q(\tau,z)
+(1+\tau)\,f_q(x(1+\tau))\Delta_2^q(\tau)
\bigg]
\biggr\} \nn
\,,\\ \label{cBg}
\cB_g^\ns &=
\sum_f Q_f^2 \frac{\as T_F}{2\pi} 
\biggl\{
\Theta_0 \bigg[-\frac{1}{\tau}
\int^1_{\tfrac{1}{1+\tau}} \frac{dz}{z} f_g\bigg(\frac{x}{z}\bigg)\,P_{qg}(z)
+ \int_x^{\tfrac{1}{1+\tau}}\frac{dz}{z} f_g\bigg(\frac{x}{z}\bigg) R^g(\tau,z)
\nn\\& \hspace{3.5cm}
-\delta(\tau-1)\int_x^{1/2}\frac{dz}{z}f_g\bigg(\frac{x}{z}\bigg)\,\Delta^g(z)
+ \frac{f_g(x(1+\tau))}{1+\tau}\Delta^g (\tfrac{1}{1+\tau})
\bigg]
\nn\\&\hspace{2.7cm}
-\frac{\Theta_1+\Theta_2}{\tau}
\int^1_{x} \frac{dz}{z} f_g\bigg(\frac{x}{z}\bigg)\,P_{qg}(z)
\biggr\}
\,,\end{align}
\end{subequations}
and the antiquark contributions $\cA_{\bar q}^{\text{ns}}$ and $\cB_{\bar q}^{\text{ns}}$ are given by the replacement $q\to\bar q$ in \eqs{cAq}{cBq}. Recall that $\Theta_0=\theta(\tau)\theta(1-\tau)\theta[(1-x)/x-\tau]$.
In \eq{cAicBi} we defined two additional theta functions
\be
\label{Theta12}
\Theta_1=\theta(-x+1/2)\,\theta(\tau-1)\,,\qquad  
\Theta_2=\theta(x-1/2)\,\theta(\tau)\,\theta(\tau-\tfrac{1-x}{x})\,.
\ee
These theta functions turn on only beyond the physical region of
$\tau$ defined by \eq{tmax}, and multiply terms that cancel the part
of the singular terms beyond $\taumax$.   The
functions $N_{0,1}$ in \eq{cBq} are the nonsingular parts of the functions
$S_{0,1}^q(\tau,x)$ in \eq{coeffq}
\begin{align}\label{intN01}
N_1 (\tau,x)
&=-4\frac{\ln \tau}{\tau}\,
\bigg\{
\Theta_0 \,\bigg[(1+\tau/2)f_q\big(x(1+\tau)\big)-f_q(x)\bigg]
-(\Theta_1+\Theta_2)\,f_q(x)
\bigg\}
\,,\nn\\
N_{0}(\tau,x)
&=\frac{\Theta_0}{\tau} \,\bigg\{-\frac{3}{2} \,\bigg[(1+\tau)f_q\big(x(1+\tau)\big)-f_q(x)\bigg] 
+2\ln\frac{\tau}{1+\tau} \bigg[f_q\big(x(1+\tau)\big)-f_q(x)\bigg]
\nn\\&\hspace{5cm} 
- \int^1_{\tfrac{1}{1+\tau}} dz\bigg[ 2\frac{f_q\big(\frac{x}{z}\big)-f_q(x)}{1-z}
+f_q\bigg(\frac{x}{z}\bigg)\frac{1-z}{z}\bigg]
\bigg\}
\nn\\&\quad
-\frac{\Theta_1+\Theta_2}{\tau} 
\,\bigg\{
\bigg[ -\frac32+2\ln(1-x)\bigg] f_q(x)
+\int_x^1 dz \bigg[2\frac{f_q\big(\frac{x}{z}\big)-f_q(x)}{1-z}
+f_q\bigg(\frac{x}{z}\bigg)\frac{1-z}{z}\bigg]
\bigg\}
\,.\end{align}
Note that the terms with $1/\tau$ and $(\ln \tau)/\tau$ are multiplied
by a term proportional to $\tau$ in the limit $\tau\to 0$ or by
$\Theta_{1,2}$ which turn off for small $\tau$, thus $N_{0,1}$ is not
singular.  For the same reason, the term with a $1/\tau$ in \eq{Bg} is
nonsingular.  The functions $R^{g,q}$ and $\Delta^{q,g}$ are given in
\eqs{coeffq}{coeffg}. The $\delta(\tau-1)$ terms in \eq{cAicBi}
correspond to the events where all final particles go to the beam
hemisphere as described in \sec{Fi}.

The cumulative version $A_i$ and $B_i$ of $\cA_i$ and $\cB_i$ can be
defined in the same way as \eqs{ppW}{gW} by integrating both sides
over $\tau$.  Their explicit expressions are given in
\eqs{AiBi}{AiBi-sing} and the delta functions in \eq{cAicBi} give rise
to discontinuities in the cumulative versions at the maximum value of
$\tau$ in \eq{AiBi}, as illustrated in \fig{disc}.

\eqs{Bq}{Bg} for $B_{q,g}$ can be re-expressed as sums of terms which
are all individually explicitly nonsingular by writing:
\begin{subequations}
\label{BNS}
\begin{align}
\label{BqNS}
B_q^\ns &= \sum_f Q_f^2 \frac{\as C_F}{4\pi} 
 \Biggl( \Theta_0 \biggl\{ \int_{x}^{\frac{1}{1+\tau}} \frac{dz}{z} \frac{f_q(x/z)}{1-z} \Bigl[ (1-4z) z\tau - (1+z^2) \ln(1-z\tau)\Bigr] \\
&\qquad\qquad\qquad + \int_{\frac{1}{1+\tau}}^1 \frac{dz}{z}\frac{1}{1-z}\biggl[ f_q(x/z)\Bigl( \frac{1-4z}{2} - (1+z^2)\ln z\tau \Bigr) + f_q(x) \Bigl(\frac{3}{2} + 2\ln\tau\Bigr)\biggr] \biggr\} \nn \\
&\qquad\qquad + (\Theta_1+\Theta_2) \biggl\{ \int _x^1 \frac{dz}{z}\frac{1}{1-z} \biggl[ f_q(x/z) \Bigl( \frac{1-4z}{2} - (1+z^2)\ln z\tau\Bigr) + f_q(x) \Bigl( \frac{3}{2}+2\ln\tau\Bigr) \biggr] \nn \\
&\qquad \qquad\qquad + f_q(x)  \Bigl( \frac{3}{2}+2\ln\tau\Bigr) \ln\frac{\tau x}{1-x} \biggr\}\Biggr) \,,\nn \\
\label{BgNS}
B_{g}^\ns &=
 \sum_f Q_f^2 \frac{\as T_F}{2\pi} \biggl( -\Theta_0\biggl\{  \int_x^{\frac{1}{1+\tau}}\frac{dz}{z} f_g(x/z) [2z\tau + P_{qg}(z)\ln(1-z\tau)]   \\
 &\quad +  \int_{\frac{1}{1+\tau}}^1 \frac{dz}{z}f_g(x/z) \bigl[ 1 + P_{qg}(z)\ln (z\tau)\bigr] \biggr\}  - (\Theta_1+\Theta_2) \int_x^1 \frac{dz}{z}f_g(x/z) \bigl[ 1 + P_{qg}(z)\ln (z\tau)\bigr] \biggr) \nn
\,.\end{align}
\end{subequations}
These forms can be more useful for numerical evaluation.

\section{Profile Function}
\label{app:profile}

The concept of profile functions was introduced in
Refs.~\cite{Ligeti:2008ac,Abbate:2010xh}. An additional complication
in DIS is that the transition between regions encoded in the profile
functions also involves dependence on $x$. Here we present the profile
function for DIS that are used for the jet, beam, and soft scales to
obtain the resummed $\tau$ cross section that is discussed in
\sec{results}.

The scales $\mu_{H,B,J,S}$ are parameterized in terms of the overall
renormalization scale $\mu$ and and a function $\mu_\text{run}(\tau)$
as
\begin{align}
 \mu_H   &=\mu\,,\nn\\
 \mu_{B,J}(\tau)&=[1+e_{B,J}\, g(\tau)]\,\sqrt{\mu\, \mu_\text{run}(\tau)}\,,\nn\\
 \mu_S(\tau) &=[1+e_S\, g(\tau)]\,\mu_\text{run}(\tau)\,.
\label{mui}
\end{align}
The parameters $e_{B,J,S}$ in \eq{mui} are used to perform variations
of the scales $\mu_{B,J,S}$ to estimate uncertainties from omitted
higher-order corrections to beam, jet, and soft functions. By default
$e_{B,J,S}=0$, and are varied away from zero according to
\eq{scalevariations} below. The function
$g(\tau)=\theta(t_3-\tau)\,(1-\tau/t_3)^2$ is designed to go to zero
beyond $\tau=t_3$, where the resummation is turned off with
$\mu_H=\mu_B=\mu_J=\mu_S=\mu_{\rm ns}$, and it no longer makes sense
to have an individual variation of the scales $\mu_{B,J,S}$.  This
parameterization maintains the relations $\mu_{J}=\sqrt{\mu_H \mu_S}$
and $\mu_{B}=\sqrt{\mu_H \mu_S}$ for the default values
($e_{B,J,S}=0$).

Theoretically the function $\mu_\text{run}(\tau)$ must be chosen to
satisfy several key properties to ensure the proper treatment of
different regions of $\tau$:
\begin{enumerate}
\item In the region $\ln\tau\gtrsim \as^{-1}$ where logs of $\tau$
  need to be resummed, it follows ``canonical'' scaling $\mu_S\sim
  Q\tau$ and $\mu_{B,J}\sim Q\sqrt{\tau}$.
\item For very small $\tau\sim\Lqcd/Q$ it reaches a plateau at a
  constant value $\mu_0$ where $\mu_0\gtrsim 1\,{\rm GeV}$ (above
  $\Lqcd$). This is the nonperturbative regime where a shape function
  becomes necessary.
\item For larger $\tau\sim 1$ (where $\tau<1$) it becomes equal to a
  constant value $\mu$ independent of $\tau$. This is the region where
  the resummation is turned off and the prediction reverts to
  fixed-order.
\item It must smoothly interpolate between each pair of regions.
\end{enumerate}
Various parameters are varied to account for the residual ambiguity in
satisfying these criteria.  One choice that satisfies these criteria
is the profile function,
\begin{align}
\label{murun}
& \mu_\mathrm{run}(\tau) =
\begin{cases}
\mu_0 + \alpha \tau^\beta\, \mu & \tau \leq t_1
\,,\\
r\, \tau \mu & t_1 \leq \tau \leq t_2
\,,\\
\zeta(\tau,t_2,t_3)\,\mu & t_2 \leq \tau \leq t_3
\,,\\
\mu & \tau > t_3
\,.\end{cases}
\end{align}
This is what we use in the singular part of the cross section in
\eq{xsection2}, with the corresponding $\mu_S(\tau)$ illustrated in
\fig{profile}.  Other choices for the profile function are also
possible, see, \emph{e.g.}, \cite{Kang:2013nha}.  The function
$\mu_\mathrm{run}(\tau)$ in \eq{murun} is linear in $\tau$ with a
slope $r$ from $t_1$ to $t_2$ so that the value of $\mu_\mathrm{run}$
sets $\mu_{B,J,S}$ to be canonical via \eq{mui}.  The function
approaches $\mu_0$ below $t_1$, and $\mu$ above $t_2$ via a smoothly
rising function $\zeta$.  The requirement of continuity for
$\mu_\mathrm{run}(\tau,\mu)$ and its first derivative at $t_1,\,t_2,$
and $t_3$, determine the parameters $\alpha$, $\beta$ and constrain
the function $\zeta(\tau,t_2,t_3)$ at $t_2$ and $t_3$, for which we
choose two connected quadratic polynomials:
\begin{align}
\label{murun-param}
&
\beta=\bigg(1-\frac{\mu_0}{r t_1\mu}\bigg)^{-1}
\,, \qquad
\alpha= \frac{r}{\beta t_1^{\beta-1}}\,,
\nn \\[.3cm]
&
\zeta(\tau,t_2,t_3)=
\begin{cases}
a+b\,\tau+c\,\tau^2 & t_2 \leq \tau \leq (t_2+t_3)/2
\,,\nn\\
a'+b'\,\tau+c'\,\tau^2 &  (t_2+t_3)/2\leq\tau \leq t_3
\,,\end{cases}
\nn \\&
c =2\frac{1-r(t_2+3t_3)/4}{(t_3-t_2)^2}
\,, \qquad\quad
b = r-2ct_2\,,\qquad
a= (r-b)t_2-c t_2^2
\,, 
\nn \\
&
c' =-2\frac{1-r(3t_2+t_3)/4}{(t_3-t_2)^2}
\,, \qquad
b' = -2c't_3\,,\qquad \ 
a' = 1-b' t_3-c' t_3^2
\,,
\end{align}

The default central values of the parameters that we choose are:
\be
\label{profileparam}
\begin{split}
\mu&=Q\,,\quad \mu_0=2~\mathrm{GeV}\,,\quad r=1\,,\quad e_{B,J,S}=0\,,
\\
t_1&=\frac{3~\mathrm{GeV}}{\mu}\,,\quad t_2=0.5(0.8-x)\,,\quad t_3=0.8\, \taumax \,.
\end{split}
\ee
The central values of structure function and cross section results
plotted in \sec{results} correspond to the use of these parameters.
Above $t_2$ the resummation effect is being gradually turned off, and
near $t_3$ the fixed order contribution dominates.  We choose $t_3$ to
be roughly the size of $\taumax$.  For $t_2$ we require that it 
well separated from $t_3$ by more than 0.3 for smooth turn-off of the
resummation, and that it be close to the region where the nonsingular
and fixed-order singular parts are of the same size. The value of
$t_2$ determined in this way depends on $x$, and is well approximated
for $x\lesssim 0.7$ with a linear fit as in \eq{profileparam}.

To estimate theoretical uncertainties in the cross section
\eq{xsection2} due to missing higher order terms in fixed-order and
resummed perturbation theory, the scales $\mu_H$, $\mu_{B,J}$, and
$\mu_S$ are varied by changing $\mu$ and $e_{B,J,S}$ \eq{mui}.  We
also vary the points $t_1$, $t_2$, and $t_3$ and $\mu_0$.  Each
parameter is separately varied one by one while keeping the others at
their default values.  The variations we perform around the central
values are as follows:
\begin{subequations}
\label{scalevariations}
\begin{align}
 \quad \delta\mu &= (2^{\pm 1} -1)Q\,,
\quad  \delta\mu_0 = \pm 0.5 \GeV\,,
 \quad \delta e_{B,J} = \pm \frac{1}{3},\pm\frac{1}{6}\,,
 \quad \delta e_S = \pm\frac{1}{3},\pm\frac{1}{6}\,,\\
  \quad \delta t_1&=\pm 0.8\GeV/\mu\,,\quad 
    \delta t_2=\pm 0.1 \,(0.8-x)\,,\quad t_3=\pm 0.1\,\taumax\,.
\end{align}
\end{subequations}
The deviations in the cross section \eq{xsection2} due to each of
these variations and the nonsingular scale variation in \eq{muNS} are
summed in quadrature to obtain the uncertainty bands in \fig{sig}.

\section{Resummed Singular Cross Section}
\label{app:resum}

Here, we collect expressions for the resummed singular part of the
cross section in \eq{xsection2} that were obtained in
\cite{Kang:2013nha} using SCET. We provide the expressions that are
necessary to obtain the resummed results in \sec{results} at NLL$'$
accuracy.  For further details on the factorization and resummation
procedure see Ref.~\cite{Kang:2013nha}.

The factorization theorem for $\tau\ll 1$ has been derived in
\cite{Kang:2013nha} and is expressed in terms of hard, jet, beam and
soft functions.  Those functions depend on the factorization scale
$\mu$ and contain large logs of $\mu^2/Q^2$, $\mu^2/(\tau Q^2)$, or
$\mu^2/(\tau^2 Q^2)$.  The large logarithms, $\ln(\tau)$, can be
resummed by evolving the functions from their natural scale
$\mu_{H,J,B,S}$ where the logs are minimized, to the scale $\mu$.  The
result of this procedure, which gives the resummed singular part of
the cross section in \eqs{xsection}{xsection2}, can be written for the
cumulative distribution as:
\begin{align}
\nn
&\hat\sigma^c_\sing(x,Q^2,\tau;\mu_H,\mu_J,\mu_B,\mu_S)
\\\nn&\qquad = 
\frac{ e^{\cK-\gamma_E\Omega}}{\Gamma(1+\Omega)}
\left(\frac{Q}{\mu_H}\right)^{\eta_H(\mu_H,\mu)}
\left(\frac{\tau\,Q^2}{\mu_{B}^2}\right)^{\eta_{B}(\mu_B,\mu)}
\left(\frac{\tau\,Q^2}{\mu_{J}^2}\right)^{\eta_{J}(\mu_J,\mu)}
\left(\frac{\tau\,Q}{\mu_{S}}\right)^{2\eta_{S}(\mu_S,\mu)}
\nn \\&
\quad\times \bigg[ \sum_j Q_f^2\, \int_x^1 \, \frac{dz}{z} \,
  f_j(x/z,\mu_B)\,\left[ W_{qj}(z,\tau)+ \Delta W_{qj}(z)\right] 
  +(q\leftrightarrow \bar{q}) 
  \bigg]
\,,\label{resummedtaumB}
\end{align}
where the cross section is normalized as in \eq{rescale}.  Here $j$
sums over quark flavors and gluons, and the $+(q\leftrightarrow \bar
q)$ includes the term for photon coupling to an antiquark.  In
\eq{resummedtaumB}, the exponential and gamma functions on the first
line on the right-hand side contain the RG evolution kernels
$\cK,\Omega$, and the terms $W_{qj}$ and $\Delta W_{qj}$ on the last
line are fixed-order factors arising from convolution of the jet,
beam, and soft functions.  For NLL$'$ accuracy, we need the evolution
kernels at NLL accuracy and the fixed-order factors at $\cO(\as)$.

The evolution kernels $\mathcal{K}$ and $\Omega$ are the sum of
kernels for each function.
\begin{subequations}\label{kernel}
\begin{align} \label{Kappa}
\cK &\equiv \mathcal{K}(\mu,\mu_H,\mu_J,\mu_B,\mu_S) = K_H(\mu_H,\mu) + K_J(\mu_J,\mu) + K_B(\mu_B,\mu) +2 K_S(\mu_S,\mu) \\
\label{Omega}
\Omega &  \equiv \Omega(\mu_J,\mu_B,\mu_S) = \eta_J(\mu_J,\mu) + \eta_B(\mu_B,\mu) + 2\eta_S(\mu_S,\mu)\,,
\end{align}
\end{subequations}
where the individual evolution kernels $K_H$, $K_J=K_B$, $K_S$,
$\eta_J=\eta_B$, and $\eta_S$ are obtained by solving RG equations for
hard, jet/beam, and soft functions and are given by integrals over
their anomalous dimensions.  Their explicit expressions can be
obtained
from~\cite{Balzereit:1998yf,Bauer:2000yr,Manohar:2003vb,Bauer:2003di,Neubert:2004dd,
  Fleming:2007xt, Ligeti:2008ac},
\begin{align} \label{Hrun}
K_i(\mu_0,\mu) &= n_i \,K_{\Gamma^q}(\mu_0,\mu) + K_{\gamma_i}(\mu_0,\mu)
\,,\nn\\
\eta_i(\mu_0,\mu) &= m_i\,\eta_{\Gamma^q}(\mu_0,\mu)
\,,\end{align}
where $n_i=\{-4,4,4,-2\}$ and $m_i=\{4,-2,-2,2\}$ for $i=\{H,B,J,S\}$
and the subscripts $\Gamma^q$ and $\gamma_i$ indicate cusp and
non-cusp parts of the anomalous dimensions.  The evolution kernels in
\eq{Hrun} at NLL are given by the expressions
\begin{align} \label{Keta}
K_\Gamma(\mu_0, \mu) &= -\frac{\Gamma_0}{4\beta_0^2}\,
\biggl\{ \frac{4\pi}{\alpha_s(\mu_0)}\, \Bigl(1 - \frac{1}{r} - \ln r\Bigr)
   + \biggl(\frac{\Gamma_1 }{\Gamma_0 } - \frac{\beta_1}{\beta_0}\biggr) (1-r+\ln r)
   + \frac{\beta_1}{2\beta_0} \ln^2 r
 \biggr\}
\,, \nn\\
\eta_\Gamma(\mu_0, \mu) &=
 - \frac{\Gamma_0}{2\beta_0}\, \biggl[ \ln r
 + \frac{\alpha_s(\mu_0)}{4\pi}\, \biggl(\frac{\Gamma_1 }{\Gamma_0 }
 - \frac{\beta_1}{\beta_0}\biggr)(r-1)
   \biggr]
\,, \nn\\
K_\gamma(\mu_0, \mu) &=
 - \frac{\gamma_0}{2\beta_0}\,\ln r
\,.\end{align}
Here, $r = \alpha_s(\mu)/\alpha_s(\mu_0)$, and $\as$ is evaluated
using the two-loop running coupling,
\begin{align} \label{alphas}
\frac{1}{\alpha_s(\mu)} &= \frac{X}{\alpha_s(\mu_0)}
  +\frac{\beta_1}{4\pi\beta_0}  \ln X
 \,,\end{align}
where $X\equiv 1+\alpha_s(\mu_0)\beta_0 \ln(\mu/\mu_0)/(2\pi)$.  The
kernels in \eq{Keta} are written in terms of the coefficients in the
expansion of the anomalous dimensions and beta function,
\be \label{F}
\Gamma^q(\as) = \sum_{n=0}^{\infty}\Gamma^q_n\,\left(\frac{\as}{4\pi}\right)^{n+1} ,\quad \gamma_i(\as) = \sum_{n=0}^{\infty}\gamma_{i\,n}\,\left(\frac{\as}{4\pi}\right)^{n+1} ,\quad \beta(\as) = -2\as\sum_{n=0}^\infty \beta_n\left(\frac{\as}{4\pi}\right)^{n+1}\,.
\ee
At NLL, we only need $\gamma_i$ to one loop and $\Gamma^q$ to two
loops \cite{Korchemsky:1987wg}, as well as the two-loop beta function
$\beta$. In the $\MSbar$ scheme the coefficients in \eq{F} used in
\eq{Keta} are given by
\begin{align} \label{Gacuspexp}
\scriptsize
\beta_0 &= \frac{11}{3}\,C_A -\frac{4}{3}\,T_F\,n_f
\,, & \Gamma^q_0 &= 4C_F\,,\qquad  \gamma^q_{H\,0}  = -2 \gamma_{B\,0}^q = -2 \gamma_{J\,0}^q = -12 C_F \nn \\
\beta_1 &= \frac{34}{3}\,C_A^2  - \Bigl(\frac{20}{3}\,C_A\, + 4 C_F\Bigr)\, T_F\,n_f \,,
& \Gamma^q_1 &= 4C_F \Bigl[\Bigl( \frac{67}{9} -\frac{\pi^2}{3} \Bigr)\,C_A  -
   \frac{20}{9}\,T_F\, n_f \Bigr]  
\,.\end{align}
The anomalous dimension for the soft function is obtained from the
consistency relation $\gamma_S = -\gamma_H^q/2 - \gamma_B^q$.

In the cross section \eq{resummedtaumB}, individual factors on the
right-hand side depend on the overall factorization scale $\mu$, but
in the combination of all terms, this depends cancel out completely at
any fixed order in either fixed-order or resummed perturbation
theory. In contrast, the dependence of \eq{resummedtaumB} on $\mu_H$,
$\mu_B$, $\mu_J$, and $\mu_S$ only cancels out order-by-order in
resummed perturbation theory. So at any given order there is always
residual dependence on these four variables that is cancelled by
higher-order terms.  This residual dependence is utilized as a measure
of the remaining theoretical uncertainty.

When all these scales are set to be same $\mu_H=\mu_J=\mu_B=\mu_S$,
\eq{kernel} reduces to zero, the resummation factors on the first line
of the right-hand side of \eq{resummedtaumB} become unity, and
\eq{resummedtaumB} reduces to the fixed-order singular part which is
given in \eq{AiBi-sing}.
The fixed-order parts in \eq{resummedtaumB} are given by
\begin{subequations}
\label{W}
\begin{align}
W_{qj}(z,\tau) &=
H(Q^2,\mu_H) \, 
\sum_{ \substack{ n_1,n_2, \\ n_3=-1 } } ^{1}
 J_{n_1}\Big[\alpha_s(\mu_J),\frac{\tau Q^2}{\mu_J^2}\Big]\,
 I^{qj}_{n_2}\Big[\as(\mu_B), z, \frac{\tau Q^2}{\mu_B^2} \Big]
\,\,
S_{n_3}\Big[\alpha_s(\mu_{S}),\frac{\tau Q}{\mu_{S}} \Big] 
\nn\\&\quad\times
\sum_{\ell_1=-1}^{n_1+n_2+1}\sum_{\ell_2=-1}^{\ell_1+n_3+1}
V_{\ell_1}^{n_1 n_2} V_{\ell_2}^{\ell_1 n_3}  \,
 V^{\ell_2}_{-1}(\Omega) \,
\,, \\[5pt]
\Delta W_{qj}(z) &
=\frac{\as(\mu_B) }{2\pi} \left[ \delta_{jq} C_F P_{qq}(z) +\delta_{jg} T_F P_{qg}(z)\right]\, \ln z 
\,,
\end{align}
\end{subequations}
where $H(Q^2,\mu_H)$ is hard function and $J_{n}$, $I^{qq}_{n}$,
$I^{qg}_{n}$, $S_{n}$ are the coefficients of jet, beam, and soft
functions and we need the function and coefficients at $\cO(\as)$.
Note that the coefficients functions contain logarithms of their last
argument and the hard function also depends on the logarithm
$\ln(Q^2/\mu_H^2)$. The logs in these fixed-order factors are
minimized by choosing the canonical scales
\be \label{canonicalscales}
 \mu_H = Q\,,\quad \mu_J = \mu_B = Q\sqrt{\tau}\,,\quad \mu_S = Q\tau\,.
\ee
Large logs of ratios of the above scales are then resummed to all
orders in $\as$ by RG evolution to the scale $\mu$, given by the
evolution kernels $\cK$ and $\Omega$ in \eq{kernel}.  The choices in
\eq{canonicalscales} are appropriate in the tail region, and
correspond to the result used with the profile \eq{murun} in the
region between $t_1$ and $t_2$.

The hard function at $\cO(\as)$ \cite{Bauer:2003di,Manohar:2003vb} is
given by
\be
\label{scalarH}
H(Q^2,\mu) = 1 + \frac{\as(\mu)C_F}{2\pi} \left(- \ln^2\frac{\mu^2}{Q^2} - 3\ln\frac{\mu^2}{Q^2} - 8 + \frac{\pi^2}{6}\right)\,. \nn
\ee
The soft, jet, and beam functions can be decomposed into a sum of plus
distributions $\cL_n$,
\be
\label{Gseries1}
G(t,\mu)=\frac{1}{\mu^{n_G}} \sum_{n=-1}^{1} G_n[\alpha_s(\mu)] \, \cL_n\left(\frac{t}{\mu^{n_G}}\right)
\,.\ee
where $G(t,\mu)$ represents the soft function $S(k,\mu)$, the jet
function $J(t,\mu)$, or the matching coefficient $I^{qq,qg}(t,z,\mu)$
onto PDFs in the beam function \cite{Stewart:2009yx,Stewart:2010qs}.
The index $n_G=\{1,2,2\}$ for $G=\{S,J,I\}$. Thus the variable $t$ has
dimension $+2$ for $J$ and $I$, and has dimension $+1$ for $S$.  The
coefficients $G_n$ in \eq{Gseries1} for the three functions are $S_n$,
$J_n$, and $I_n^{qq,qg}$.  These coefficients are given at order $\as$
by
\begin{subequations}\label{SJn}
\begin{align}
&S_{-1}(\as)=1+\frac{\as C_F}{4\pi} \frac{\pi^2}{3}\,,
&S_0(\as)&= 0\,, 
&S_1(\as)&= \frac{\as C_F}{4\pi} (-16)\,,
\label{Sn}\\[.2cm]
&J_{-1}(\as)=1+\frac{\as C_F}{\pi} \left(\frac74 -\frac{\pi^2}{4}  \right)\,,
&J_0(\as)&= -\frac{\as C_F}{\pi} \frac{3}{4}\,, 
&J_1(\as)&= \frac{\as C_F}{\pi}\,,
\label{Jn}
\end{align}
\end{subequations}
and
\begin{align}\label{In}
&I^{qq}_{-1}(\as,z)=\cL_{-1}(1\minus z)+\frac{\as C_F}{2\pi} \bigg[ \cL_1(1\minus z) (1\plus z^2)
\minus\frac{\pi^2}{6} \cL_{-1}(1\minus z)+\theta(1 \minus z)\Bigl( 1\minus z \minus \frac{1\plus z}{1 \minus z}\ln z\Bigr)
\bigg]\,, \nn
\\
&I^{qq}_0(\as,z)=\frac{\as C_F}{2\pi} \theta (z) \left( P_{qq}(z) -\frac32 \cL_{-1}(1-z)\right)\,, 
\hspace{.5cm}
I^{qq}_1(\as,z)= \frac{\as C_F}{2\pi}\, 2\cL_{-1}(1-z)\,,
\nn\\[.2cm]
&I^{qg}_{-1}(\as,z)=\frac{\as T_F}{2\pi}\theta(z)\! \Bigl[P_{qg}(z)\ln\frac{1 \minus z}{z} \plus 2\theta(1\minus z) z(1 \minus z)   \!\Bigr] \,, \quad 
I^{qg}_0(\as,z)= \frac{\as T_F}{2\pi}\theta(z) P_{qg}(z)\,,
\end{align}
where coefficients not listed above are zero at $\cO(\as)$.

The argument of the plus distributions $\cL_n$ in \eq{Gseries1} can be
rescaled by $\lambda$ and rewritten as
\be
\label{Fseries2}
G(t,\mu)=\frac{1}{\lambda\mu^{n_G}} \sum_{n=-1}^{1} G_n[\as (\mu),\lambda] \, \cL_n\left(\frac{\lambda^{-1} t}{\mu^{n_G}}\right)
\,,\ee
where the coefficients $G_n(\as,\lambda)$ are expressed in terms of
$G_n(\as)$ in \eq{Gseries1} as
\begin{align}
\label{Gn}
G_{-1}(\as,\lambda) &= G_{-1}(\as) +\sum_{n=0}^{\infty} G_n(\as)\frac{\ln^{n+1} \lambda}{n+1}\,,
\nn\\
G_n(\as,\lambda) &=\sum_{k=0}^{\infty}\frac{(n+k)!}{n!\, k!} G_{n+k}(\as) \ln^k \lambda\,,
\end{align}
where $G_n = \left\{S_n,J_n,I_n^{qq,qg}\right\}$.  Explicit
expressions for $S_n(\as,\lambda)$, $J_n(\as,\lambda)$, and
$I^{qq,qg}_n(\as,\lambda)$ are obtained by inserting \eqs{SJn}{In}
into \eq{Gn}.

The coefficients $V_k^{mn}$ and $V_k^n(\Omega)$ in \eq{W} are produced
by convolutions of plus distributions in jet, beam, and soft
functions.  The coefficients $V_k^n(a)$ and $V_k^{mn}$ are obtained
from the Taylor series expansion of $V(a,b)$ around $a = 0$ and $a = b
= 0$, where $V(a,b)$ is defined by
\be \label{Vab}
V(a,b)=\frac{\Gamma(a) \Gamma(b)}{\Gamma(a+b)} -\frac{1}{a}-\frac{1}{b}\,,
\ee
which satisfies $V(0,0) = 0$.  The $V_k^n(a)$ for $n\ge 0$ are

\be \label{eq:Vkna-def}
V_k^n(a) =
 \begin{cases}
  a\, \frac{\df^n}{\df b^n}\,\frac{V(a,b)}{a+b}\bigg\vert_{b = 0}\,,
  &   k=-1\,, \\[10pt]
  a\, \binom{n}{k}   \frac{\df^{n-k}}{\df b^{n-k}}\, V(a,b)
  \bigg\vert_{b = 0} + \delta_{kn} \,, \qquad
  & 0\le k\le n \,,  \\[10pt]
   \frac{a}{n+1} \,,
  & k=n+1  \,.
\end{cases}
\ee
The $V_k^{mn}$ are symmetric in $m$ and $n$, and for $m,n\ge 0$ they
are
\be
\label{eq:Vkmn-def}
V_k^{mn}=
 \begin{cases}
 \displaystyle \frac{\df^m}{\df a^m}\, \frac{\df^n}{\df b^n}\,\frac{V(a,b)}{a+b}\bigg\vert_{a = b = 0} \,,
   & k=-1\,, \\[15pt]
\displaystyle  \sum_{p = 0}^m \sum_{q = 0}^n \delta_{p+q,k}\,\binom{m}{p} \binom{n}{q}
\frac{\df^{m-p}}{\df a^{m-p}}\, \frac{\df^{n-q}}{\df b^{n-q}} \ V(a,b)
  \bigg\vert_{a = b = 0}\,,  & 0\le k \le m+n \,,\\[15pt]
 \displaystyle  \frac{1}{m+1} + \frac{1}{n+1}\,, & k=m+n+1 \,.
\end{cases}
\ee
For the cases $n = -1$ or $m = -1$,
\be
\begin{split}
& V_{-1}^{-1}(a) = 1
\,,\qquad
V_0^{-1}(a) = a \,,\qquad
V_{k \geq 1}^{-1}(a) = 0
\,,\qquad\\
& V^{-1,n}_k = V^{n,-1}_k = \delta_{nk}
\,.
\end{split}
\ee

The resummed differential distribution can be written in similar
pattern to \eq{resummedtaumB}, which we do not write out explicitly
here.  Alternatively, the differential distribution can be obtained by
numerically differentiating the cumulant in \eq{resummedtaumB}
\be
\label{cumulantderivative}
\frac{d\hat\sigma_\sing}{d \tau}
= \lim_{\e\to 0} \frac{\hat\sigma^c_\sing(\tau+\e;\mu_i(\tau))-\hat\sigma^c_\sing(\tau-\e;\mu_i(\tau))}{2\e}
\,,\ee
which corresponds to differentiating the explicit $\tau$ dependence in
$\hat\sigma^c$ but not the dependence inside $\mu_i(\tau)$. See
footnote \ref{footnote} on why we choose this procedure.


\bibliography{DIS}

\providecommand{\href}[2]{#2}\begingroup\raggedright\begin{thebibliography}{10}

\bibitem{Catani:1991hj}
S.~Catani, Y.~L. Dokshitzer, M.~Olsson, G.~Turnock, and B.~R. Webber, {\it {New
  clustering algorithm for multi - jet cross-sections in $e^+ e^-$
  annihilation}},  {\em Phys. Lett.} {\bf B269} (1991) 432--438.

\bibitem{Catani:1993hr}
S.~Catani, Y.~L. Dokshitzer, M.~H. Seymour, and B.~R. Webber, {\it
  {Longitudinally invariant $k_\perp$ clustering algorithms for hadron hadron
  collisions}},  {\em Nucl. Phys.} {\bf B406} (1993) 187--224.

\bibitem{Ellis:1993tq}
S.~D. Ellis and D.~E. Soper, {\it Successive combination jet algorithm for
  hadron collisions},  {\em Phys. Rev.} {\bf D48} (1993) 3160--3166,
  [\href{http://xxx.lanl.gov/abs/hep-ph/9305266}{{\tt hep-ph/9305266}}].

\bibitem{Dokshitzer:1997in}
Y.~L. Dokshitzer, G.~D. Leder, S.~Moretti, and B.~R. Webber, {\it Better jet
  clustering algorithms},  {\em JHEP} {\bf 08} (1997) 001,
  [\href{http://xxx.lanl.gov/abs/hep-ph/9707323}{{\tt hep-ph/9707323}}].

\bibitem{Salam:2007xv}
G.~P. Salam and G.~Soyez, {\it A practical {S}eedless {I}nfrared-{S}afe {C}one
  jet algorithm},  {\em JHEP} {\bf 05} (2007) 086,
  [\href{http://xxx.lanl.gov/abs/0704.0292}{{\tt arXiv:0704.0292}}].

\bibitem{Cacciari:2008gp}
M.~Cacciari, G.~P. Salam, and G.~Soyez, {\it The anti-{$k_t$} jet clustering
  algorithm},  {\em JHEP} {\bf 04} (2008) 063,
  [\href{http://xxx.lanl.gov/abs/0802.1189}{{\tt arXiv:0802.1189}}].

\bibitem{Dasgupta:2003iq}
M.~Dasgupta and G.~P. Salam, {\it {Event shapes in $e^+ e^-$ annihilation and
  deep inelastic scattering}},  {\em J.Phys.G} {\bf G30} (2004) R143,
  [\href{http://xxx.lanl.gov/abs/hep-ph/0312283}{{\tt hep-ph/0312283}}].

\bibitem{Farhi:1977sg}
E.~Farhi, {\it A {QCD} test for jets},  {\em Phys. Rev. Lett.} {\bf 39} (1977)
  1587--1588.

\bibitem{GehrmannDeRidder:2007bj}
A.~{Gehrmann-De Ridder}, T.~Gehrmann, E.~W.~N. Glover, and G.~Heinrich, {\it
  {Second-order QCD corrections to the thrust distribution}},  {\em Phys. Rev.
  Lett.} {\bf 99} (2007) 132002, [\href{http://xxx.lanl.gov/abs/0707.1285}{{\tt
  arXiv:0707.1285}}].

\bibitem{GehrmannDeRidder:2007hr}
A.~{Gehrmann-De Ridder}, T.~Gehrmann, E.~W.~N. Glover, and G.~Heinrich, {\it
  {NNLO corrections to event shapes in $e^+e^-$ annihilation}},  {\em JHEP}
  {\bf 12} (2007) 094, [\href{http://xxx.lanl.gov/abs/0711.4711}{{\tt
  arXiv:0711.4711}}].

\bibitem{Weinzierl:2008iv}
S.~Weinzierl, {\it {NNLO corrections to 3-jet observables in electron-positron
  annihilation}},  {\em Phys. Rev. Lett.} {\bf 101} (2008) 162001,
  [\href{http://xxx.lanl.gov/abs/0807.3241}{{\tt arXiv:0807.3241}}].

\bibitem{Weinzierl:2009ms}
S.~Weinzierl, {\it {Event shapes and jet rates in electron-positron
  annihilation at NNLO}},  {\em JHEP} {\bf 06} (2009) 041,
  [\href{http://xxx.lanl.gov/abs/0904.1077}{{\tt arXiv:0904.1077}}].

\bibitem{Becher:2008cf}
T.~Becher and M.~D. Schwartz, {\it A precise determination of {$\alpha_s$} from
  {LEP} thrust data using effective field theory},  {\em JHEP} {\bf 07} (2008)
  034, [\href{http://xxx.lanl.gov/abs/0803.0342}{{\tt arXiv:0803.0342}}].

\bibitem{Abbate:2010xh}
R.~Abbate, M.~Fickinger, A.~H. Hoang, V.~Mateu, and I.~W. Stewart, {\it {Thrust
  at N$^3$LL with Power Corrections and a Precision Global Fit for
  alphas(mZ)}},  {\em Phys. Rev.} {\bf D83} (2011) 074021,
  [\href{http://xxx.lanl.gov/abs/1006.3080}{{\tt arXiv:1006.3080}}].

\bibitem{Lee:2006nr}
C.~Lee and G.~Sterman, {\it Momentum flow correlations from event shapes:
  Factorized soft gluons and {Soft-Collinear Effective Theory}},  {\em Phys.
  Rev.} {\bf D75} (2007) 014022,
  [\href{http://xxx.lanl.gov/abs/hep-ph/0611061}{{\tt hep-ph/0611061}}].

\bibitem{Mateu:2012nk}
V.~Mateu, I.~W. Stewart, and J.~Thaler, {\it {Power Corrections to Event Shapes
  with Mass-Dependent Operators}},  {\em Phys.Rev.} {\bf D87} (2013) 014025,
  [\href{http://xxx.lanl.gov/abs/1209.3781}{{\tt arXiv:1209.3781}}].

\bibitem{Chien:2010kc}
Y.-T. Chien and M.~D. Schwartz, {\it {Resummation of heavy jet mass and
  comparison to LEP data}},  {\em JHEP} {\bf 1008} (2010) 058,
  [\href{http://xxx.lanl.gov/abs/1005.1644}{{\tt arXiv:1005.1644}}].

\bibitem{Adloff:1997gq}
{\bf H1 Collaboration} Collaboration, C.~Adloff et~al., {\it {Measurement of
  event shape variables in deep inelastic e p scattering}},  {\em Phys.Lett.}
  {\bf B406} (1997) 256--270,
  [\href{http://xxx.lanl.gov/abs/hep-ex/9706002}{{\tt hep-ex/9706002}}].

\bibitem{Adloff:1999gn}
{\bf H1 Collaboration} Collaboration, C.~Adloff et~al., {\it {Investigation of
  power corrections to event shape variables measured in deep inelastic
  scattering}},  {\em Eur.Phys.J.} {\bf C14} (2000) 255--269,
  [\href{http://xxx.lanl.gov/abs/hep-ex/9912052}{{\tt hep-ex/9912052}}].

\bibitem{Aktas:2005tz}
{\bf H1 Collaboration} Collaboration, A.~Aktas et~al., {\it {Measurement of
  event shape variables in deep-inelastic scattering at HERA}},  {\em
  Eur.Phys.J.} {\bf C46} (2006) 343--356,
  [\href{http://xxx.lanl.gov/abs/hep-ex/0512014}{{\tt hep-ex/0512014}}].

\bibitem{Breitweg:1997ug}
{\bf ZEUS Collaboration} Collaboration, J.~Breitweg et~al., {\it {Event shape
  analysis of deep inelastic scattering events with a large rapidity gap at
  HERA}},  {\em Phys.Lett.} {\bf B421} (1998) 368--384,
  [\href{http://xxx.lanl.gov/abs/hep-ex/9710027}{{\tt hep-ex/9710027}}].

\bibitem{Chekanov:2002xk}
{\bf ZEUS Collaboration} Collaboration, S.~Chekanov et~al., {\it {Measurement
  of event shapes in deep inelastic scattering at HERA}},  {\em Eur.Phys.J.}
  {\bf C27} (2003) 531--545,
  [\href{http://xxx.lanl.gov/abs/hep-ex/0211040}{{\tt hep-ex/0211040}}].

\bibitem{Chekanov:2006hv}
{\bf ZEUS Collaboration} Collaboration, S.~Chekanov et~al., {\it {Event shapes
  in deep inelastic scattering at HERA}},  {\em Nucl.Phys.} {\bf B767} (2007)
  1--28, [\href{http://xxx.lanl.gov/abs/hep-ex/0604032}{{\tt hep-ex/0604032}}].

\bibitem{Antonelli:1999kx}
V.~Antonelli, M.~Dasgupta, and G.~P. Salam, {\it {Resummation of thrust
  distributions in DIS}},  {\em JHEP} {\bf 0002} (2000) 001,
  [\href{http://xxx.lanl.gov/abs/hep-ph/9912488}{{\tt hep-ph/9912488}}].

\bibitem{Dasgupta:2002dc}
M.~Dasgupta and G.~P. Salam, {\it {Resummed event shape variables in DIS}},
  {\em JHEP} {\bf 0208} (2002) 032,
  [\href{http://xxx.lanl.gov/abs/hep-ph/0208073}{{\tt hep-ph/0208073}}].

\bibitem{Dasgupta:2001sh}
M.~Dasgupta and G.~P. Salam, {\it Resummation of non-global {QCD} observables},
   {\em Phys. Lett.} {\bf B512} (2001) 323--330,
  [\href{http://xxx.lanl.gov/abs/hep-ph/0104277}{{\tt hep-ph/0104277}}].

\bibitem{Stewart:2010tn}
I.~W. Stewart, F.~J. Tackmann, and W.~J. Waalewijn, {\it {N-Jettiness: An
  Inclusive Event Shape to Veto Jets}},  {\em Phys. Rev. Lett.} {\bf 105}
  (2010) 092002, [\href{http://xxx.lanl.gov/abs/1004.2489}{{\tt
  arXiv:1004.2489}}].

\bibitem{Kang:2013nha}
D.~Kang, C.~Lee, and I.~W. Stewart, {\it {Using 1-Jettiness to Measure 2 Jets
  in DIS 3 Ways}},  {\em Phys.Rev.} {\bf D88} (2013) 054004,
  [\href{http://xxx.lanl.gov/abs/1303.6952}{{\tt arXiv:1303.6952}}].

\bibitem{Kang:2012zr}
Z.-B. Kang, S.~Mantry, and J.-W. Qiu, {\it {N-Jettiness as a Probe of Nuclear
  Dynamics}},  {\em Phys.Rev.} {\bf D86} (2012) 114011,
  [\href{http://xxx.lanl.gov/abs/1204.5469}{{\tt arXiv:1204.5469}}].

\bibitem{Kang:2013wca}
Z.-B. Kang, X.~Liu, S.~Mantry, and J.-W. Qiu, {\it {Probing nuclear dynamics in
  jet production with a global event shape}},  {\em Phys.Rev.} {\bf D88} (2013)
  074020, [\href{http://xxx.lanl.gov/abs/1303.3063}{{\tt arXiv:1303.3063}}].

\bibitem{Bauer:2000ew}
C.~W. Bauer, S.~Fleming, and M.~E. Luke, {\it Summing {S}udakov logarithms in
  {$ B\to X_s \gamma$} in effective field theory},  {\em Phys. Rev.} {\bf D63}
  (2000) 014006, [\href{http://xxx.lanl.gov/abs/hep-ph/0005275}{{\tt
  hep-ph/0005275}}].

\bibitem{Bauer:2000yr}
C.~W. Bauer, S.~Fleming, D.~Pirjol, and I.~W. Stewart, {\it An effective field
  theory for collinear and soft gluons: Heavy to light decays},  {\em Phys.
  Rev.} {\bf D63} (2001) 114020,
  [\href{http://xxx.lanl.gov/abs/hep-ph/0011336}{{\tt hep-ph/0011336}}].

\bibitem{Bauer:2001ct}
C.~W. Bauer and I.~W. Stewart, {\it Invariant operators in collinear effective
  theory},  {\em Phys. Lett.} {\bf B516} (2001) 134--142,
  [\href{http://xxx.lanl.gov/abs/hep-ph/0107001}{{\tt hep-ph/0107001}}].

\bibitem{Bauer:2001yt}
C.~W. Bauer, D.~Pirjol, and I.~W. Stewart, {\it Soft-collinear factorization in
  effective field theory},  {\em Phys. Rev.} {\bf D65} (2002) 054022,
  [\href{http://xxx.lanl.gov/abs/hep-ph/0109045}{{\tt hep-ph/0109045}}].

\bibitem{Bauer:2002nz}
C.~W. Bauer, S.~Fleming, D.~Pirjol, I.~Z. Rothstein, and I.~W. Stewart, {\it
  Hard scattering factorization from effective field theory},  {\em Phys. Rev.}
  {\bf D66} (2002) 014017, [\href{http://xxx.lanl.gov/abs/hep-ph/0202088}{{\tt
  hep-ph/0202088}}].

\bibitem{Kang:2013lga}
Z.-B. Kang, X.~Liu, and S.~Mantry, {\it {The 1-Jettiness DIS event shape: NNLL
  + NLO results}},  \href{http://xxx.lanl.gov/abs/1312.0301}{{\tt
  arXiv:1312.0301}}.

\bibitem{Accardi:2012hwp}
A.~Accardi, J.~Albacete, M.~Anselmino, N.~Armesto, E.~Aschenauer, et~al., {\it
  {Electron Ion Collider: The Next QCD Frontier - Understanding the glue that
  binds us all}},  \href{http://xxx.lanl.gov/abs/1212.1701}{{\tt
  arXiv:1212.1701}}.

\bibitem{Manohar:2003vb}
A.~V. Manohar, {\it Deep inelastic scattering as {$x \to 1$} using
  {Soft-Collinear Effective Theory}},  {\em Phys. Rev.} {\bf D68} (2003)
  114019, [\href{http://xxx.lanl.gov/abs/hep-ph/0309176}{{\tt
  hep-ph/0309176}}].

\bibitem{Chay:2005rz}
J.~Chay and C.~Kim, {\it {Deep inelastic scattering near the endpoint in
  soft-collinear effective theory}},  {\em Phys.Rev.} {\bf D75} (2007) 016003,
  [\href{http://xxx.lanl.gov/abs/hep-ph/0511066}{{\tt hep-ph/0511066}}].

\bibitem{Becher:2006mr}
T.~Becher, M.~Neubert, and B.~D. Pecjak, {\it Factorization and momentum-space
  resummation in deep-inelastic scattering},  {\em JHEP} {\bf 01} (2007) 076,
  [\href{http://xxx.lanl.gov/abs/hep-ph/0607228}{{\tt hep-ph/0607228}}].

\bibitem{Chen:2006vd}
P.-y. Chen, A.~Idilbi, and X.-d. Ji, {\it {QCD Factorization for Deep-Inelastic
  Scattering At Large Bjorken $x_B \sim 1 -
  \mathcal{O}(\Lambda_{\text{QCD}}/Q)$}},  {\em Nucl.Phys.} {\bf B763} (2007)
  183--197, [\href{http://xxx.lanl.gov/abs/hep-ph/0607003}{{\tt
  hep-ph/0607003}}].

\bibitem{Fleming:2012kb}
S.~Fleming and O.~Zhang, {\it {Rapidity Divergences and Deep Inelastic
  Scattering in the Endpoint Region}},
  \href{http://xxx.lanl.gov/abs/1210.1508}{{\tt arXiv:1210.1508}}.

\bibitem{Sveshnikov:1995vi}
N.~A. Sveshnikov and F.~V. Tkachov, {\it Jets and quantum field theory},  {\em
  Phys. Lett.} {\bf B382} (1996) 403--408,
  [\href{http://xxx.lanl.gov/abs/hep-ph/9512370}{{\tt hep-ph/9512370}}].

\bibitem{Cherzor:1997ak}
P.~S. Cherzor and N.~A. Sveshnikov, {\it Jet observables and energy-momentum
  tensor},  \href{http://xxx.lanl.gov/abs/hep-ph/9710349}{{\tt
  hep-ph/9710349}}.

\bibitem{Belitsky:2001ij}
A.~V. Belitsky, G.~P. Korchemsky, and G.~Sterman, {\it Energy flow in {QCD} and
  event shape functions},  {\em Phys. Lett.} {\bf B515} (2001) 297--307,
  [\href{http://xxx.lanl.gov/abs/hep-ph/0106308}{{\tt hep-ph/0106308}}].

\bibitem{Bauer:2008dt}
C.~W. Bauer, S.~Fleming, C.~Lee, and G.~Sterman, {\it Factorization of
  {$e^+e^-$} event shape distributions with hadronic final states in {Soft
  Collinear Effective Theory}},  {\em Phys. Rev.} {\bf D78} (2008) 034027,
  [\href{http://xxx.lanl.gov/abs/0801.4569}{{\tt arXiv:0801.4569}}].

\bibitem{Stewart:2010qs}
I.~W. Stewart, F.~J. Tackmann, and W.~J. Waalewijn, {\it {The Quark Beam
  Function at NNLL}},  {\em JHEP} {\bf 09} (2010) 005,
  [\href{http://xxx.lanl.gov/abs/1002.2213}{{\tt arXiv:1002.2213}}].

\bibitem{Ellis:1991qj}
R.~K. Ellis, W.~J. Stirling, and B.~Webber, {\it {QCD and collider physics}},
  {\em Camb.Monogr.Part.Phys.Nucl.Phys.Cosmol.} {\bf 8} (1996) 1--435.

\bibitem{Almeida:2014uva}
L.~G. Almeida, S.~D. Ellis, C.~Lee, G.~Sterman, I.~Sung, and J.~R. Walsh, {\it
  {Comparing and counting logs in direct and effective methods of QCD
  resummation}},  {\em JHEP} {\bf 1404} (2014) 174,
  [\href{http://xxx.lanl.gov/abs/1401.4460}{{\tt arXiv:1401.4460}}].

\bibitem{Gaunt:2014xga}
J.~R. Gaunt, M.~Stahlhofen, and F.~J. Tackmann, {\it {The Quark Beam Function
  at Two Loops}},  {\em JHEP} {\bf 1404} (2014) 113,
  [\href{http://xxx.lanl.gov/abs/1401.5478}{{\tt arXiv:1401.5478}}].

\bibitem{N3LL}
D.~Kang, C.~Lee, and I.~W. Stewart \emph{in preparation}, 2014.

\bibitem{Martin:2009iq}
A.~Martin, W.~Stirling, R.~Thorne, and G.~Watt, {\it {Parton distributions for
  the LHC}},  {\em Eur.Phys.J.} {\bf C63} (2009) 189--285,
  [\href{http://xxx.lanl.gov/abs/0901.0002}{{\tt arXiv:0901.0002}}].

\bibitem{Ligeti:2008ac}
Z.~Ligeti, I.~W. Stewart, and F.~J. Tackmann, {\it Treating the {$b$} quark
  distribution function with reliable uncertainties},  {\em Phys. Rev.} {\bf
  D78} (2008) 114014, [\href{http://xxx.lanl.gov/abs/0807.1926}{{\tt
  arXiv:0807.1926}}].

\bibitem{Berger:2010xi}
C.~F. Berger, C.~Marcantonini, I.~W. Stewart, F.~J. Tackmann, and W.~J.
  Waalewijn, {\it {Higgs Production with a Central Jet Veto at NNLL+NNLO}},
  {\em JHEP} {\bf 1104} (2011) 092,
  [\href{http://xxx.lanl.gov/abs/1012.4480}{{\tt arXiv:1012.4480}}].

\bibitem{Stewart:2014nna}
I.~W. Stewart, F.~J. Tackmann, and W.~J. Waalewijn, {\it {Dissecting Soft
  Radiation with Factorization}},
  \href{http://xxx.lanl.gov/abs/1405.6722}{{\tt arXiv:1405.6722}}.

\bibitem{Hoang:2007vb}
A.~H. Hoang and I.~W. Stewart, {\it Designing gapped soft functions for jet
  production},  {\em Phys. Lett.} {\bf B660} (2008) 483--493,
  [\href{http://xxx.lanl.gov/abs/0709.3519}{{\tt arXiv:0709.3519}}].

\bibitem{Sterman:1994ce}
G.~F. Sterman, {\em {An Introduction to quantum field theory}}.
\newblock Cambridge University Press, 1994.

\bibitem{Bauer:2003di}
C.~W. Bauer, C.~Lee, A.~V. Manohar, and M.~B. Wise, {\it Enhanced
  nonperturbative effects in {Z} decays to hadrons},  {\em Phys. Rev.} {\bf
  D70} (2004) 034014, [\href{http://xxx.lanl.gov/abs/hep-ph/0309278}{{\tt
  hep-ph/0309278}}].

\bibitem{Balzereit:1998yf}
C.~Balzereit, T.~Mannel, and W.~Kilian, {\it {Evolution of the light-cone
  distribution function for a heavy quark}},  {\em Phys. Rev.} {\bf D58} (1998)
  114029, [\href{http://xxx.lanl.gov/abs/hep-ph/9805297}{{\tt
  hep-ph/9805297}}].

\bibitem{Neubert:2004dd}
M.~Neubert, {\it {Renormalization-group improved calculation of the $B \to
  X_s\gamma$ branching ratio}},  {\em Eur.Phys.J.} {\bf C40} (2005) 165--186,
  [\href{http://xxx.lanl.gov/abs/hep-ph/0408179}{{\tt hep-ph/0408179}}].

\bibitem{Fleming:2007xt}
S.~Fleming, A.~H. Hoang, S.~Mantry, and I.~W. Stewart, {\it Top jets in the
  peak region: Factorization analysis with {NLL} resummation},  {\em Phys.
  Rev.} {\bf D77} (2008) 114003, [\href{http://xxx.lanl.gov/abs/0711.2079}{{\tt
  arXiv:0711.2079}}].

\bibitem{Korchemsky:1987wg}
G.~P. Korchemsky and A.~V. Radyushkin, {\it Renormalization of the {W}ilson
  loops beyond the leading order},  {\em Nucl. Phys.} {\bf B283} (1987)
  342--364.

\bibitem{Stewart:2009yx}
I.~W. Stewart, F.~J. Tackmann, and W.~J. Waalewijn, {\it {Factorization at the
  LHC: From PDFs to Initial State Jets}},  {\em Phys. Rev.} {\bf D81} (2010)
  094035, [\href{http://xxx.lanl.gov/abs/0910.0467}{{\tt arXiv:0910.0467}}].

\end{thebibliography}\endgroup

\end{document}